\documentclass[a4paper,11pt]{article}

\usepackage{cite}
\usepackage{amsmath}
\usepackage{amssymb}
\usepackage{bm}
\usepackage{color}
\usepackage{colortbl}

\usepackage{ulem}
\newcommand\mt{\mathfrak{t}}
\newcommand{\nin}{\hspace{0.15cm}\backslash\hspace{-0.35cm}\in\hspace{0.01cm}}
\newcommand{\ninnotdisplay}{\hspace{0.15cm}\backslash\hspace{-0.2cm}\in\hspace{0.05cm}}

\usepackage{url}
\usepackage[dvipdfmx]{graphicx}
\usepackage{tikz}

\numberwithin{equation}{section}

\usepackage{geometry}
\numberwithin{equation}{section}
\allowdisplaybreaks

\begin{document}

\newcommand\Tr{\mathrm{Tr}}

\newcommand{\aff}[1]{${}^{#1}$}
\renewcommand{\thefootnote}{\fnsymbol{footnote}}

\begin{titlepage}
\vskip1.5cm
\begin{center}
{
\Large\bf
Asymptotic degeneracies of M2-brane SCFTs
}\\
\bigskip\bigskip
\bigskip\bigskip
{\large Hirotaka Hayashi,\footnote{\tt h.hayashi@tokai.ac.jp}}\aff{1}
{\large Tomoki Nosaka\footnote{\tt nosaka@yukawa.kyoto-u.ac.jp}}\aff{2}
{\large and Tadashi Okazaki\footnote{\tt tokazaki@seu.edu.cn}}\aff{3}\\
\bigskip\bigskip
\aff{1} {\small
\it Department of Physics, School of Science, Tokai University, 4-1-1 Kitakaname, Hiratsuka-shi, Kanagawa 259-1292, Japan
}\\
\aff{2} {\small
\it Kavli Institute for Theoretical Sciences, University of Chinese Academy of Sciences, Beijing, 100190, China
}\\
\aff{3} {\small
\it Shing-Tung Yau Center of Southeast University, Yifu Architecture Building, No.2 Sipailou, Xuanwu district, Nanjing, Jiangsu, 210096, China
}\\

\bigskip
\end{center}
\bigskip
\bigskip
\begin{abstract}
We study the asymptotic growth of the degeneracy of the BPS local operators with scaling dimension $n/2$ in the three-dimensional superconformal field theories describing $N$ M2-branes. 
From the large $N$ supersymmetric indices we obtain the asymptotic formulas for degeneracies of the M2-brane SCFTs according to the Meinardus theorem. 
We observe an intriguing universal $n^{2/3}$ growth of the degeneracies in various theories of M2-brane SCFTs.
We also determine the coefficients of $n^{2/3}$ growth as well as further corrections in these theories explicitly.
\end{abstract}

\bigskip\bigskip\bigskip

\end{titlepage}

\setcounter{footnote}{0}
\renewcommand{\thefootnote}{$\dagger$\arabic{footnote}}

\tableofcontents

\section{Introduction and summary}
\label{sec_intro}
The asymptotic behavior of the degeneracy of the states has attracted significant attention 
in conformal field theories \cite{Cardy:1986ie,Kani:1989im,Cardy:1991kr,Pal:2020wwd,Lin:2022dhv}, 
superconformal field theories (SCFTs) \cite{Feng:2007ur,Dolan:2007rq,Lucietti:2008cv,Ramgoolam:2018epz,Murthy:2022tbj,Okazaki:2022sxo,Hatsuda:2022xdv}, 
the $T\overline{T}$ deformed theories \cite{Hartman:2018tkw,Datta:2018thy,LeFloch:2019wlf}, 
effective field theories \cite{Henning:2017fpj,Melia:2020pzd} 
and quantum field theories with a finite group global symmetry \cite{Harlow:2021trr,Cao:2021euf,Kang:2022orq}. 
In particular, the superconformal field theories which have holographically dual descriptions 
are expected to capture the state densities of $p$-branes in string theory/M-theory as well as the BPS black holes. 
The asymptotic degeneracy of the states is important in the study of their thermodynamical properties 
through a microscopic description of entropy by means of the counting of microstates. 

In this paper, we study the asymptotic degeneracies $d_n$ of states corresponding to the half-BPS local operators with scaling dimension $n/2$ in three-dimensional $\mathcal{N}\ge 4$ superconformal field theories 
describing a stack of $N$ coincident M2-branes in M-theory (M2-brane SCFTs). 
We analyze the large $N$ limit of three kinds of supersymmetric indices: Coulomb indices (a.k.a.~Coulomb branch Hilbert series), 
Higgs indices (a.k.a.~Higgs branch Hilbert series) and full indices (a.k.a.~superconformal indices) \cite{Bhattacharya:2008zy,Bhattacharya:2008bja,Kim:2009wb,Imamura:2011su,Kapustin:2011jm,Dimofte:2011py}, 
which play a role of counting functions for the Coulomb branch operators, the Higgs branch operators and the full half-BPS local operators respectively. 

The large $N$ indices can be expressed in terms of the infinite product of the form 
\begin{align}
\label{abcd_inf}
\prod_{n=1}^{\infty}
\frac{1}{(1-u^{\mathfrak{c}n+\mathfrak{d}})^{\mathfrak{a}n+\mathfrak{b}}}, 
\end{align}
where $u$ is some power of the fugacity counting the scaling dimension, and $\mathfrak{a}$, $\mathfrak{b}$, $\mathfrak{c}$ and $\mathfrak{d}$ are some real numbers. 
Applying the refined version of the Meinardus theorem in the presence of multiple poles in the associated Dirichlet series \cite{MR2958955}, 
we present an asymptotic formula for the coefficient of the infinite product (\ref{abcd_inf}). 
By means of the asymptotic formula and the convolution theorem of the infinite series, we derive asymptotic formulas for the degeneracies of states in the M2-brane SCFTs. 
The degeneracies turn out to have a common leading behavior with 
\begin{align}
\log d_n&\sim n^{2/3}. 
\end{align}

For the $U(N)_k\times U(N)_{-k}$ ABJM theory the limit where $N$ is large with $\lambda=N/k$ fixed corresponds to the 't Hooft limit 
so that the holographic dual configuration is compactified to type IIA string theory on $AdS_4$ $\times$ $\mathbb{CP}^3$ \cite{Aharony:2008ug}. 
We observe that the large $N$ supersymmetric index of the ABJM theory admits the large $k$ limit 
for which the asymptotic degeneracies of the large $N$ indices exhibit a string-like behavior
\begin{align}
\label{string_n12}
\log d_n&\sim n^{1/2}. 
\end{align}
Also the Coulomb indices of the M2-brane SCFTs parametrizing the probed geometries $\mathbb{C}^2/\mathbb{Z}_l$ and $\mathbb{C}^2/\widehat{D}_l$ permit the large $l$ limit, 
which again leads to the asymptotic behavior (\ref{string_n12}).\footnote{The similar limit for the ADHM theory in which $l$ is large while $N/l$ is kept is discussed for the three sphere partition function in \cite{Grassi:2014vwa}.}

The rest of the paper is organized as follows.
In section \ref{sec_multiplepoles1} we start by reviewing the Meinardus theorem \cite{MR62781,MR1634067} and its generalization \cite{MR2958955}. 
We derive a convolution theorem and a new asymptotic formula for the coefficient $d_n$ 
of the infinite series (\ref{abcd_inf}). 
By means of the theorems, we study in section \ref{sec_largeN} the asymptotic degeneracies of the BPS local operators 
with large scaling dimensions from the large $N$ limit of the Coulomb indices, Higgs indices and full indices of the M2-brane SCFTs. 
Analytic expressions of the degeneracies with universal $n^{2/3}$ leading behavior are obtained together with the subleading corrections in $n$.
Finally, in section \ref{sec_semi1} we discuss potential physical interpretations of the leading $n^{2/3}$ behavior, 
including the asymptotic degeneracies of a membrane obtained from the semiclassical quantization and the resulting thermodynamic functions \cite{Fubini:1972mf, Dethlefsen:1974dr,Strumia:1975rd,Alvarez:1991qs,Harms:1992jt}.
In appendix \ref{app_HZformulas} we list the formulas for Hurwitz zeta function $\zeta(s,a)$ which we use to simplify the expressions for asymptotic degeneracies.
In appendix \ref{app_numerical} we list more examples for the supersymmetric indices and their asymptotic degeneracies supplementing those in section \ref{sec_largeN}.

\section{Meinardus theorem and generalizations}
\label{sec_multiplepoles1}

Before proceeding, we present several formulas associated with the Meinardus theorem \cite{MR62781,MR1634067} which allow us to analyze the asymptotic behaviors of the supersymmetric indices. 


\subsection{Asymptotic degeneracy by Meinardus theorem}
\label{sec_semi3}

Given an infinite product of the form 
\begin{align}
f(u)&=\prod_{n=1}^{\infty} \frac{1}{(1-u^n)^{b_n}}=
\sum_{n=0}^{\infty} d_n u^n,
\end{align}
where $u=e^{-\beta}$ and $\mathrm{Re}[\beta]>0$, we introduce an auxiliary Dirichlet series 
\begin{align}
D(s)&=\sum_{n=1}^{\infty}\frac{b_n}{n^s}.\label{D_series}
\end{align}
Assume that ($s=\sigma+i\tau$)
\begin{enumerate}
\item $D(s)$ is analytic except for a pole of order $1$ at $s=s_0$ with residue $R_{0}$,
\item $D(s)$ converges for $\sigma>s_0$, a positive number and has an analytic continuation in the region $\sigma\ge -C_0$ with $0<C_0<1$, 
\item $D(s)\rightarrow\mathcal{O}(|\tau|^{C_1})$ as $|\tau|\rightarrow \infty$ in $\sigma\ge -C_0$ for a fixed positive number $C_1$. 
\end{enumerate}
Then we have 
\begin{align}
d_n&\sim 
Cn^{\kappa} \exp 
\left[
n^{\frac{s_0}{s_0+1}}
\left(
1+\frac{1}{s_0} 
\right)
\left(
R_{0}\Gamma(s_0+1) \zeta(s_0+1)
\right)^{\frac{1}{s_0+1}}
\right]& 
&\textrm{as $n\rightarrow \infty$},
\label{singlepoleMein}
\end{align}
where 
\begin{align}
C&=e^{D'(0)}\left[2\pi(1+s_0) \right]^{-\frac12}
\left[
R_{0}\Gamma(s_0+1)\zeta(s_0+1)
\right]^{(1-2D(0))/(2+2s_0)}
\end{align}
and 
\begin{align}
\kappa&=\frac{D(0)-1-\frac{s_0}{2}}{1+s_0}. 
\end{align}
The proof of the theorem stems from the saddle point method for 
\begin{align}
d_n=\oint\frac{du}{2\pi iu^{n+1}}f(u)
=\int_{-\pi i}^{\pi i}\frac{d\beta}{2\pi i} e^{n\beta+\log f(e^{-\beta})}
\end{align}
at $n\rightarrow\infty$, with $\log f(e^{-\beta})$ evaluated by the rewriting via the Mellin transform of the Gamma function
\begin{align}
e^{-x}&=\frac{1}{2\pi i}
\int_{y-i\infty}^{y+i\infty}
x^{-s}\Gamma(s)ds
\end{align}
as
\begin{align}
\log f(e^{-\beta})
&=\frac{1}{2\pi i}\int_{y-i\infty}^{y+i\infty} 
\beta^{-s}\Gamma(s)\zeta(s+1)D(s)ds,
\label{logfMellinD}
\end{align}
where $y$ is any positive number and $D(s)$ is the Dirichlet series \eqref{D_series}, and plugging \eqref{logfMellinD} into the Cauchy's integral.
For the detail, see \cite{MR1634067}.

\subsection{Multiple poles}
\label{sec_multiplepoles2}
When the auxiliary Dirichlet series (\ref{D_series}) has additional poles at $s=s_i$ ($<s_0$), the subleading terms generally appear. 
The generalized Meinardus theorem that is applicable for multiple poles is presented in \cite{MR2958955}. 
Applying the generalized Meinardus theorem to an infinite product of the form
\begin{align}
\prod_{n=1}^{\infty} \frac{1}{(1-u^n)^{\mathfrak{a}n+\mathfrak{b}}}
&=\sum_{n\ge 0}d_n(\mathfrak{a},\mathfrak{b}) u^n,\label{Meinardus1}
\end{align}
whose Dirichlet series $D(s)$ has poles at $s=1,2$, we find the asymptotic behavior of the coefficient
\begin{align}
d_n(\mathfrak{a},\mathfrak{b})&\sim 
d_n^{\textrm{asymp}}(\mathfrak{a},\mathfrak{b})=
C(\mathfrak{a},\mathfrak{b}) n^{\kappa(\mathfrak{a},\mathfrak{b})} \exp\left[
\alpha(\mathfrak{a}) n^{\frac{2}{3}}+\beta(\mathfrak{a},\mathfrak{b}) n^{\frac{1}{3}}+\gamma(\mathfrak{a},\mathfrak{b})
\right],\label{Meinardus_thm}
\end{align}
where 
\begin{align}
\alpha(\mathfrak{a})&=\frac{3 \left( \mathfrak{a}\zeta(3) \right)^{\frac13}}{2^{\frac23}}, \\
\beta(\mathfrak{a},\mathfrak{b})&=\frac{\mathfrak{b}\pi^2}{3\cdot 2^{\frac43}(\mathfrak{a}\zeta(3))^{\frac13}}, \\
\gamma(\mathfrak{a},\mathfrak{b})&=\frac{\mathfrak{a}}{12}-\frac{\mathfrak{b}^2\pi^4}{432\mathfrak{a}\zeta(3)}, \\
C(\mathfrak{a},\mathfrak{b})&=
\frac{(\mathfrak{a}\zeta(3))^{\frac{\mathfrak{a}}{36}+\frac{\mathfrak{b}}{6}+\frac16}}{A^{\mathfrak{a}}2^{-\frac{\mathfrak{a}}{36}+\frac{\mathfrak{b}}{3}+\frac13 }3^{\frac12}\pi^{\frac{\mathfrak{b}+1}{2}}}, \\
\kappa(\mathfrak{a},\mathfrak{b})&=-\frac{\mathfrak{a}}{36}-\frac{\mathfrak{b}}{6}-\frac23.
\end{align}
Here $A=1.2824271\cdots$ is the Glaisher-Kinkelin constant.

More generally, we consider an infinite product of the form
\begin{align}
\label{inf_prod_abcd}
\prod_{n=1}^{\infty}\frac{1}{(1-u^{\mathfrak{c}n+\mathfrak{d}})^{\mathfrak{a}n+\mathfrak{b}}}
&=\sum_{n\ge0} d_n(\mathfrak{a},\mathfrak{b},\mathfrak{c},\mathfrak{d}) u^n,
\end{align}
for which the associated Dirichlet series $D(s)$ is given by the Hurwitz zeta functions $\zeta(s,a)$ (see appendix \ref{app_HZformulas} for the definition) as
\begin{align}
D(s)=\sum_{n=1}^\infty\frac{\mathfrak{a}n+\mathfrak{b}}{(\mathfrak{c}n+\mathfrak{d})^s}=\frac{1}{\mathfrak{c}^s}\Bigl[
\mathfrak{a}\zeta\Bigl(s-1,1+\frac{\mathfrak{d}}{\mathfrak{c}}\Bigr)
+\Bigl(\mathfrak{b}-\frac{\mathfrak{a}\mathfrak{d}}{\mathfrak{c}}\Bigr)\zeta\Bigl(s,1+\frac{\mathfrak{d}}{\mathfrak{c}}\Bigr)\Bigr].
\end{align}
According to the generalized Meinardus theorem \cite{MR2958955}, we obtain the asymptotic coefficient as $n\rightarrow \infty$\footnote{
Note that the infinite product (\ref{inf_prod_abcd}) can be written as 
\begin{align}
\prod_{n\in \mathbb{Z}_{\ge 1}+\frac{\mathfrak{d}}{\mathfrak{c}}}
\frac{1}{(1-u^{\mathfrak{c}n})^{\mathfrak{a}n+\mathfrak{b}-\frac{\mathfrak{a}\mathfrak{d}}{\mathfrak{c}}}},
\end{align}
hence we can obtain a rough estimate for the asymptotics of $d_n(\mathfrak{a},\mathfrak{b},\mathfrak{c},\mathfrak{d})$ from this expression by replacing $\mathbb{Z}_{\ge 1}+\frac{\mathfrak{d}}{\mathfrak{c}}$ with $\mathbb{Z}_{\ge 1}$ and applying the formula \eqref{Meinardus_thm}, as
\begin{align}
\label{asy_coeffsub}
d_{\frac{n}{\mathfrak{c}}}^{\textrm{asymp}}\left(\mathfrak{a},\mathfrak{b}-\frac{\mathfrak{a}\mathfrak{d}}{\mathfrak{c}}\right).
\end{align}
This reproduces the leading coefficient (\ref{alpha1}) and subleading coefficient (\ref{beta1}) correctly.
}
\begin{align}
&d_n(\mathfrak{a},\mathfrak{b},\mathfrak{c},\mathfrak{d})
\nonumber\\
&=C(\mathfrak{a},\mathfrak{b},\mathfrak{c},\mathfrak{d}) n^{\kappa(\mathfrak{a},\mathfrak{b},\mathfrak{c},\mathfrak{d})}
\exp\left[
\alpha(\mathfrak{a},\mathfrak{c})n^{\frac23}
+\beta(\mathfrak{a},\mathfrak{b},\mathfrak{c},\mathfrak{d})n^{\frac13}
+\gamma(\mathfrak{a},\mathfrak{b},\mathfrak{c},\mathfrak{d})
\right],\label{asy_coeff0}
\end{align}
with
\begin{align}
\alpha(\mathfrak{a},\mathfrak{c})&=\frac{3 (\mathfrak{a}\zeta(3))^{\frac13}}{(2\mathfrak{c})^{\frac23}},\label{alpha1} \\
\beta(\mathfrak{a},\mathfrak{b},\mathfrak{c},\mathfrak{d})&=\frac{(\mathfrak{b}\mathfrak{c}-\mathfrak{a}\mathfrak{d}) \pi^2}{6\mathfrak{a}^{\frac13}\mathfrak{c}^{\frac43} (2\zeta(3))^{\frac13}},\label{beta1} \\
\gamma(\mathfrak{a},\mathfrak{b},\mathfrak{c},\mathfrak{d})&=
\mathfrak{a}\zeta' \Bigl(-1,\frac{\mathfrak{c}+\mathfrak{d}}{\mathfrak{c}}\Bigr)
-\frac{\pi ^4 (\mathfrak{b}\mathfrak{c}-\mathfrak{a}\mathfrak{d})^2}{432\mathfrak{a}\mathfrak{c}^2 \zeta (3)},\label{gamma1} \\
C(\mathfrak{a},\mathfrak{b},\mathfrak{c},\mathfrak{d})&=\frac{\text{GCD}(\mathfrak{c},\mathfrak{d})}{(6\pi \mathfrak{c})^{1/2}}(2\mathfrak{a}\mathfrak{c}\zeta(3))^{\frac{1}{36}(-\frac{6\mathfrak{a}\mathfrak{d}^2}{\mathfrak{c}^2}+\mathfrak{a}+6(\frac{2\mathfrak{b}\mathfrak{d}}{\mathfrak{c}}+\mathfrak{b}+1))}\Bigl(\frac{\sqrt{2\pi}}{\Gamma(1+\frac{\mathfrak{d}}{\mathfrak{c}})}\Bigr)^{\frac{\mathfrak{a}\mathfrak{d}-\mathfrak{b}\mathfrak{c}}{\mathfrak{c}}},\label{C1} \\
\kappa(\mathfrak{a},\mathfrak{b},\mathfrak{c},\mathfrak{d})&=
\frac{1}{36} \left(\frac{6 \mathfrak{a} \mathfrak{d}^2}{\mathfrak{c}^2}-\mathfrak{a}-6 \left(\frac{2 \mathfrak{b} \mathfrak{d}}{\mathfrak{c}}+\mathfrak{b}+4\right)\right).\label{kappa1}
\end{align}
Here $\zeta'(s,a)$ $:=$ $(\partial/\partial s)\zeta(s,a)$.

\subsection{Convolution theorem}
\label{sec_convolution}
Suppose
\begin{align}
f_1(u)=\sum_{n\in p_1\mathbb{Z}_{\ge 0}} d_n^{(1)} u^n,\quad 
f_2(u)=\sum_{n\in p_2\mathbb{Z}_{\ge 0}} d_n^{(2)} u^n
\label{f1f2}
\end{align}
have the asymptotic behaviors of the coefficients $d_n^{(1)}$ and $d_n^{(2)}$
\begin{align}
d_n^{(1)}\sim C_1 n^{\kappa_1} \exp\left[
\alpha_1 n^{\frac23} +\beta_1 n^{\frac13} +\gamma_1
\right],\quad
d_n^{(2)}\sim C_2 n^{\kappa_2} \exp\left[
\alpha_2 n^{\frac23} +\beta_2 n^{\frac13} +\gamma_2
\right].
\label{d1d2asym}
\end{align}
Here $p_1$ and $p_2$ are integers.
Let us also denote $\text{GCD}(p_1,p_2)$ by $P$.
Then the asymptotic behavior of the coefficients of $f_1(u)f_2(u)$
\begin{align}
f_1(u)f_2(u)
=\sum_{n\in P\mathbb{Z}_{\ge 0}}d_n u^n
\end{align}
is given as 
\begin{align}
d_n\sim C n^\kappa e^{\alpha n^{2/3}+\beta n^{1/3}+\gamma},
\label{conv_thm}
\end{align}
with
\begin{align}
&\alpha=(\alpha_1^3+\alpha_2^3)^{\frac{1}{3}},\label{conv_thm_alpha} \\
&\beta=\frac{\alpha_1\beta_1+\alpha_2\beta_2}{(\alpha_1^3+\alpha_2^3)^{\frac{1}{3}}},\\
&\gamma=\frac{(\alpha_2^2\beta_1-\alpha_1^2\beta_2)^2}{4\alpha_1\alpha_2(\alpha_1^3+\alpha_2^3)}+\gamma_1+\gamma_2,\\
&\kappa=\kappa_1+\kappa_2+\frac{2}{3},\\
&C=\frac{3P}{p_1p_2}C_1C_2\pi^{\frac{1}{2}}(\alpha_1^3+\alpha_2^3)^{-\kappa_1-\kappa_2-\frac{7}{6}}\alpha_1^{3\kappa_1+\frac{3}{2}}\alpha_2^{3\kappa_2+\frac{3}{2}}.\label{conv_thm_C}
\end{align}

To obtain the convolution formula \eqref{conv_thm}, first notice that $d_n$ is given by $d_n^{(1)}$ and $d_n^{(2)}$ as
\begin{align}
d_n=\sum_{j=0}^nd_{n-j}^{(1)}d_j^{(2)}.
\label{dn}
\end{align}
Then the asymptotics of $d_n$ can be obtained by the saddle point approximation for $j$, which goes as follows.
Let us assume that the summation \eqref{dn} is dominated by the contributions from $j$ such that $j\gg 1$ and $n-j\gg 1$, and write \eqref{dn} using the asymptotics of $d_n^{(1)}$ and $d_n^{(2)}$ \eqref{d1d2asym} as
\begin{align}
d_n\sim C_1C_2n^{\kappa_1+\kappa_2}e^{\gamma_1+\gamma_2}\sum_je^{g(m)},
\label{d1*d2largen}
\end{align}
where $m=\frac{j}{n}$ and
\begin{align}
g(m)=g_1(m)n^{\frac{2}{3}}+g_2(m)n^{\frac{1}{3}}+g_3(m),
\end{align}
with
\begin{align}
&g_1(m)=\alpha_1(1-m)^{\frac{2}{3}}+\alpha_2m^{\frac{2}{3}},\quad
g_2(m)=\beta_1(1-m)^{\frac{1}{3}}+\beta_2m^{\frac{1}{3}},\nonumber \\
&g_3(m)=\kappa_1\log(1-m)+\kappa_2\log m.
\end{align}
Now we apply the saddle point approximation to \eqref{d1*d2largen}: denote the saddle $m_*$ such that $\frac{dg}{dm}|_{m_*}=0$, write $j$ as $j=nm_*+x$ and expand $g(m)$ to the second order in $x$.
As a result we obtain
\begin{align}
d_n&\sim \frac{P^2}{p_1p_2}\int_{-\infty}^\infty dx C_1C_2n^{\kappa_1+\kappa_2}e^{\gamma_1+\gamma_2}e^{g(m_*)
+\frac{1}{2}[\frac{d^2g}{dm^2}|_{m_*}]n^{-2}x^2
}.
\label{Gaussian}
\end{align}
Here we have approximated the discrete summation over $x$ with the continuous integration.
The overall factor $\frac{P^2}{p_1p_2}$ comes from the sparsity of the summation \eqref{dn} due to $d^{(1)}_{n-j\ninnotdisplay p_1\mathbb{Z}_{\ge 0}}=0$ and $d^{(2)}_{j\ninnotdisplay p_2\mathbb{Z}_{\ge 0}}=0$.
Performing the Gaussian integration \eqref{Gaussian} we obtain
\begin{align}
d_n\sim \frac{P^2}{p_1p_2}C_1C_2n^{\kappa_1+\kappa_2}e^{\gamma_1+\gamma_2}
\biggl(\frac{2\pi n^2}{-[\frac{d^2g}{dm^2}|_{m_*}]}\biggr)^{\frac{1}{2}}e^{g(m_*)}.
\label{saddlefinal}
\end{align}
To find the asymptotics of $d_n$ in the form of \eqref{conv_thm} it is sufficient to calculate $g(m_*)$ to the order $n^0$ (i.e.~ignore all terms with negative power of $n$) and $\frac{d^2g}{dm^2}|_{m_*}$ only at its leading order in the large $n$ limit.
For the first ingredient, since $g(m)\sim n^{2/3}$ it is sufficient to know the large $n$ expansion of $m_*$ only to the order $n^{-2/3}$.
For the second ingredient we only need the leading part of $m_*$ in the large $n$ limit.
After all, we only need $m_*$ in the large $n$ expansion to the order $n^{-2/3}$, which we can calculate straightforwardly by substituting $m_*=m_*^{(0)}+m_*^{(1)}n^{-\frac{1}{3}}+m_*^{(2)}n^{-\frac{2}{3}}$ to $\frac{dg}{dm}|_{m_*}=0$ and solving the equation order by order in $n$, as
\begin{align}
&m_*=m_*^{(0)}+m_*^{(1)}n^{-\frac{1}{3}}+m_*^{(2)}n^{-\frac{2}{3}}+\cdots,\nonumber \\
&m_*^{(0)}=\frac{\alpha_2^3}{\alpha_1^3+\alpha_2^3},\quad
m_*^{(1)}=\frac{3\alpha_1\alpha_2(\alpha_1^2\beta_2-\alpha_2^2\beta_1)}{2(\alpha_1^3+\alpha_2^3)^{\frac{5}{3}}},\nonumber \\
&m_*^{(2)}=\frac{3(\alpha_2^2\alpha_2^3\beta_1^2-\alpha_1\alpha_2(\alpha_1^3-\alpha_2^3)\beta_1\beta_2-\alpha_1^3\alpha_2^2\beta_2^2)}{2(\alpha_1^3+\alpha_2^3)^{\frac{7}{3}}}
-\frac{9(\alpha_2^3\kappa_1-\alpha_1^3\kappa_2)}{2(\alpha_1^3+\alpha_2^3)^{\frac{4}{3}}}.
\end{align}
With these results, we find $g(m_*)$ and $\frac{d^2g}{dm^2}|_{m_*}$ as
\begin{align}
&g(m_*)=(\alpha_1^3+\alpha_2^3)^{\frac{1}{3}}n^{\frac{2}{3}}+\frac{\alpha_1\beta_1+\alpha_2\beta_2}{(\alpha_1^3+\alpha_2^3)^{\frac{1}{3}}}n^{\frac{1}{3}}+\frac{(\alpha_2^2\beta_1-\alpha_1^2\beta_2)^2}{4\alpha_1\alpha_2(\alpha_1^3+\alpha_2^3)}+\kappa_1\log\frac{\alpha_1^3}{\alpha_1^3+\alpha_2^3}\nonumber \\
&\quad\quad\quad\quad +\kappa_2\log\frac{\alpha_2^3}{\alpha_1^3+\alpha_2^3}+\cdots,\nonumber \\
&\frac{d^2g}{dm^2}\Bigr|_{m_*}=-\frac{2\alpha_1^{-3}\alpha_2^{-3}(\alpha_1^3+\alpha_2^3)^{\frac{7}{3}}}{9}n^{\frac{2}{3}}+\cdots.
\end{align}
Substituting these into \eqref{saddlefinal} we finally obtain \eqref{conv_thm} with \eqref{conv_thm_alpha}-\eqref{conv_thm_C}.

The convolution formula \eqref{conv_thm} can also be generalized to the cases with more than two components ($i=1,2,\cdots,L$)
\begin{align}
f_i(u)=\sum_{n\in p_i\mathbb{Z}_{\ge 0}}d_n^{(i)}u^n,\quad 
d_n^{(i)}
\sim 
C_in^{\kappa_i}\exp\left[
\alpha_i n^{\frac23}+\beta_i n^{\frac13}+\gamma_i
\right],
\end{align}
as
\begin{align}
\prod_{i=1}^{L} f_i(u)
=\sum_{n\in P\mathbb{Z}_{\ge 0}}d_n u^n,\quad
d_n\sim 
C n^{\kappa}\exp
\left[
\alpha n^{\frac23}+\beta n^{\frac13}+\gamma
\right],\label{Lconv}
\end{align}
with $P=\text{GCD}(p_1,\cdots,p_L)$ and
\begin{align}
\label{coeff_repeat1}
\alpha&=\left(
\sum_{i=1}^{L} \alpha_i^3
\right)^{\frac13},\\
\label{coeff_repeat2}
\beta&=
\frac{
\sum_{i=1}^{L}\alpha_i \beta_i
}
{
\left(
\sum_{i=1}^{L} \alpha_i^3
\right)^{\frac13}
}, \\
\label{coeff_repeat3}
\gamma&=\sum_{i=1}^L\Bigl(\gamma_i+\frac{\beta_i^2}{4\alpha_i}\Bigr)-\frac{(\sum_{i=1}^L\alpha_i\beta_i)^2}{4\sum_{i=1}^L\alpha_i^3},\\
\label{coeff_repeat4}
C
&=
\frac{P^L}{\prod_{i=1}^Lp_i}
\cdot 3^{L-1} \pi^{\frac{L-1}{2}}
\frac{\prod_{i=1}^L C_i \alpha_i^{3\kappa_i+\frac32}}
{
\left(
\sum_{i=1}^L \alpha_i^3
\right)^{\sum_{i=1}^{L}\kappa_i+\frac23L-\frac16}
}, 
\\
\label{coeff_repeat5}
\kappa&=\sum_{i=1}^{L}\kappa_i +\frac{2(L-1)}{3}. 
\end{align}
Note that the resulting coefficients (\ref{coeff_repeat1})-(\ref{coeff_repeat5}) are independent of the ordering of the convolutions.
Also note that although in general $\gamma_i$ contains $\zeta'(-1,a)$ with $a\nin \mathbb{N}$, in all of the examples we study in section \ref{sec_largeN} and in appendix \ref{app_numerical} we observe that $\zeta'(-1,a)$ in the final expression for $\gamma$ after the convolution formula always appear in a summation over $a$ such that they can be rewritten by using the formulas in appendix \ref{app_HZformulas} without using Hurwitz zeta functions explicitly.

\subsection{Examples}
Let us close this section by providing several examples which confirm the validity of the new formulas \eqref{asy_coeff0} and \eqref{conv_thm}.
First, if we apply the convolution \eqref{conv_thm} to
\begin{align}
f_1(u)&=\prod_{n=1}^{\infty} \frac{1}{(1-u^{2n})^{2n}}, \qquad
f_2(u)=\prod_{n=1}^{\infty} \frac{1}{(1-u^{2n-1})^{2n-1}}, 
\end{align}
we find the asymptotics of the coefficients in $f_1(u)f_2(u)=\sum_{n=1}^\infty d_nu^n$ as
\begin{align}
d_n\sim \frac{3^{\frac{1}{2}}}{2^{\frac{2}{9}}\cdot \pi^{\frac{1}{2}}}\zeta(3)^{\frac{7}{36}}e^{\frac{3(\zeta(3))^{\frac{1}{3}}}{2^{\frac{2}{3}}}n^{\frac{2}{3}}+2(\zeta'(-1,\frac{1}{2})+\zeta'(-1,1))}.
\end{align}
Evaluating $\zeta'(-1,1)$ and $\zeta'(-1,\frac{1}{2})$ by the formulas \eqref{HZrelation1} and \eqref{HZformula2}, we obtain
\begin{align}
d_n\sim 
\frac{\zeta(3)^{\frac{7}{36}}
\exp\left[
\frac{3 \zeta(3)^{\frac13}}{2^{\frac23}}n^{\frac23}
+\frac{1}{12}
\right]}
{2^{\frac{11}{36}}3^{\frac12} \pi A n^{\frac{25}{36}}}. 
\label{f1f2_ex1a}
\end{align}
On the other hand, since we have 
\begin{align}
f_1(u)f_2(u)=
\prod_{n=1}^{\infty}
\frac{1}{(1-u^n)^n},\label{pp_gene}
\end{align}
we can also obtain the asymptotics of $d_n$ directly by the formula \eqref{asy_coeff0} as
\begin{align}
\label{f1f2_ex1b}
d_n&\sim 
\frac{\zeta(3)^{\frac{7}{36}}
\exp\left[
\frac{3 \zeta(3)^{\frac13}}{2^{\frac23}}n^{\frac23}
+\frac{1}{12}
\right]}
{2^{\frac{11}{36}}3^{\frac12} \pi A n^{\frac{25}{36}}},
\end{align}
which is the asymptotic behavior for the plane partition of $n$ \cite{Wright1931ASYMPTOTICPF}. 
We see that \eqref{f1f2_ex1a} and \eqref{f1f2_ex1b} perfectly agree with each other. 
Similarly, we can find the asymptotic degeneracy of \eqref{pp_gene} by applying $(L-1)$ convolutions to
\begin{align}
\label{f1_fk}
f_{1}(u) \cdots f_{L}(u)=\prod_{n=1}^\infty\frac{1}{(1-u^n)^n},\quad
f_k(u)&=\prod_{n=1}^{\infty}\frac{1}{(1-u^{Ln-k+1})^{Ln-k+1}}.
\end{align}
In this case, $\gamma$ obtained after the convolution formula involves the following derivatives of Hurwitz zeta function
\begin{align}
\gamma=L\sum_{k=1}^L\zeta'\Bigl(-1,\frac{L-k+1}{L}\Bigr)+\cdots,
\end{align}
which can be simplified by using the formula \eqref{HZrelation3}.
We finally obtain the following asymptotics of the coefficient $d_n$
\begin{align}
d_n&\sim 
\frac{\zeta(3)^{\frac{7}{36}}
\exp\left[
\frac{3 \zeta(3)^{\frac13}}{2^{\frac23}}n^{\frac23}
+\frac{1}{12}
\right]}
{2^{\frac{11}{36}}3^{\frac12} \pi A n^{\frac{25}{36}}}, 
\end{align}
which again coincides with (\ref{f1f2_ex1b}). 

As another example, let us consider $f_1(u)f_2(u)$ with
\begin{align}
f_1(u)=\prod_{n=1}^\infty\frac{1}{(1-u^{2n})^n},\quad
f_2(u)=\prod_{n=1}^\infty\frac{1}{(1-u^{3n})^n}.
\end{align}
By applying the formulas \eqref{asy_coeff0} and \eqref{conv_thm} we obtain the asymptotics of the coefficients of $f_1(u)f_2(u)=\sum_{n=1}^\infty d_nu^n$ as
\begin{align}
&d_n\sim d_n^{\text{asym}}=Cn^\kappa e^{\alpha n^{2/3}+\beta n^{1/3}+\gamma},\nonumber \\
&\alpha=\frac{(39\zeta(3))^{\frac{1}{3}}}{2^{\frac{4}{3}}},\quad
\beta=0,\quad
\gamma=\frac{1}{6},\quad
C=\frac{1}{6}\times \frac{2^{\frac{13}{36}}\cdot 3^{\frac{5}{36}}\cdot (13\zeta(3))^{\frac{2}{9}}}{A^2\pi^{\frac{1}{2}}},\quad
\kappa=-\frac{13}{18}.
\end{align}
This asymptotics excellently agrees with the exact values of $d_n$:
\begin{align}
\begin{array}{c|c|c|c} 
n&d_n                       &d_n^{\text{asym}}     &d_n/d_n^{\text{asym}} \\ \hline 
10  &33                     &27.0501               &1.21996\\
100 &1.66783\times 10^{11}  &1.63756\times 10^{11} &1.01848\\
1000&1.76484\times 10^{59}  &1.77171\times 10^{59} &0.996122\\
10000&6.80833\times 10^{284}&6.81402\times 10^{284}&0.999165
\end{array}.
\end{align}

These calculations strongly support the validity of the asymptotic formula (\ref{asy_coeff0}) and the convolution theorem (\ref{conv_thm}). 

In section \ref{sec_largeN} we analytically derive the asymptotic degeneracies for the large $N$ indices of the M2-brane SCFTs by making use of the asymptotic formula (\ref{asy_coeff0}) and the convolution theorem (\ref{conv_thm}). 

\section{Large $N$ indices}
\label{sec_largeN}

\subsection{Coulomb indices}

The Coulomb index (a.k.a. Hilbert series for the Coulomb branch) counts the number $d_n$ of the Coulomb branch operators with the scaling dimension $n/2$ in 3d $\mathcal{N}\ge 4$ sueprsymmetric gauge theory. 
For the M2-brane SCFTs the Coulomb index encodes the geometry $\mathbb{C}^2/\Gamma$ probed by M2-branes 
where $\Gamma$ is a discrete subgroup of $SU(2)$. 
Let ${\cal I}^{(C)}_N(\mathfrak{t})$ be the Coulomb index of the theories of $N$ M2-branes.
The Coulomb index at $N\rightarrow\infty$, ${\cal I}^{(C)}_{N=\infty}(\mathfrak{t})$, can be obtained from ${\cal I}^{(C)}_{N=1}(\mathfrak{t})$ by using the following formula \cite{Benvenuti:2006qr,Feng:2007ur}:
\begin{align}
{\cal I}_{N=\infty}^{(C)}(z;\mathfrak{t})=\text{PE}[{\cal I}_{N=1}(z;\mathfrak{t})-1],
\label{BHF}
\end{align}
where $z$ collectively denotes the extra fugacities and PE stands for the plethystic expontential given by
\begin{align}
\text{PE}[f(z;\mt)]=\exp\left(\sum_{n=1}^{\infty}\frac{1}{n}f(z^n;\mt^n)\right).
\end{align}
In the following we set $z=1$ for simplicity.
Note that ${\cal I}_{N=1}^{(C)}(z=1)$ for $\Gamma={\widehat A},{\widehat D},{\widehat E}$ are also given in \cite[eq(3.9)]{Benvenuti:2006qr}.

For $\mathbb{C}^2/\mathbb{Z}_l$, the formula \eqref{BHF} can also be obtained from the index of $U(N)$ ADHM theory by showing that the grand canonical index
\begin{align}
\Xi(\mu;\mathfrak{t})=\sum_{N=0}^\infty \mu^N{\cal I}^{(C)}_N(\mathfrak{t}),
\label{XiC}
\end{align}
takes the following form \cite{Hayashi:2022ldo}:
\begin{align}
&\Xi(\mu;\mathfrak{t})=\frac{1}{1-\mu}\prod_{n=1}^\infty
\frac{1}{(1-\mu\mathfrak{t}^{cn})^{b_n}},
\label{XiCgeneralstructure}
\end{align}
with $c$ some positive rational number and $b_n$ some non-negative integers.
That is,
\begin{itemize}
\item [(i)] there is an overall factor $\frac{1}{1-\mu}$,
\item [(ii)] the power of $\mu$ is $1$ for all factors.
\end{itemize}
Note that the first property guarantees that the small $\mathfrak{t}$ expansion of ${\cal I}_{N=1}^{(C)}(\mathfrak{t})$ starts with $1$ for all $N$.
From \eqref{XiCgeneralstructure}, we obtain ${\cal I}_{N=1}^{(C)}(\mathfrak{t})$ by expanding the right-hand side and collecting the terms of order $\mu$ as
\begin{align}
{\cal I}_{N=1}^{(C)}(\mathfrak{t})=1+\sum_{n=1}^\infty b_n\mathfrak{t}^{cn}.
\end{align}
On the other hand, we can obtain a formal expansion for ${\cal I}_{N}^{(C)}$ by evaluating the inversion formula of \eqref{XiC}
\begin{align}
{\cal I}_N^{(C)}(\mathfrak{t})=\oint_{|\mu|<1}\frac{d\mu}{2\pi i\mu}\mu^{-N}\Xi(\mu;\mathfrak{t}),
\end{align}
by assuming $|\mathfrak{t}|<1$ and picking the poles in $|\mu|\ge 1$, as
\begin{align}
{\cal I}_N^{(C)}(\mathfrak{t})=\prod_{n=1}^\infty\frac{1}{(1-\mathfrak{t}^{cn})^{b_n}}+\sum_{n=1}^\infty \text{Res}\Bigl[\frac{1}{2\pi i\mu}\mu^{-N}\Xi(\mu;\mathfrak{t}),\mu\rightarrow \mathfrak{t}^{-cn}\Bigr].
\end{align}
Here the first term is the contribution from the pole at $\mu=1$.
Note that the second terms are proportional to $\mathfrak{t}^{cN}$.
Since $c>0$ and $|\mathfrak{t}|<1$, these contributions can be ignored in the large $N$ limit.\footnote{
These contributions correspond to the giant gravitons, which we would like to study in detail elsewhere.
}
Hence we obtain ${\cal I}_{N=\infty}(\mathfrak{t})$ as
\begin{align}
{\cal I}_{N=\infty}^{(C)}(\mathfrak{t})=\prod_{n=1}^\infty\frac{1}{(1-\mathfrak{t}^{cn})^{b_n}}.
\end{align}
This precisely coincides with $\text{PE}[{\cal I}^{(C)}_{N=1}(\mathfrak{t})-1]$ \eqref{BHF}.

\subsubsection{$\mathbb{C}^2/\mathbb{Z}_l$}
The Coulomb index of the 3d $\mathcal{N}=4$ $U(N)$ ADHM theory with $l$ flavors is identified with the Hilbert series for the $N$-th symmetric product of the $\mathbb{C}^2/\mathbb{Z}_l$ which is probed by $N$ M2-branes. In the large $N$ limit, the Coulomb index can be computed from \eqref{BHF} as\footnote{
The same Hilbert series is also obtained as the Coulomb(=Higgs) limit of the $U(N)_k\times U(N)_{-k}$ ABJM theory with $k=l$ and the Coulomb limit of the $U(N)_k\times U(N)_0^{\otimes (p-1)}\times U(N)_{-k}$ Chern-Simons matter theory (called $(p,1)_k$ model in section \ref{sec_p1kH}) with $pk=l$.
See \cite{Hayashi:2022ldo} for more detail.
}
\begin{align}
{\cal I}^{{\widehat A}_{l-1}}_{\infty}(\mathfrak{t})=\mathcal{I}^{U(\infty)+\textrm{adj}-[l] (C)}(\mathfrak{t}) = \text{PE}\left[\frac{1-\mt^{2l}}{(1-\mt^2)(1-\mt^l)^2}-1\right].\label{inf_U_PE1}
\end{align}
In order to use the Meinardus theorem, we rewrite \eqref{inf_U_PE1} in the form of infinite products \eqref{Meinardus1}. Note that the expression in the bracket in \eqref{inf_U_PE1} can be written as
\begin{align}
\frac{1-\mt^{2l}}{(1-\mt^2)(1-\mt^l)^2}-1=\frac{\mt^l}{(1-\mt^l)} +\frac{\mt^l }{(1-\mt^l)^2} +  \sum_{m=1}^{l-1}\frac{\mt^{2m}}{(1-\mt^l)^2}.
\end{align}
Then, using 
\begin{align}
\text{PE}\left[\frac{\mt^{l+m}}{(1-\mt^l)}\right] = \prod_{n=1}^{\infty}\frac{1}{\left(1-\mt^{ln+m}\right)},\quad 
\text{PE}\left[\frac{\mt^{l+m}}{(1-\mt^l)^2}\right]= \prod_{n=1}^{\infty}\frac{1}{\left(1-\mt^{ln+m}\right)^n},\label{PE_prod}
\end{align}
\eqref{inf_U_PE1} can be written as
\begin{align}\label{inf_U}
{\cal I}^{{\widehat A}_{l-1}}_{\infty}(\mathfrak{t})=\prod_{n=1}^{\infty}\frac{1}{\left(1-\mt^{ln}\right)^{n+1}}\prod_{m=1}^{l-1}\frac{1}{\left(1-\mt^{ln-(2m-l)}\right)^{n}}.
\end{align}

For odd $l$ the product for $m$ in \eqref{inf_U} can be decomposed into two parts, which gives rise to
\begin{align}
\label{inf_U_Codd}
{\cal I}^{{\widehat A}_{l-1}}_{\infty}(\mathfrak{t})
&=\prod_{n=1}^{\infty}
\frac{1}{(1-\mathfrak{t}^{ln})^{n+1}}
\prod_{m=1}^{\frac{l-1}{2}}
\frac{1}{(1-\mathfrak{t}^{ln-2m+1})^n (1-\mathfrak{t}^{ln+2m-1})^n}.
\end{align}
For example, for $l=1$, $3$ and $5$ we have
\begin{align}
{\cal I}^{{\widehat A}_0}_{\infty}(\mathfrak{t})
&=\prod_{n=1}^{\infty}
\frac{1}{(1-\mathfrak{t}^n)^{n+1}}, \\
{\cal I}^{{\widehat A}_2}_{\infty}(\mathfrak{t})
&=\prod_{n=1}^{\infty}
\frac{1}{(1-\mathfrak{t}^{3n-1})^{n} (1-\mathfrak{t}^{3n})^{n+1} (1-\mathfrak{t}^{3n+1})^{n}},\\
{\cal I}^{{\widehat A}_4}_{\infty}(\mathfrak{t})
&=\prod_{n=1}^{\infty}
\frac{1}{(1-\mathfrak{t}^{5n-1})^n (1-\mathfrak{t}^{5n-3})^n (1-\mathfrak{t}^{5n})^{n+1} (1-\mathfrak{t}^{5n+1})^n (1-\mathfrak{t}^{5n+3})^n}.
\end{align}
The infinite product (\ref{inf_U_Codd}) contains $l$ sets of infinite product over the integer $n$. 
Accordingly, we can find the asymptotic coefficients by applying $(l-1)$ convolutions (\ref{conv_thm}) to the large $N$ index (\ref{inf_U_Codd}) and the formulas for $\zeta'(-1,a)$ in appendix \ref{app_HZformulas}
\begin{align}
\label{asymp_U_Codd}
&{\cal I}^{{\widehat A}_{l-1}}_\infty(\mathfrak{t})=\sum_{n\ge 0}d_n\mathfrak{t}^n,\nonumber \\
&d_n\sim 
\exp\left[
3
\left(
\frac{\zeta(3)}{4 l}
\right)^{\frac13}
n^{\frac23}
+
\frac{\pi^2}{6 (2\zeta(3) l^2)^{\frac13}}
n^{\frac13}
+\frac{1}{12l}-\frac{\pi^4}{432l\zeta(3)}
\right]
\nonumber\\
&\quad \quad \times 
\frac{n^{\frac{1}{36}(l+\frac{4}{l}-36)}}
{2^{\frac{l^2+18l+4}{36l}} 3^{\frac12} \pi A^{\frac{1}{l}} \zeta(3)^{\frac{1}{36}(l+\frac{4}{l}-18)} l^{\frac{1}{36} (2l+\frac{11}{l}-18)}}
\prod_{m=1}^{\frac{l-1}{2}}
\left(
\frac{
\Gamma\left(\frac{2m-1}{l}\right)
}
{
\Gamma\left(1-\frac{2m-1}{l}\right)
}
\right)^{\frac{1-2m}{l}}. 
\end{align}
For example, when $l=3$ we have\footnote{
See \cite{Okazaki:2022sxo} for $l=1$.
}
\begin{align}
d_n&\sim 
\exp\left[
\frac{3^{\frac23} \zeta(3)^{\frac13}}{2^{\frac23}}n^{\frac23}
+\frac{\pi^2}{6\cdot 3^{\frac23} (2\zeta(3))^{\frac13}}n^{\frac13}
+\frac{1}{36}-\frac{\pi^4}{1296\zeta(3)}
\right]
\nonumber\\
&\times 
\frac{\zeta(3)^{\frac{41}{108}}}
{2^{\frac{43}{108}}\cdot 3^{\frac{29}{108}} \pi^{\frac56}A^{\frac13}}
\frac{1}{n^{\frac{95}{108}}}
\frac{1}{\Gamma\left(\frac16 \right)^{\frac13}},
\end{align}
where we have used the relations $\Gamma(x)\Gamma(1-x)$ $=$ $\pi/\sin(\pi x)$ 
and $\Gamma(2x)$ $=$ $2^{2x-1}$ $\pi^{-1/2}$ $\Gamma(x)\Gamma(x+1/2)$. 
The actual coefficients and the analytic values are given by
\begin{align}
\label{U_C3_asy_table}
\begin{array}{c|c|c|c} 
n&d_n&d_n^{\textrm{asym}}&d_n/d_n^{\textrm{asym}} \\ \hline 
10   &32                &35.4599                &0.902428 \\ 
100  &3.48886\times 10^{11} &3.43110\times 10^{11} &1.01683\\ 
1000 &3.23093\times 10^{59}&3.17900\times 10^{59}&1.01634\\ 
10000&8.61729\times 10^{281}&8.53641\times 10^{281}&1.00947\\ 
\end{array}.
\end{align}
See Appendix \ref{app:Coulomb_C2Zl} for the comparison for $l=5, 7, 9$.

When $l$ is even, the factors in \eqref{inf_U} can be further put together as 
\begin{align}
&{\cal I}^{{\widehat A}_{l-1}}_{\infty}(\mathfrak{t})\nonumber \\
&=\prod_{n=1}^{\infty}\frac{1}{\left(1-\mt^{ln}\right)^{n+1}}\left(\prod_{m=1}^{\frac{l}{2}-1}\frac{1}{\left(1-\mt^{ln-(2m-l)}\right)^{n}}\right)\frac{1}{\left(1-\mt^{ln}\right)^{n}}\left(\prod_{m=\frac{l}{2}+1}^{l-1}\frac{1}{\left(1-\mt^{ln-(2m-l)}\right)^{n}}\right)\nonumber \\
&=\prod_{n=1}^{\infty}\frac{1}{\left(1-\mt^{ln}\right)^{2n+1}}\left(\prod_{m=1}^{\frac{l}{2}-1}\frac{1}{\left(1-\mt^{l(n+1)-2m}\right)^{n}}\right)\left(\prod_{m=1}^{\frac{l}{2}-1}\frac{1}{\left(1-\mt^{ln-2m}\right)^{n}}\right)\cr
&=\prod_{n=1}^{\infty} 
\frac{1}{(1-\mathfrak{t}^{ln})^{2n+1}}
\prod_{m=1}^{\frac{l}{2}-1}
\frac{1}{(1-\mathfrak{t}^{ln-2m})^{2n-1}}.\label{inf_U_Ceven}
\end{align}
For example, when $l=2$, $4$ and $6$ 
\begin{align}
\label{inf_U_C2}
{\cal I}^{{\widehat A}_1}_{\infty}(\mathfrak{t})
&=\prod_{n=1}^{\infty}
\frac{1}{(1-\mathfrak{t}^{2n})^{2n+1}}, \\
\label{inf_U_C4}
{\cal I}^{{\widehat A}_3}_{\infty}(\mathfrak{t})
&=\prod_{n=1}^{\infty}
\frac{1}{(1-\mathfrak{t}^{4n-2})^{2n-1} (1-\mathfrak{t}^{4n})^{2n+1}}, \\
\label{inf_U_C6}
{\cal I}^{{\widehat A}_5}_{\infty}(\mathfrak{t})
&=\prod_{n=1}^{\infty}
\frac{1}{(1-\mathfrak{t}^{6n-2})^{2n-1} (1-\mathfrak{t}^{6n-4})^{2n-1} (1-\mathfrak{t}^{6n})^{2n+1}}. 
\end{align}
By performing $(\frac{l}{2}-1)$ convolutions (\ref{conv_thm}), 
we obtain the asymptotic coefficient of the large $N$ index (\ref{inf_U_Ceven})
\begin{align}
&{\cal I}^{{\widehat A}_{l-1}}_\infty(\mathfrak{t})=\sum_{n\in 2\mathbb{Z}_{\ge 0}}d_n\mathfrak{t}^n,\nonumber \\
&d_n\sim 
\exp\left[
3
\left(
\frac{\zeta(3)}{4 l}
\right)^{\frac13}
n^{\frac23}
+
\frac{\pi^2}{6 (2\zeta(3) l^2)^{\frac13}}
n^{\frac13}
+\frac{1}{3l}-\frac{\pi^4}{432\zeta(3)l}
\right]
\nonumber\\
&\quad \quad \times 
\frac{
n^{\frac{1}{36}(l+\frac{4}{l}-36)}
}
{
2^{\frac{l^2-18l-8}{36l}} 3^{\frac12} \pi A^{\frac{4}{l}} \zeta(3)^{\frac{1}{36}(l+\frac{4}{l}-18)} 
l^{\frac{1}{36}(2l+\frac{20}{l}-18)}
}
\prod_{m=1}^{\frac{l}{2}-1}
\Gamma\left(1-\frac{2m}{l} \right)^{\frac{4m}{l}-1}.\label{asymp_U_Ceven}
\end{align}
The product of the gamma functions can be also written as
\begin{align}
\prod_{m=1}^{\frac{l}{2}-1}
\Gamma\left(1-\frac{2m}{l} \right)^{\frac{4m}{l}-1}
&=
\prod_{m=1}^{\frac{l}{2}-1}
\left(
\frac{
\Gamma\left(\frac{2m}{l} \right)
}
{
\Gamma\left(1-\frac{2m}{l} \right)
}
\right)^{\frac{1-2m}{l}}, 
\end{align}
which takes the same form as (\ref{asymp_U_Codd}). 
For example, for $l=4$ we find 
\begin{align}
d_n\sim 
\exp\left[
\frac32 \left(\frac{\zeta(3)}{2}\right)^{\frac13} n^{\frac23}
+\frac{\pi^2}{12\cdot 2^{\frac23}\zeta(3)^{\frac13}}n^{\frac13}
+\frac{1}{12}-\frac{\pi^4}{1728\zeta(3)}
\right]
\times 
\frac{2^{\frac{13}{18}} \zeta(3)^{\frac{13}{36}}}
{3^{\frac12}\pi A} \frac{1}{n^{\frac{31}{36}}}, 
\end{align}
and the actual coefficients and the analytic values are given as follows: 
\begin{align}
\label{U_C4_asy_table}
\begin{array}{c|c|c|c} 
n&d_n&d_n^{\textrm{asym}}&d_n/d_n^{\textrm{asym}} \\ \hline 
10   &32                &36.7238                &0.871370\\ 
100  &3.33565 \times 10^{10} &3.33170\times 10^{10} &1.00119\\ 
1000 &8.53804 \times 10^{53}&8.46068\times 10^{53}&1.00914\\ 
10000&5.15739\times 10^{255}&5.12572\times 10^{255}&1.00618\\ 
\end{array}. 
\end{align}
See Appendix \ref{app:Coulomb_C2Zl} for the comparison for $l=6,8$.

We see that the asymptotic behaviors (\ref{asymp_U_Codd}) for odd $l$ and (\ref{asymp_U_Ceven}) for even $l$ 
coincide up to the constant overall factor. 

Since the scaling dimension of monopole operator is proportional to $l$, 
the number of the Coulomb branch operators with fixed scaling dimension should decrease when $l$ increases. 
In fact, we see from (\ref{asymp_U_Codd}) and (\ref{asymp_U_Ceven}) that the leading coefficient decreases as $l$ increases. 
In the large $l$ limit, the large $N$ Coulomb indices (\ref{inf_U_Codd}) and (\ref{inf_U_Ceven}) can be written as
\begin{align}
\label{inf_U_l}
\lim_{l\rightarrow\infty}{\cal I}^{{\widehat A}_{l-1}}_{\infty}(\mathfrak{t})
&=
\prod_{n\ge1}\frac{1}{1-\mathfrak{t}^{2n}}. 
\end{align}
The index (\ref{inf_U_l}) is a generating function for the partitions. 
The asymptotic expansion is given by the celebrated Hardy-Ramanujan formula \cite{MR2280879}
\begin{align}
{\cal I}^{{\widehat A}_\infty}_\infty(\mathfrak{t})=\sum_{n\in 2\mathbb{Z}_{\ge 0}}d_n\mathfrak{t}^n,\quad
d_n&\sim 
\frac{\exp
\left[
(2\zeta(2))^{\frac12}
n^{\frac12}
\right]
}{2\sqrt{3}n}.\label{asymp_U_largel}
\end{align}
In the large $l$ limit the asymptotic degeneracy turns into the growth for a string $\log d_n\sim n^{\frac{1}{2}}$ rather than a membrane $\log d_n\sim n^{\frac{2}{3}}$.

\subsubsection{$\mathbb{C}^2/\widehat{D}_l$}
Next consider the 3d $\mathcal{N}=4$ gauge theories with orthogonal and symplectic gauge groups, 
a single rank-$2$ tensor hypermultiplet (i.e.~antisymmetric (A) and symmetric (S) hypermultiplets) 
and $2l$ or $2(l\pm2)$ fundamental hypermultiplets. 
The Coulomb indices of these theories agree with the Hilbert series 
of the $N$-th symmetric product of the singularity $\mathbb{C}^2/\widehat{D}_{l}$ probed by M2-branes. In the large $N$ limit, the Coulomb index can be calculated from \eqref{BHF}
\begin{align}
{\cal I}^{{\widehat D}_l}_\infty(\mathfrak{t})&=
\mathcal{I}^{USp(\infty)+\textrm{A}-[2(l+2)] (C)}(\mathfrak{t})
=
\mathcal{I}^{USp(\infty)+\textrm{S}-[2(l)] (C)}(\mathfrak{t})
=\mathcal{I}^{O(\infty)+\textrm{A}-[2(l)] (C)}(\mathfrak{t})\nonumber \\
&=\mathcal{I}^{O(\infty)+\textrm{S}-[2(l-2)] (C)}(\mathfrak{t})\cr
&=\text{PE}\left[\frac{1+\mt^{2l+2}}{(1-\mt^4)(1-\mt^{2l})}-1\right]\cr
&=\text{PE}\left[\frac{\left(\sum_{k=1}^{l-1}\mt^{4k}\right) + \left(\mt^{2l}+\mt^{2l+2}+\mt^{4l+2}\right)\left(\sum_{k=0}^{l-1}\mt^{4k}\right)}{(1-t^{4l})^2} + \frac{\mt^{4l}}{(1-\mt^{4l})^2} + \frac{\mt^{4l}}{1-\mt^{4l}}\right]\cr
&=\prod_{n=1}^{\infty}\prod_{k=1}^{l-1}\frac{1}{\left(1-\mt^{4ln+4k}\right)^n}\prod_{n=1}^{\infty}\prod_{k=1}^{l}\frac{1}{\left(1-\mt^{4ln+2l+4k-4}\right)^n}\prod_{n=1}^{\infty}\prod_{k=1}^{l}\frac{1}{\left(1-\mt^{4ln+2l+4k-2}\right)^n}\cr
&\times\prod_{n=1}^{\infty}\prod_{k=1}^{l}\frac{1}{\left(1-\mt^{4ln+4l+4k-4}\right)^n}\prod_{n=1}^{\infty}\frac{1}{\left(1-\mt^{4ln}\right)^{n+1}}.
\label{inf_OSp_PE1}
\end{align}

For odd $l$, it is possible to rewrite \eqref{inf_OSp_PE1} as
\begin{align}
&{\cal I}^{{\widehat D}_l}_\infty(\mathfrak{t})
\nonumber\\
&=\text{PE}\left[\frac{1}{(1-\mt^{4l})^{2}}\left\{\left(2\sum_{k=1}^{\frac{l-1}{2}}\mt^{4l-4k}\right) + \left(2\sum_{k=1}^{\frac{l-1}{2}}\mt^{6l-4k}\right)+ \left(\sum_{k=1}^{\frac{l-1}{2}}\mt^{2l-4k+2}\right)\right.\right.\nonumber \\
&\quad \left.\left.+ \left(\sum_{k=1}^{\frac{l-1}{2}}\mt^{6l-4k+2}\right)+ \left(\sum_{k=1}^{\frac{l+1}{2}}\mt^{4l-4k+2}\right)+ \left(\sum_{k=1}^{\frac{l+1}{2}}\mt^{8l-4k+2}\right) + 2\mt^{4l}\right\} + \frac{\mt^{4l}}{1-\mt^{4l}}\right]\cr
&=
\prod_{n=1}^{\infty}\prod_{k=1}^{\frac{l+1}{2}}
\frac{1}{(1-\mathfrak{t}^{4ln-4k+2})^{2n-1}}
\prod_{n=1}^{\infty}\prod_{k=1}^{\frac{l-1}{2}}
\frac{1}{(1-\mathfrak{t}^{4ln-4k+2l})^{2n}}
\prod_{n=1}^{\infty}\prod_{k=1}^{\frac{l-1}{2}}
\frac{1}{(1-\mathfrak{t}^{4ln-4k})^{2n}}\nonumber \\
&\quad \times 
\prod_{n=1}^{\infty}\prod_{k=1}^{\frac{l-1}{2}}
\frac{1}{(1-\mathfrak{t}^{4ln-4k-2l+2})^{2n-1}}
\prod_{n=1}^{\infty}
\frac{1}{(1-\mathfrak{t}^{4ln})^{2n+1}}. 
\label{inf_OSp_Codd}
\end{align}
For $l=1,3$ and $5$ we have
\begin{align}
\label{inf_OSp_C3}
{\cal I}^{{\widehat D}_1}_\infty(\mathfrak{t})
&=\prod_{n=1}^{\infty}
\frac{1}{(1-\mathfrak{t}^{4n-2})^{2n-1} (1-\mathfrak{t}^{4n})^{2n+1}},\\
\label{inf_OSp_C5}
{\cal I}^{{\widehat D}_3}_\infty(\mathfrak{t})
&=\prod_{n=1}^{\infty}
\frac{1}{(1-\mathfrak{t}^{12n-8})^{2n-1} (1-\mathfrak{t}^{12n-6})^{2n-1} (1-\mathfrak{t}^{12n-4})^{2n}}
\nonumber\\
&\quad \times 
\frac{1}
{(1-\mathfrak{t}^{12n-2})^{2n-1} (1-\mathfrak{t}^{12n})^{2n+1} (1-\mathfrak{t}^{12n+2})^{2n} },\\
\label{inf_OSp_C7}
{\cal I}^{{\widehat D}_5}_\infty(\mathfrak{t})
&=\prod_{n=1}^{\infty}
\frac{1}{(1-\mathfrak{t}^{20n-4})^{2n} (1-\mathfrak{t}^{20n-8})^{2n} (1-\mathfrak{t}^{20n+2})^{2n} (1-\mathfrak{t}^{20n+6})^{2n}}
\nonumber\\
&\quad \times 
\frac{1}{(1-\mathfrak{t}^{20n-2})^{2n-1} (1-\mathfrak{t}^{20n-6})^{2n-1} (1-\mathfrak{t}^{20n-10})^{2n-1} (1-\mathfrak{t}^{20n-12})^{2n-1}}
\nonumber\\
&\quad \times 
\frac{1}{(1-\mathfrak{t}^{20n-16})^{2n-1} (1-\mathfrak{t}^{20n})^{2n+1}}. 
\end{align}
Note that the index (\ref{inf_OSp_C3}) is equal to the large $N$ index of the $U(N)$ ADHM theory ${\cal I}^{{\widehat A}_{l-1}}_\infty(\mathfrak{t})$ \eqref{inf_U_C4} with $l=4$ as $\widehat{D}_1=\mathbb{Z}_4$. 
The $2l-1$ convolutions lead to the asymptotic coefficients for odd $l$ 
\begin{align}
\label{asymp_OSp_Codd}
&{\cal I}^{{\widehat D}_l}_\infty(\mathfrak{t})=\sum_{n\in 2\mathbb{Z}_{\ge 0}}d_n\mathfrak{t}^n,\nonumber \\
&d_n\sim 
\exp\left[
3
\left(
\frac{\zeta(3)}{4 \cdot (4l)}
\right)^{\frac13}
n^{\frac23}
+\frac{\pi^2}{6(2\zeta(3)\cdot (4l)^2)^{\frac13}}
n^{\frac13}
+\frac{1}{12l}-\frac{\pi^4}{1728\zeta(3)l}
\right]
\nonumber\\
&\quad\quad \times 
\frac{
n^{\frac{1}{36} (l+\frac{1}{l}-33)}
}{
2^{\frac{1}{36}(23l+\frac{8}{l}-57)}
3^{\frac12} \pi
A^{\frac{1}{l}} 
\zeta(3)^{\frac{1}{36} (l+\frac{1}{l}-15)}
l^{\frac{1}{36}(11l+\frac{14}{l}-30)}}
\Bigl(\frac{H_{l-1}}{H_{(l-1)/2}^2}\Bigr)^{\frac{1}{l}}\nonumber \\
&\quad \times \prod_{m=1}^{\frac{l+1}{2}}
\Gamma\left(1-\frac{2m-1}{2l} \right)^{-1+\frac{2m-1}{l}}
\nonumber\\
&\quad\quad \times 
\prod_{m=1}^{\frac{l-1}{2}}
\Gamma\left(1-\frac{2m+l-1}{2l} \right)^{-1+\frac{2m+l-1}{l}}
\Gamma\left(1-\frac{2m-l}{2l} \right)^{-1+\frac{2m}{l}}
\Gamma\left(1-\frac{m}{l} \right)^{\frac{2m}{l}},
\end{align}
where $H_n$ is hyperfactorial $H_n=\prod_{k=1}^nk^k$.
For example, for $l=3$ we find 
\begin{align}
d_n&\sim 
\exp\left[
\frac{3^{\frac23}}{2}
\left(\frac{\zeta(3)}{2} \right)^{\frac13}n^{\frac23}
+\frac{\pi^2}{12\cdot 6^{\frac23} \zeta(3)^{\frac13}}n^{\frac13}
+\frac{1}{36}-\frac{\pi^4}{5184\zeta(3)}
\right]
\nonumber\\
&\quad \times 
\frac{2^{\frac{1}{27}} \zeta(3)^{\frac{35}{108}}}{3^{\frac{77}{108}} \pi^{\frac{4}{3}} A^{\frac13}}
\frac{1}{n^{\frac{89}{108}}}
\frac{\Gamma\left(\frac{1}{3} \right)}
{\Gamma\left(\frac{7}{6} \right)^{\frac13}},
\end{align}
and the actual coefficients and the analytic values are shown as follows: 
\begin{align}
\label{OSp_C3_asy_table}
\begin{array}{c|c|c|c} 
n&d_n&d_n^{\textrm{asym}}&d_n/d_n^{\textrm{asym}} \\ \hline 
10   &2                &4.04125                &0.494896 \\ 
100  &2.91943 \times 10^{6} &3.00975 \times 10^{6} &0.969993\\ 
1000 &1.26968 \times 10^{36}&1.27521\times 10^{36}&0.995660\\ 
10000&1.85071\times 10^{175}&1.85013\times 10^{175}&1.00031\\ 
\end{array}.
\end{align}
See Appendix \ref{app:Coulomb_C2Dl} for the comparison for $l=5, 7, 9$. 

For even $l$, the expression \eqref{inf_OSp_PE1} can be written as
\begin{align}
{\cal I}^{{\widehat D}_l}_\infty(\mathfrak{t})
&=\text{PE}\left[\frac{\sum_{k=1}^{\frac{l}{2}-1}\mt^{2l-4k} + \sum_{k=1}^{\frac{l}{2}}\mt^{2l+4k-2}}{(1-\mt^{2l})^2}+\frac{\mt^{2l}}{(1-\mt^{2l})^2}+\frac{\mt^{2l}}{(1-\mt^{2l})}\right]\cr
&=
\prod_{n=1}^{\infty}
\frac{1}{(1-\mathfrak{t}^{2ln})^{n+1}}
\prod_{n=1}^{\infty}
\prod_{k=1}^{\frac{l}{2}-1}
\frac{1}{(1-\mathfrak{t}^{2ln-4k})^n}
\prod_{n=1}^{\infty}
\prod_{k=1}^{\frac{l}{2}}
\frac{1}{(1-\mathfrak{t}^{2ln+4k-2})^n}.\label{inf_OSp_Ceven}
\end{align}
For example, when $l=2,4$ and $6$ the large $N$ Coulomb indices are 
\begin{align}
\label{inf_OSp_C2}
{\cal I}^{{\widehat D}_2}_\infty(\mathfrak{t})
&=\prod_{n=1}^{\infty}
\frac{1}{(1-\mathfrak{t}^{4n})^{n+1} (1-\mathfrak{t}^{4n+2})^{n} }, \\
\label{inf_OSp_C4}
{\cal I}^{{\widehat D}_4}_\infty(\mathfrak{t})
&=\prod_{n=1}^{\infty}
\frac{1}{(1-\mathfrak{t}^{8n})^{n+1} (1-\mathfrak{t}^{8n-4})^{n} (1-\mathfrak{t}^{8n+2})^{n} (1-\mathfrak{t}^{8n+6})^{n}},\\
\label{inf_OSp_C6}
{\cal I}^{{\widehat D}_6}_\infty(\mathfrak{t})
&=\prod_{n=1}^{\infty}
\frac{1}{(1-\mathfrak{t}^{12n})^{n+1} (1-\mathfrak{t}^{12n-4})^{n} (1-\mathfrak{t}^{12n-8})^{n} }
\nonumber\\
&\quad \times 
\frac{1}{(1-\mathfrak{t}^{12n+2})^n (1-\mathfrak{t}^{12n+6})^{n} (1-\mathfrak{t}^{12n+10})^n}.
\end{align}
The asymptotic coefficients can be obtained by performing $(l-1)$ convolutions (\ref{conv_thm}) 
\begin{align}
\label{asymp_OSp_Ceven}
&{\cal I}^{{\widehat D}_l}_\infty(\mathfrak{t})=\sum_{n\in 2\mathbb{Z}_{\ge 0}}d_n\mathfrak{t}^n,\nonumber \\
&d_n\sim 
\exp\left[
3
\left(
\frac{\zeta(3)}{4 \cdot (4l)}
\right)^{\frac13}
n^{\frac23}
+\frac{\pi^2}{6(2\zeta(3)\cdot (4l)^2)^{\frac13}}
n^{\frac13}
+\frac{1}{3l}-\frac{\pi^4}{1728\zeta(3)l}
\right]
\nonumber\\
&\quad\quad \times 
\frac{
n^{\frac{1}{36} (l+\frac{1}{l}-33)}
(H_{l/2-1})^{\frac{2}{l}}
}
{2^{-\frac{7l^2+21l+1}{36l}} 3^{\frac12} \pi^{\frac34} A^{\frac{4}{l}} 
\zeta(3)^{\frac{1}{36}(l+\frac{1}{l}-15)} l^{\frac{1}{36} (11l-12+\frac{5}{l})}}
\nonumber\\
&\quad\quad \times 
\prod_{m=1}^{\frac{l}{2}-1}
\Gamma\left(1-\frac{2m}{l} \right)^{\frac{2m}{l}}
\prod_{m=1}^{\frac{l}{2}}
\Gamma\left(1+\frac{2m-1}{l} \right)^{-\frac{2m-1}{l}}. 
\end{align}
For example, for $l=2$ we get
\begin{align}
d_n&\sim 
\exp\left[
\frac{3\zeta(3)^{\frac13}}{2\cdot 2^{\frac23}}n^{\frac23}
+\frac{\pi^2}{24(2\zeta(3))^{\frac13}}n^{\frac13}
+\frac{1}{24}-\frac{\pi^4}{3456\zeta(3)}
\right]
\nonumber\\
&\quad \times 
\frac{2^{\frac{43}{72}} \zeta(3)^{\frac{25}{72}}}
{3^{\frac12}\pi A^{\frac12}}\frac{1}{n^{\frac{61}{72}}},
\end{align}
and the actual coefficients and the analytic values are 
\begin{align}
\label{OSp_C2_asy_table}
\begin{array}{c|c|c|c} 
n&d_n&d_n^{\textrm{asym}}&d_n/d_n^{\textrm{asym}} \\ \hline 
10   &4                &7.78140                &0.514046 \\ 
100  &5.52928 \times 10^{7} &5.63457\times 10^{7} &0.981312\\ 
1000 &7.11358 \times 10^{41}&7.10866\times 10^{41}&1.00069\\ 
10000&2.77884\times 10^{201}&2.77173\times 10^{201}&1.00257\\ 
\end{array}. 
\end{align}
See Appendix \ref{app:Coulomb_C2Dl} for the comparision for $l=4, 6, 8$. 
The asymptotic behavior \eqref{asymp_OSp_Ceven} for even $l$ agrees with (\ref{asymp_OSp_Codd}) for odd $l$ up to the overall constant factor. 

The asymptotic degeneracy of Coulomb branch operators decreases as the number $l$ of flavors increases 
as seen from (\ref{asymp_OSp_Codd}) and (\ref{asymp_OSp_Ceven}). 
We find that when $l$ goes to $\infty$, the large $N$ indices (\ref{inf_OSp_Codd}) and (\ref{inf_OSp_Ceven}) become 
\begin{align}
{\cal I}^{{\widehat D}_\infty}_{\infty}(\mathfrak{t})=\prod_{n=1}^{\infty}
\frac{1}{1-\mathfrak{t}^{4n}}.
\end{align}
The asymptotic growth of the coefficients is given by \cite{MR2280879}
\begin{align}
{\cal I}^{{\widehat D}_\infty}_{\infty}(\mathfrak{t})=\sum_{n\in 4\mathbb{Z}_{\ge 0}}^\infty d_n\mathfrak{t}^n,\quad 
d_n\sim 
\frac{
\exp\left[ \left(2\cdot \frac{\zeta(2)}{2}\right)^{\frac12} n^{\frac12} \right]
}
{\sqrt{3} n},\label{asymp_OSp_largel}
\end{align}
which is analogous to the asymptotics (\ref{asymp_U_largel}) for $\mathbb{C}^2/\mathbb{Z}_l$.

\subsubsection{$\mathbb{C}^2/\widehat{E}_l$}
While
a Lagrangian describing the $N$ M2-branes probing $\mathbb{C}^2/\widehat{E}_{l}$ singularity is unknown, the mirror theory is given by a quiver theory whose shape is given by the Dynkin diagram of the affine $\mathfrak{e}_l$ algebra. 
The large $N$ index of the Coulomb index (or the Higgs index for the mirror theory) should be given by the large $N$ limit of the Hilbert series for the $N$-th symmetric product of $\mathbb{C}^2/\widehat{E}_l$. 

For $l=6$, the singularity is given by binary tetrahedral group. 
The large $N$ limit of the Coulomb index is given by
\begin{align}
\mathcal{I}^{\widehat{E}_6}_{\infty}(\mathfrak{t})&=\text{PE}\left[\frac{1-\mt^4+\mt^8}{1-\mt^4-\mt^6+\mt^{10}}-1\right]\cr
&=\text{PE}\left[\frac{\mt^6+\mt^8+\mt^{12}+\mt^{14}+\mt^{16}+\mt^{22}}{(1-\mt^{12})^2} + \frac{\mt^{12}}{1-\mt^{12}}\right]\cr
&=\prod_{n=1}^{\infty}
\frac{1}{(1-\mathfrak{t}^{12n})^{n+1} (1-\mathfrak{t}^{12n+2})^n (1-\mathfrak{t}^{12n+4})^n}
\nonumber\\
&\quad\times \frac{1}{(1-\mathfrak{t}^{12n-6})^n (1-\mathfrak{t}^{12n-4})^n (1-\mathfrak{t}^{12n+10})^n}. 
\end{align}
The asymptotic behavior of the coefficient is obtained by applying the convolutions five times. 
We find 
\begin{align}
&{\cal I}^{{\widehat E}_6}_\infty(\mathfrak{t})=\sum_{n\in 2\mathbb{Z}_{\ge 0}}d_n\mathfrak{t}^n,\nonumber \\
&d_n\sim 
\exp\left[
3
\left(
\frac{\zeta(3)}{4 \cdot (24)}
\right)^{\frac13}
n^{\frac23}
+\frac{\pi^2}{6(2\zeta(3)\cdot (24)^2)^{\frac13}}
n^{\frac13}
+\frac{1}{72}-\frac{\pi^4}{10368\zeta(3)}
\right]
\nonumber\\
&\quad\quad \times 
\frac{2^{\frac{71}{216}} \zeta(3)^{\frac{65}{216}} }
{3^{\frac{89}{216}} \pi^{\frac23} A^{\frac16}}
\frac{1}{n^{\frac{173}{216}}}
\frac{1}{
\Gamma\left(\frac{1}{6}\right)^{\frac16} 
\Gamma\left(\frac{5}{6}\right)^{\frac12} 
}. 
\end{align}
The actual coefficients and the analytic values are 
\begin{align}
\label{E_C6_asy_table}
\begin{array}{c|c|c|c} 
n&d_n&d_n^{\textrm{asym}}&d_n/d_n^{\textrm{asym}} \\ \hline 
12   &3                     &1.98252               &1.51323\\ 
100  &41794                 &43935.7               &0.951254\\
1000 &8.27152\times 10^{27} &8.35027\times 10^{27} &0.990569\\
10000&1.08249\times 10^{138}&1.08437\times 10^{138}&0.998268\\
\end{array}.
\end{align}

For $l=7$ the discrete subgroup $\widehat{E}_7$ is the binary octahedral group. 
We find that the large $N$ index is expressed as
\begin{align}
\mathcal{I}^{\widehat{E}_7}_{\infty}(\mathfrak{t})&=\text{PE}\left[\frac{1-\mt^6+\mt^{12}}{1-\mt^6-\mt^8+\mt^{14}}-1\right]\cr
&=\text{PE}\left[\frac{\mt^8+\mt^{12}+\mt^{16}+\mt^{18}+\mt^{20}+\mt^{24}+\mt^{26}+\mt^{28}+\mt^{30}+\mt^{34}+\mt^{38}+\mt^{46}}{(1-\mt^{24})^2}\right.\nonumber \\
&\quad\quad\quad\quad \left. + \frac{\mt^{24}}{1-\mt^{24}}\right]\cr
&=\prod_{n=1}^{\infty}
\frac{1}{(1-\mathfrak{t}^{24n})^{n+1} (1-\mathfrak{t}^{24n+2})^n  (1-\mathfrak{t}^{24n+4})^n  (1-\mathfrak{t}^{24n+6})^n  (1-\mathfrak{t}^{24n-16})^n}
\nonumber\\
&\quad \times 
\frac{1}{ (1-\mathfrak{t}^{24n+10})^n  (1-\mathfrak{t}^{24n-12})^n  (1-\mathfrak{t}^{24n+14})^n  (1-\mathfrak{t}^{24n-8})^n  (1-\mathfrak{t}^{24n-6})^n}
\nonumber\\
&\quad \times 
\frac{1}{ (1-\mathfrak{t}^{24n-4})^n  (1-\mathfrak{t}^{24n+22})^n}. 
\end{align}
The asymptotic coefficient can be obtained by using the $11$ convolutions. 
It leads to
\begin{align}
&{\cal I}^{{\widehat E}_7}_\infty(\mathfrak{t})=\sum_{n\in 2\mathbb{Z}_{\ge 0}}d_n\mathfrak{t}^n,\nonumber \\
&d_n\sim 
\exp\left[
3
\left(
\frac{\zeta(3)}{4 \cdot (48)}
\right)^{\frac13}
n^{\frac23}
+\frac{\pi^2}{6(2\zeta(3)\cdot (48)^2)^{\frac13}}
n^{\frac13}
+\frac{1}{144}-\frac{\pi^4}{20736\zeta(3)}
\right]
\nonumber\\
&\quad\quad \times 
\frac{
2^{\frac{7}{18}} \zeta(3)^{\frac{119}{432}}
}
{
3^{\frac{197}{432}} \pi^{\frac12} A^{\frac{1}{12}}
}
\frac{1}{n^{\frac{335}{432}}}
\frac{
\Gamma\left( \frac13 \right)^{\frac23}
\Gamma\left( \frac23 \right)^{\frac13}
\Gamma\left( \frac34 \right)^{\frac14}
\Gamma\left( \frac56 \right)^{\frac16}
}
{
\Gamma\left( \frac{1}{12} \right)^{\frac{1}{12}}
\Gamma\left( \frac16 \right)^{\frac16}
\Gamma\left( \frac14 \right)^{\frac14}
\Gamma\left( \frac{5}{12} \right)^{\frac{5}{12}}
\Gamma\left( \frac{7}{12} \right)^{\frac{7}{12}}
\Gamma\left( \frac{11}{12} \right)^{\frac{11}{12}}
}. 
\end{align}
The actual coefficients and the analytic values are 
\begin{align}
\label{E_C7_asy_table}
\begin{array}{c|c|c|c} 
n&d_n&d_n^{\textrm{asym}}&d_n/d_n^{\textrm{asym}} \\ \hline 
12   &1                &0.834326                &1.19857\\ 
100&1667&1648.37&1.01130\\
1000&3.10235\times 10^{21}&3.14424\times 10^{21}&0.986677\\
10000&4.28876\times 10^{108}&4.30249\times 10^{108}&0.996809\\
\end{array}.
\end{align}

For $l=8$ the discrete subgroup $\widehat{E}_8$ is the binary icosahedral group. 
The large $N$ index reads
\begin{align}
\mathcal{I}^{\widehat{E}_8}_{\infty}(\mathfrak{t})&=\text{PE}\left[\frac{1+\mt^2-\mt^6-\mt^8-\mt^{10}+\mt^{14}+\mt^{16}}{1+\mt^2-\mt^6-\mt^8-\mt^{10}-\mt^{12}+\mt^{16}+\mt^{18}}-1\right]\cr
&=\text{PE}\left[\frac{1}{(1-\mt^{60})^2}\left(\mt^{12}+\mt^{20}+\mt^{24}+\mt^{30}+\mt^{32}+\mt^{36}+\mt^{40}+\mt^{42}+\mt^{44}+\mt^{48}+\mt^{50}\right.\right.\cr
&\hspace{1.5cm}\left.+\mt^{52}+\mt^{54}+\mt^{56}+\mt^{60}+\mt^{62}+\mt^{64}+\mt^{66}+\mt^{68}+\mt^{70}+\mt^{74}+\mt^{76}+\mt^{78}\right.\cr
&\hspace{5cm}\left.\left.+\mt^{82}+\mt^{86}+\mt^{88}+\mt^{94}+\mt^{98}+\mt^{106}+\mt^{118}\right) + \frac{\mt^{60}}{1-\mt^{60}}\right]\cr
&=\prod_{n=1}^{\infty}
\frac{1}{(1-\mathfrak{t}^{60n})^{n+1} (1-\mathfrak{t}^{60n+2})^{n} (1-\mathfrak{t}^{60n+4})^{n} (1-\mathfrak{t}^{60n+6})^{n} (1-\mathfrak{t}^{60n+8})^{n}}
\nonumber\\
&\quad \times 
\frac{1}{(1-\mathfrak{t}^{60n+10})^{n} (1-\mathfrak{t}^{60n-48})^{n} (1-\mathfrak{t}^{60n+14})^{n} (1-\mathfrak{t}^{60n+16})^{n} (1-\mathfrak{t}^{60n+18})^{n} }
\nonumber\\
&\quad \times 
\frac{1}{(1-\mathfrak{t}^{60n-40})^{n} (1-\mathfrak{t}^{60n+22})^{n} (1-\mathfrak{t}^{60n-36})^{n} (1-\mathfrak{t}^{60n+26})^{n} (1-\mathfrak{t}^{60n+28})^{n} }
\nonumber\\
&\quad \times 
\frac{1}{(1-\mathfrak{t}^{60n-30})^{n} (1-\mathfrak{t}^{60n-28})^{n} (1-\mathfrak{t}^{60n+34})^{n} (1-\mathfrak{t}^{60n-24})^{n} (1-\mathfrak{t}^{60n+38})^{n} }
\nonumber\\
&\quad \times 
\frac{1}{(1-\mathfrak{t}^{60n-20})^{n} (1-\mathfrak{t}^{60n-18})^{n} (1-\mathfrak{t}^{60n-16})^{n} (1-\mathfrak{t}^{60n+46})^{n} (1-\mathfrak{t}^{60n-12})^{n} }
\nonumber\\
&\quad \times 
\frac{1}{(1-\mathfrak{t}^{60n-10})^{n} (1-\mathfrak{t}^{60n-8})^{n} (1-\mathfrak{t}^{60n-6})^{n} (1-\mathfrak{t}^{60n-4})^{n} (1-\mathfrak{t}^{60n+58})^{n} }. 
\end{align}
By means of $29$ convolutions, 
we get the asymptotic behavior of the large $N$ index
\begin{align}
&{\cal I}^{{\widehat E}_8}_\infty(\mathfrak{t})=\sum_{n\in 2\mathbb{Z}_{\ge 0}}d_n\mathfrak{t}^n,\nonumber \\
&d_n\sim 
\exp\left[
3
\left(
\frac{\zeta(3)}{4 \cdot (120)}
\right)^{\frac13}
n^{\frac23}
+\frac{\pi^2}{6(2\zeta(3)\cdot (120)^2)^{\frac13}}
n^{\frac13}
+\frac{1}{360}-\frac{\pi^4}{51840\zeta(3)}
\right]
\nonumber\\
&\quad\quad \times 
\frac{2^{\frac{263}{1080}} \zeta(3)^{\frac{269}{1080}}}
{3^{\frac{109}{216}} 5^{\frac{1}{216}} \pi^{\frac14} A^{\frac{1}{30}}}
\frac{1}{n^{\frac{809}{1080}}}
\prod_{m=1}^{29}
\frac{1}
{\Gamma\left(
\frac{m}{30}
\right)^{\frac{m}{30}}}
\nonumber\\
&\quad\quad \times 
\Gamma\left(\frac{1}{5} \right)
\Gamma\left(\frac{1}{3} \right)
\Gamma\left(\frac{2}{5} \right)
\Gamma\left(\frac{8}{15} \right)
\Gamma\left(\frac{3}{5} \right)
\Gamma\left(\frac{2}{3} \right)
\Gamma\left(\frac{7}{10} \right)
\Gamma\left(\frac{11}{15} \right)
\Gamma\left(\frac{4}{5} \right)
\nonumber\\
&\quad\quad \times 
\Gamma\left(\frac{5}{6} \right)
\Gamma\left(\frac{13}{15} \right)
\Gamma\left(\frac{9}{10} \right)
\Gamma\left(\frac{14}{15} \right). 
\end{align}
The actual coefficients and the analytic values are 
\begin{align}
\label{E_C8_asy_table}
\begin{array}{c|c|c|c} 
n&d_n&d_n^{\textrm{asym}}&d_n/d_n^{\textrm{asym}} \\ \hline 
12   &1                &0.339036                &2.94954 \\ 
100&59&59.7722&0.987081\\
1000&1.04334\times 10^{15}&1.06168\times 10^{15}&0.982726\\
10000&9.08859\times 10^{78}&9.12853\times 10^{78}&0.995625\\
\end{array}. 
\end{align}
 
To summarize, we find that the asymptotic coefficient of the large $N$ Coulomb index for 
M2-brane SCFTs for the quotient singularity $\mathbb{C}^2/\Gamma$, where $\Gamma$ is a discrete subgroup of $SU(2)$, takes the form
\begin{align}
&{\cal I}^{\Gamma}_\infty(\mathfrak{t})=\sum_nd_n\mathfrak{t}^n,\nonumber \\
&
d_n=
n^{\frac{1}{36}\left(\frac{4}{|\Gamma|}+\textrm{rank}(\mathfrak{g}) + 1\right)-1}\exp\left[
3
\left(
\frac{\zeta(3)}{4 |\Gamma|}
\right)^{\frac13}
n^{\frac23}
+\frac{\pi^2}{6(2\zeta(3) |\Gamma|^2)^{\frac13}}
n^{\frac13}+\cdots
\right],
\end{align}
up to the overall constants ($C$ and $e^\gamma$) whose unified expression we could not identify.
Here
$|\Gamma|$ is the order of the discrete subgroup $\Gamma$ and $\textrm{rank}(\mathfrak{g})$ is the rank of the root system $\mathfrak{g}$ associated to $\Gamma$.
The range of the summation over $n$ depends on the choice of $\Gamma$.
The data of the geometry probed by M2-branes are summarized as follows: 
\begin{align}
\begin{array}{c|c|c|c|c}
\textrm{probed geometry}&\textrm{ root system $\mathfrak{g}$}&\textrm{rank}&\textrm{ finite subgroup $\Gamma$ }&\textrm{ order $|\Gamma|$ }\\ \hline
\mathbb{C}^2/\mathbb{Z}_l&A_{l-1}&l-1&\textrm{ cyclic group $\mathbb{Z}_l$ }&l\\
\mathbb{C}^2/\widehat{D}_l&D_{l+2}&l+2&\textrm{ binary dihedral group $\widehat{D}_l$ }&4l\\
\mathbb{C}^2/\widehat{E}_6&E_{6}&6&\textrm{ binary tetrahedral group $\widehat{E}_6$ }&24\\
\mathbb{C}^2/\widehat{E}_7&E_{7}&7&\textrm{ binary octahedral group $\widehat{E}_7$ }&48\\
\mathbb{C}^2/\widehat{E}_8&E_{8}&8&\textrm{ binary icosahedral group $\widehat{E}_8$ }&120\\
\end{array}.
\end{align}

\subsection{Higgs indices}
\label{sec_Higgs}
The Higgs index (a.k.a. Hilbert series for the Higgs branch) enumerates the number $d_n$ of the Higgs branch operators with the scaling dimension $n/2$ in 3d $\mathcal{N}\ge 4$ sueprsymmetric gauge theory. 
While the Higgs indices are equivalent to the Coulomb indices for the ABJ(M) theories as they are self-mirror, 
they are distinguished for the ADHM theories so that the Higgs indices encode the moduli space of instantons on $\mathbb{R}^4$. 

\subsubsection{$U(N)$}
The Higgs branch of the $U(N)$ ADHM theory with $l$ flavors is the moduli space of $N$ $SU(l)$ instantons on $\mathbb{R}^4$. 
The large $N$ Higgs index of the $U(N)$ ADHM theory with $l$ flavors can be written as \cite{Crew:2020psc}
\begin{align}
\label{inf_U_H}
\mathcal{I}^{U(\infty)+\textrm{adj}-[l] (H)}(\mathfrak{t})
&=\prod_{n=1}^{\infty}
\frac{1}{(1-\mathfrak{t}^n)^{l^2(n-1)+2}}. 
\end{align}
For example, for $l=1,2,3$ and $4$ we have
\begin{align}
\label{inf_U_H1}
\mathcal{I}^{U(\infty)+\textrm{adj}-[1] (H)}(\mathfrak{t})
&=\prod_{n=1}^{\infty}
\frac{1}{(1-\mathfrak{t}^{n})^{n+1}},\\
\label{inf_U_H2}
\mathcal{I}^{U(\infty)+\textrm{adj}-[2] (H)}(\mathfrak{t})
&=\prod_{n=1}^{\infty}
\frac{1}{(1-\mathfrak{t}^{n})^{4n-2}},\\
\label{inf_U_H3}
\mathcal{I}^{U(\infty)+\textrm{adj}-[3] (H)}(\mathfrak{t})
&=\prod_{n=1}^{\infty}
\frac{1}{(1-\mathfrak{t}^{n})^{9n-7}},\\
\label{inf_U_H4}
\mathcal{I}^{U(\infty)+\textrm{adj}-[4] (H)}(\mathfrak{t})
&=\prod_{n=1}^{\infty}
\frac{1}{(1-\mathfrak{t}^{n})^{16n-14}}. 
\end{align}
When $l=1$ it coincides with the large $N$ Coulomb index (\ref{inf_U_Codd}) as the theory is self-mirror. 
Applying the Meinardus theorem (\ref{Meinardus_thm}), we obtain the asymptotic behavior 
\begin{align}
\label{asymp_U_H}
&{\cal I}^{U(\infty)+\text{adj-}[l](H)}(\mathfrak{t})=\sum_{n\ge 0}d_n\mathfrak{t}^n,\nonumber \\
&d_n\sim 
\frac{
\exp\left[
3
\left(
\frac{\zeta(3)l^2}{4}
\right)^{\frac13}
n^{\frac23}
-\frac{(l^2-2) \pi^2}{6(2\zeta(3) l^2)^{\frac13}}
n^{\frac13}
+\frac{l^2}{12}-\frac{(l^2-2)^2\pi^4}{432l^2\zeta(3)}
\right]
}
{
2^{1-\frac{13l^2}{36}}3^{\frac12} \pi^{\frac12(3-l^2)} A^{l^2} \zeta(3)^{\frac{5l^2}{36}-\frac12} l^{\frac{5}{18}l^2-l} n^{1-\frac{5l^2}{36}}
}
. 
\end{align}
For example, the exact numbers $d_{n}$ and the values $d_n^{\textrm{asym}}$ obtained from the formula (\ref{asymp_U_H}) for $l=2,3$ are
\begin{align}
&\begin{array}{c|c|c|c} 
\multicolumn{4}{c}{l=2}\\ \hline
n    &d_n                    &d_n^{\textrm{asym}}    &d_n/d_n^{\textrm{asym}}\\ \hline
10   &18880                  &18843.7                &1.00193\\
100  &3.79367\times 10^{25}  &3.76035\times 10^{25}  &1.00886\\
1000 &1.65367\times 10^{130} &1.64455\times 10^{130} &1.00554\\
10000&3.08810\times 10^{626} &3.07934\times 10^{626} &1.00285\\
20000&1.97098\times 10^{1000}&1.96647\times 10^{1000}&1.00230\\
\end{array},\label{U_H_asy_table2}\\
&\begin{array}{c|c|c|c} 
\multicolumn{4}{c}{l=3}\\ \hline
n    &d_n                    &d_n^{\textrm{asym}}    &d_n/d_n^{\textrm{asym}}\\ \hline 
10   &126571                 &74797.6                &1.69218 \\
100  &2.88256\times 10^{31}  &2.20572\times 10^{31}  &1.30686 \\     
1000 &2.83060\times 10^{164}  &2.48690\times 10^{164}  &1.13820 \\
10000&1.00432\times 10^{805} &9.44663\times 10^{804} &1.06315\\
20000&8.93733\times 10^{1289}&8.51192\times 10^{1289}&1.04998\\
\end{array}.\label{U_H_asy_table3}
\end{align}
See Appendix \ref{app:Higgs_C2Zl} for the comparison for $l=4,5,6,7,8$. 
Unlike the Coulomb branch, 
the degeneracy of the Higgs branch operators increases as $l$ does. 
 
\subsubsection{$USp(2N)$}
The Higgs branch of the $USp(2N)$ ADHM theory, i.e.~the $USp(2N)$ gauge theory with an antisymmetric hypermultiplet and  $(2l+\delta)$ fundamental half-hypermultiplets ($\delta=0,1$) is the moduli space of $N$ $SO(2l+\delta)$ instantons on $\mathbb{R}^4$. 
We find that the large $N$ Higgs index of the $USp(2N)$ ADHM theory with $(2l+\delta)$ fundamental half-hypers coincides with\footnote{
The expression for the large $N$ Higgs indices of $USp(N)/O(N)$ ADHM theories \eqref{inf_USp_H},\eqref{inf_O_H} are guessed from the small $\mathfrak{t}$ expansions of finite $N$ indices.
It would be interesting to find analytic derivations of these results.
}
\begin{align}
\label{inf_USp_H}
\mathcal{I}^{USp(\infty)+\textrm{A}-[2l+\delta] (H)}(\mathfrak{t})
&=\prod_{n=1}^{\infty}
\frac{1}{(1-\mathfrak{t}^n)^{\left(\frac{(2l+\delta)(2l+\delta-1)}{2}+1\right)(n-1)+2}}. 
\end{align}
While the Coulomb indices and full indices for $l$ $=$ $1$ and $2$ are not well-defined due to presence of the non-positive scaling dimensions of monopole operators as the theories are ugly or bad \cite{Gaiotto:2008ak}, the Higgs indices can be nicely computed. 
For example, for $l=1,2$ and $3$ we have
\begin{align}
\label{inf_USp_H2}
\mathcal{I}^{USp(\infty)+\textrm{A}-[2] (H)}(\mathfrak{t})
&=\prod_{n=1}^{\infty}
\frac{1}{(1-\mathfrak{t}^n)^{2n}}, \\
\label{inf_USp_H3}
\mathcal{I}^{USp(\infty)+\textrm{A}-[3] (H)}(\mathfrak{t})
&=\prod_{n=1}^{\infty}
\frac{1}{(1-\mathfrak{t}^n)^{4n-2}}, \\
\label{inf_USp_H4}
\mathcal{I}^{USp(\infty)+\textrm{A}-[4] (H)}(\mathfrak{t})
&=\prod_{n=1}^{\infty}
\frac{1}{(1-\mathfrak{t}^n)^{7n-5}}, \\
\label{inf_USp_H5}
\mathcal{I}^{USp(\infty)+\textrm{A}-[5] (H)}(\mathfrak{t})
&=\prod_{n=1}^{\infty}
\frac{1}{(1-\mathfrak{t}^n)^{11n-9}}, \\
\label{inf_USp_H6}
\mathcal{I}^{USp(\infty)+\textrm{A}-[6] (H)}(\mathfrak{t})
&=\prod_{n=1}^{\infty}
\frac{1}{(1-\mathfrak{t}^n)^{16n-14}},\\
\label{inf_USp_H7}
\mathcal{I}^{USp(\infty)+\textrm{A}-[7] (H)}(\mathfrak{t})
&=\prod_{n=1}^{\infty}
\frac{1}{(1-\mathfrak{t}^n)^{22n-20}}. 
\end{align}
The large $N$ index (\ref{inf_USp_H3}) coincides with (\ref{inf_U_H2}) as $SO(3)\cong SU(2)$ and (\ref{inf_USp_H6}) with (\ref{inf_U_H4}) as $SO(6)\cong SU(4)$. 
From the Meinardus theorem (\ref{Meinardus_thm}) the asymptotic behavior is simply obtained by replacing $l^2$ in the expression (\ref{asymp_U_H}) with $\left( \frac{(2l+\delta) (2l+\delta-1)}{2}+1 \right)$. 

\subsubsection{$O(N)$}
The Higgs branch of the $O(N)$ ADHM theory, i.e.~the $O(N)$ gauge theory with a symmetric hypermultiplet and $2l$ fundamental half-hypermultiplets is the moduli space of $N$ $USp(2l)$ instantons on $\mathbb{R}^4$. 
We find that for the $O(N)$ ADHM theory with $l$ fundamental hypers the large $N$ Higgs index takes the form
\begin{align}
\label{inf_O_H}
\mathcal{I}^{O(\infty)+\textrm{S}-[2l](H)}(\mathfrak{t})
&=\prod_{n=1}^{\infty}
\frac{1}{(1-\mathfrak{t}^n)^{(2l^2+l+1)(n-1)+2}}. 
\end{align}
For $l=1,2$ and $3$ we have
\begin{align}
\label{inf_O_H1}
\mathcal{I}^{O(\infty)+\textrm{S}-[2](H)}(\mathfrak{t})
&=\prod_{n=1}^{\infty}
\frac{1}{(1-\mathfrak{t}^n)^{4n-2}},\\
\label{inf_O_H2}
\mathcal{I}^{O(\infty)+\textrm{S}-[4](H)}(\mathfrak{t})
&=\prod_{n=1}^{\infty}
\frac{1}{(1-\mathfrak{t}^n)^{11n-9}},\\
\label{inf_O_H3}
\mathcal{I}^{O(\infty)+\textrm{S}-[6](H)}(\mathfrak{t})
&=\prod_{n=1}^{\infty}
\frac{1}{(1-\mathfrak{t}^n)^{22n-20}}. 
\end{align}
The large $N$ index (\ref{inf_O_H1}) is equal to (\ref{inf_U_H2}) and (\ref{inf_USp_H3}) as $USp(2)\cong SU(2)\cong SO(3)$ and (\ref{inf_O_H2}) to (\ref{inf_USp_H5}) as $USp(4)\cong SO(5)$. 
The Meinardus theorem (\ref{Meinardus_thm}) leads to the asymptotic growth that takes the same form as (\ref{asymp_U_H}) with $l^2$ replaced by $2l^2+l+1$. 

\subsubsection{$U(N)_k\times U(N)_0^{\otimes (p-1)}\times U(N)_{-k}$ Chern-Simons theories}
\label{sec_p1kH}
Consider the $U(N)_k\times U(N)_0^{\otimes (p-1)} \times U(N)_{-k}$ quiver Chern-Simons matter theories which describe $N$ M2-branes propagating in $(\mathbb{C}^2/\mathbb{Z}_p \times \mathbb{C}^2)/\mathbb{Z}_k$.
We shall call this theory as $(p,1)_k$ model.
In \cite{Hayashi:2022ldo} an expression for the Higgs indices of these theories without holonomy integrations was obtained as

\begin{align}
{\cal I}^{(p,1)_k}(\mathfrak{t})=\sum_{\substack{
\nu_m^{(1)},\cdots,\nu_m^{(p)}\ge 0\\
(\sum_{m=-\infty}^\infty\nu_m^{(a)}=N)}}
\frac{\mathfrak{t}^{\epsilon(\nu_m^{(a)})}}{
\prod_{m=-\infty}^\infty\prod_{a=1}^p
(\mathfrak{t}^2;\mathfrak{t}^2)_{\nu_m^{(a)}}},
\label{p1kHwithoutintegration}
\end{align}
with
\begin{align}
&(x;q)_n=\prod_{j=0}^{n-1}(1-xq^j),\nonumber \\
&\epsilon_1(\nu_m^{(a)})=
\sum_{m<m'}|m-m'|\sum_{a=1}^p(
-2\nu_m^{(a)}\nu_{m'}^{(a)}
+\nu_m^{(a)}\nu_{m'}^{(a+1)}
+\nu_m^{(a+1)}\nu_{m'}^{(a)}
)
+\sum_{m=-\infty}^\infty |km|\nu_m^{(1)},
\end{align}
where $\nu_m^{(p+1)}=\nu_m^{(1)}$.
Although the expression contains infinite summations, for relatively small $N$ the summations are tractable and we obtain the leading coefficients of the small $\mathfrak{t}$ expansion of ${\cal I}^{(H)}_N(\mathfrak{t})$ explicitly.
Namely, due to the constraints $\sum_{m=-\infty}^\infty \nu_m^{(a)}=N$, for each $a$ only a finite number (at most $N$) of the components $\nu_m^{(a)}$ can be non-zero.
Hence the configuration $\nu_m^{(a)}$ can be labeled by a set of non-zero values adding up to $N$ and the list of $m$ where these non-zero values are assigned.
For example, $\sum_{\nu_m^{(1)}}$ can be written for $N=1,2,3,4$ as
\begin{align}
&\sum_{\substack{\nu_m^{(1)}\ge 0\\ (\sum_{m=-\infty}^\infty \nu_m^{(1)}=1)}}=\sum_{m_1=-\infty}^\infty(\nu_m^{(1)}=\delta_m^{m_1}),\nonumber \\
&\sum_{\substack{\nu_m^{(1)}\ge 0\\ (\sum_{m=-\infty}^\infty \nu_m^{(1)}=2)}}=\sum_{m_1=-\infty}^\infty(\nu_m^{(1)}=2\delta_m^{m_1})
+\sum_{m_1<m_2}(\nu_m^{(1)}=\delta_m^{m_1}+\delta_m^{m_2}),\nonumber \\
&\sum_{\substack{\nu_m^{(1)}\ge 0\\ (\sum_{m=-\infty}^\infty \nu_m^{(1)}=3)}}=\sum_{m_1=-\infty}^\infty(\nu_m^{(1)}=3\delta_m^{m_1})
+\sum_{m_1\neq m_2}(\nu_m^{(1)}=2\delta_m^{m_1}+\delta_m^{m_2})\nonumber \\
&\quad\quad\quad\quad\quad\quad\quad\quad +\sum_{m_1<m_2<m_3}(\nu_m^{(1)}=\delta_m^{m_1}+\delta_m^{m_2}+\delta_m^{m_3}),\nonumber \\
&\sum_{\substack{\nu_m^{(1)}\ge 0\\ (\sum_{m=-\infty}^\infty \nu_m^{(1)}=4)}}=\sum_{m_1=-\infty}^\infty(\nu_m^{(1)}=4\delta_m^{m_1})
+\sum_{m_1\neq m_2}(\nu_m^{(1)}=3\delta_m^{m_1}+\delta_m^{m_2})\nonumber \\
&\quad\quad\quad\quad\quad\quad\quad\quad +\sum_{m_1<m_2}(\nu_m^{(1)}=2\delta_m^{m_1}+2\delta_m^{m_2})\nonumber \\
&\quad\quad\quad\quad\quad\quad\quad\quad +\sum_{m_1=-\infty}^\infty\sum_{\substack{m_2<m_3\\ (m_2,m_3\neq m_1)}}(\nu_m^{(1)}=2\delta_m^{m_1}+\delta_m^{m_2}+\delta_m^{m_3})\nonumber \\
&\quad\quad\quad\quad\quad\quad\quad\quad +\sum_{m_1<m_2<m_3<m_4}(\nu_m^{(1)}=\delta_m^{m_1}+\delta_m^{m_2}+\delta_m^{m_3}+\delta_m^{m_4}).
\end{align}
As a result, the infinite dimensional summation over $\nu_m^{(a)}$ \eqref{p1kHwithoutintegration} can be decomposed into a finite, at most $Np$-dimensional summations over the lists of $m$ for each $a$.
We also observe that as long as we are interested in the small $\mathfrak{t}$ expansion of the Higgs indices up to some finite order it is sufficient to truncate each infinite sum of $m$ both from below and above.
From the data thus obtained for finite $N$, we can guess the large $N$ Higgs indices ${\cal I}_{\infty}^{(H)}(\mathfrak{t})$ of these Chern-Simons theories.

For example, for the $(2,1)_2$ model we obtain the following results.
\begin{align}
&{\cal I}^{(2,1)_2\,(H)}_1(\mathfrak{t})
=\uwave{1 + 6 \mathfrak{t}^2} + 19 \mathfrak{t}^4 + 44 \mathfrak{t}^6 + 85 \mathfrak{t}^8 + 146 \mathfrak{t}^{10}+\cdots,\nonumber \\
&{\cal I}^{(2,1)_2\,(H)}_2(\mathfrak{t})=1+6\mathfrak{t}^2\uwave{+35\mathfrak{t}^4}+131\mathfrak{t}^6+427\mathfrak{t}^8+1151\mathfrak{t}^{10}+\cdots,\nonumber \\
&{\cal I}^{(2,1)_2\,(H)}_3(\mathfrak{t})=1+6\mathfrak{t}^2+35\mathfrak{t}^4\uwave{+162\mathfrak{t}^6}+636\mathfrak{t}^8+2193\mathfrak{t}^{10}+\cdots,\nonumber \\
&{\cal I}^{(2,1)_2\,(H)}_4(\mathfrak{t})=1+6\mathfrak{t}^2+35\mathfrak{t}^4+162\mathfrak{t}^6\uwave{+687\mathfrak{t}^8}+2578\mathfrak{t}^{10}
+\cdots,\nonumber \\
&{\cal I}^{(2,1)_2\,(H)}_5(\mathfrak{t})=1
+6\mathfrak{t}^2+35\mathfrak{t}^4+162\mathfrak{t}^6
+687\mathfrak{t}^8
+\cdots.
\end{align}
The coefficient of $\mathfrak{t}^{2n}$ saturates at $N=n$ (those underlined) and gives the coefficient in ${\cal I}^{(H)}_\infty(\mathfrak{t})$, hence we have
\begin{align}
{\cal I}^{(2,1)_2\,(H)}_\infty(\mathfrak{t})=1
+6\mathfrak{t}^2
+35\mathfrak{t}^4
+162\mathfrak{t}^6
+687\mathfrak{t}^8
+\cdots.
\label{212inftyfirstfew}
\end{align}
Assuming that ${\cal I}_\infty^{(2,1)_2\,(H)}(\mathfrak{t})$ is written in the following infinite product
\begin{align}
{\cal I}_\infty^{(2,1)_2\,(H)}(\mathfrak{t})=\prod_{n=1}^\infty\frac{1}{(1-\mathfrak{t}^{2n})^{a_n}},
\end{align}
we can read off $a_n$ from \eqref{212inftyfirstfew} as
\begin{align}
a_1=6,\quad
a_2=14,\quad
a_3=22,\quad
a_4=30,
\end{align}
which coincides with $8n-2$.
Hence we propose
\begin{align}
{\cal I}_\infty^{(2,1)_2\,(H)}(\mathfrak{t})=\prod_{n=1}^\infty\frac{1}{(1-\mathfrak{t}^{2n})^{8n-2}}.
\end{align}

For the $(2,1)_3$ model, we find ${\cal I}_\infty^{(H)}(\mathfrak{t})$ coincide with the following expressions at least up to possible ${\cal O}(\mathfrak{t}^{8+1})$ corrections:
\begin{align}
&{\cal I}^{(2,1)_3}_\infty(\mathfrak{t})=\prod_{n=1}^8\frac{1}{(1-\mathfrak{t}^n)^{a_n}}+\cdots,
\end{align}
with
\begin{align}
a_1=0,\quad
a_2=4,\quad
a_3=2,\quad
a_4=4,\quad
a_5=8,\quad
a_6=6,\quad
a_7=8,\quad
a_8=12.
\end{align}
Noticing the structure of global linear growth in $n$ combined with an ${\cal O}(1)$ oscillation with period $3$, it is not difficult to guess the following expression for $a_n$:
\begin{align}
a_n=\frac{4n}{3}+\delta_{(n\text{ mod }3)},\quad
\delta_1=-\frac{4}{3},\quad
\delta_2=\frac{4}{3},\quad
\delta_0=-2.
\end{align}
Hence we propose the following infinite produce expression for ${\cal I}^{(2,1)_3\,(H)}_\infty(\mathfrak{t})$
\begin{align}
{\cal I}^{(2,1)_3\,(H)}_\infty(\mathfrak{t})
&=
\prod_{n=1}^\infty\frac{1}{(1-\mathfrak{t}^n)^{\frac{4n}{3}+\delta_{(n\text{ mod }3)}}}\nonumber \\
&=\prod_{m=1}^\infty\frac{1}{
(1-\mathfrak{t}^{3m-2})^{4m-4}
(1-\mathfrak{t}^{3m-1})^{4m}
(1-\mathfrak{t}^{3m})^{4m-2}
}.
\end{align}

For $(2,1)_4$ model, we find (at least up to ${\cal O}(\mathfrak{t}^{4+2})$ corrections)
\begin{align}
&{\cal I}^{(2,1)_4}_{N=\infty}(\mathfrak{t})=\prod_{n=1}^4\frac{1}{(1-\mathfrak{t}^{2n})^{a_n}}+\cdots,
\end{align}
with
\begin{align}
a_1=4,\quad
a_2=6,\quad
a_3=12,\quad
a_4=14.
\end{align}
Again, assuming the superposition of a linear growth and a periodic oscillation with period $2$ we guess
\begin{align}
a_n=4n+\delta_{(n\text{ mod }4)},\quad
\delta_1=0,\quad
\delta_0=-2.
\end{align}
Hence we propose
\begin{align}
{\cal I}^{(2,1)_4\,(H)}_{N=\infty}(\mathfrak{t})=\prod_{n=1}^\infty\frac{1}{(1-\mathfrak{t}^{2n})^{4n+\delta_{(n\text{ mod }2)}}}
=
\prod_{m=1}^\infty\frac{1}{(1-\mathfrak{t}^{4m-2})^{8m-4}
(1-\mathfrak{t}^{4m})^{8m-2}}.
\end{align}

The same strategy works for the $(2,1)_k$ models with $k\ge 5$ and the $(p,1)_k$ models with $p\ge 3$.
The results for general $p$ and $k$ are summarized as
\begin{align}
{\cal I}^{(p,1)_k(H)}_{\infty}
=
\prod_{n=1}^\infty\Bigl(\prod_{j=1}^{[\frac{k}{2}]}
\frac{1}{
(1-\mathfrak{t}^{k(n-1)+2j-1})^{p^2n-p^2}
(1-\mathfrak{t}^{k(n-1)+2j})^{p^2n}
}
\Bigr)
\frac{1}{(1-\mathfrak{t}^{kn})^{p^2n-(p^2-2)}}
\label{IHp1kodd}
\end{align}
for odd $k$ and
\begin{align}
{\cal I}^{(p,1)_k(H)}_{\infty}
=
\prod_{n=1}^\infty\Bigl(\prod_{j=1}^{\frac{k}{2}-1}
\frac{1}{
(1-\mathfrak{t}^{k(n-1)+2j})^{2p^2n-p^2}
}\Bigr)
\frac{1}{(1-\mathfrak{t}^{kn})^{2p^2n-(p^2-2)}}
\label{IHp1keven}
\end{align}
for even $k$.

Now let us investigate the asymptotic degeneracies of ${\cal I}^{(p,1)_k(H)}_\infty(\mathfrak{t})$.
As the simplest non-trivial examples, let us start with $p=2$.
For $p=2$ and odd $k$, the large $N$ Higgs index takes the form 
\begin{align}
\label{inf_21_kodd}
\mathcal{I}_{\infty}^{(2,1)_k (H)}(\mathfrak{t})
&=\prod_{n=1}^{\infty}
\frac{1}{(1-\mathfrak{t}^{kn})^{4n-2}}
\prod_{m=1}^{\frac{k-1}{2}}
\frac{1}{(1-\mathfrak{t}^{kn-2m+1})^{4n} (1-\mathfrak{t}^{kn-2m})^{4n-4} }. 
\end{align}
For example, for $k=1,3$ and $5$ 
\begin{align}
\label{inf_21_1}
\mathcal{I}_{\infty}^{(2,1)_1 (H)}(\mathfrak{t})
&=\prod_{n=1}^{\infty}
\frac{1}{(1-\mathfrak{t}^n)^{4n-2}}, \\
\label{inf_21_3}
\mathcal{I}_{\infty}^{(2,1)_3 (H)}(\mathfrak{t})
&=\prod_{n=1}^{\infty}
\frac{1}{(1-\mathfrak{t}^{3n-2})^{4n-4} (1-\mathfrak{t}^{3n-1})^{4n} (1-\mathfrak{t}^{3n})^{4n-2}}, \\
\label{inf_21_5}
\mathcal{I}_{\infty}^{(2,1)_5 (H)}(\mathfrak{t})
&=\prod_{n=1}^{\infty}
\frac{1}{(1-\mathfrak{t}^{5n-4})^{4n-4} (1-\mathfrak{t}^{5n-3})^{4n} (1-\mathfrak{t}^{5n-2})^{4n-4} (1-\mathfrak{t}^{5n-1})^{4n} (1-\mathfrak{t}^{5n})^{4n-2}}. 
\end{align}
The index (\ref{inf_21_1}) coincides with the large $N$ Higgs index (\ref{inf_U_H2}) 
of the $U(N)$ ADHM theory with two flavors as these theories are dual \cite{Gang:2011xp}. 
By making use of the convolutions $(k-1)$ times, we obtain the asymptotic coefficient 
\begin{align}
\label{asymp_21Hodd}
&{\cal I}^{(2,1)_k(H)}_{\infty}(\mathfrak{t})=\sum_{n\ge 0}d_n\mathfrak{t}^n,\nonumber \\
&d_n\sim 
\exp\left[
3
\left(
\frac{\zeta(3)}{4 \left( \frac{k}{4} \right)}
\right)^{\frac13}
n^{\frac23}
-\frac{\pi^2}{6(2\zeta(3) \frac{k^2}{2})^{\frac13}}
n^{\frac13}
+\frac{1}{3k}
-\frac{\pi^4}{432\zeta(3)k}
\right]
\nonumber\\
&\quad\quad \times 
\frac{n^{\frac19 (k+\frac{4}{k}-9)}}
{2^{\frac{k}{3}+\frac{4}{3k}-2} 
3^{\frac12}
\pi^{-\frac12} 
A^{\frac{4}{k}}
\zeta(3)^{\frac{1}{18} (2k+\frac{8}{k}-9)}
k^{\frac{2k}{9}+\frac{11}{9k}-\frac12}
}
\nonumber\\
&\quad\quad \times 
\prod_{m=1}^{\frac{k-1}{2}}
\Gamma\left(1+ \frac{1-2m}{k} \right)^{-\frac{4(1-2m)}{k}}
\Gamma\left(
1-\frac{2m}{k}
\right)^{-4+\frac{8m}{k}}. 
\end{align}
For example for $k=3$ we have
\begin{align}
\label{21k3_asy_table}
\begin{array}{c|c|c|c} 
n&d_n&d_n^{\textrm{asym}}&d_n/d_n^{\textrm{asym}} \\ \hline 
10   &478                &427.308               &1.11863 \\ 
100  &3.67477\times 10^{17}&3.50195\times 10^{17}&1.04935\\ 
1000 &4.61444\times 10^{90}&4.50079\times 10^{90}&1.02525\\ 
10000&1.56077\times 10^{436}&1.54206\times 10^{436}&1.01213\\ 
\end{array}. 
\end{align}

For $p=2$ and even $k$ we find 
\begin{align}
\mathcal{I}_\infty^{(2,1)_k (H)}(\mathfrak{t})
&=\prod_{n=1}^{\infty}
\frac{1}{(1-\mathfrak{t}^{kn})^{8n-2}}
\prod_{m=1}^{\frac{k-2}{2}}
\frac{1}{(1-\mathfrak{t}^{kn-2m})^{8n-4}}. 
\end{align}
For example, for $k=2,4$ and $6$
\begin{align}
&\mathcal{I}_\infty^{(2,1)_2 (H)}(\mathfrak{t})
=\prod_{n=1}^{\infty}
\frac{1}{(1-\mathfrak{t}^{2n})^{8n-2}},\quad
\mathcal{I}_\infty^{(2,1)_4 (H)}(\mathfrak{t})
=\prod_{n=1}^{\infty}
\frac{1}{(1-\mathfrak{t}^{4n-2})^{8n-4} (1-\mathfrak{t}^{4n})^{8n-2}},\\
&\mathcal{I}_\infty^{(2,1)_6 (H)}(\mathfrak{t})
=\prod_{n=1}^{\infty}
\frac{1}{
(1-\mathfrak{t}^{6n-4})^{8n-4}
(1-\mathfrak{t}^{6n-2})^{8n-4}
(1-\mathfrak{t}^{6n})^{8n-2}}.
\end{align}
By the $k/2$ convolutions 
we obtain the asymptotic coefficient 
\begin{align}
\label{asymp_21Heven}
&{\cal I}^{(2,1)_k(H)}_\infty(\mathfrak{t})=\sum_{n\in 2\mathbb{Z}_{\ge 0}}d_n\mathfrak{t}^n,\nonumber \\
&d_n\sim 
\exp\left[
3
\left(
\frac{\zeta(3)}{4 \left( \frac{k}{4} \right)}
\right)^{\frac13}
n^{\frac23}
-\frac{\pi^2}{6(2\zeta(3) \frac{k^2}{2})^{\frac13}}
n^{\frac13}
+\frac{4}{3k}
-\frac{\pi^4}{432\zeta(3)k}
\right]
\nonumber\\
&\quad\quad \times 
\frac{n^{\frac19 (k+\frac{4}{k}-9)}}
{2^{\frac{k}{3}-3} 3^{\frac12} \pi^{-\frac12} A^{\frac{16}{k}}
\zeta(3)^{\frac{1}{18}(2k+\frac{8}{k}-9)}
k^{\frac{2k}{9}+\frac{20}{9k}-\frac12}
}
\nonumber\\
&\quad\quad \times 
\prod_{m=1}^{\frac{k-2}{2}}
\Gamma\left(1-\frac{2m}{k}\right)^{-4+\frac{16m}{k}}. 
\end{align}
This agrees with the asymptotic behavior (\ref{asymp_21Hodd}) for odd $k$ up to the constant factor. 
For example for $k=4$ we have
\begin{align}
\label{21k4_asy_table}
\begin{array}{c|c|c|c} 
n&d_n&d_n^{\textrm{asym}}&d_n/d_n^{\textrm{asym}} \\ \hline 
10   &528                &459.967              &1.14791 \\ 
100  &2.18583\times 10^{16}&2.02305\times 10^{16}&1.08046\\ 
1000 &8.25665\times 10^{82}&7.95014\times 10^{82}&1.03855\\ 
10000&1.52689\times 10^{397}&1.49982\times 10^{397}&1.01805\\ 
\end{array}.
\end{align}

For general $p$, by applying the formulas \eqref{asy_coeff0} and \eqref{Lconv} to ${\cal I}_\infty^{(p,1)_k(H)}(\mathfrak{t})$ given in \eqref{IHp1kodd} and \eqref{IHp1keven} we obtain
\begin{align}
{\cal I}^{(p,1)_k(H)}_{\infty}=
\begin{cases}
&\displaystyle \sum_{n=0}^\infty d_n^{\textrm{odd}}\mathfrak{t}^n\quad (k\text{: odd})\\
&\displaystyle \sum_{n\in 2\mathbb{Z}_{\ge 0}}^\infty d_n^{\textrm{even}}\mathfrak{t}^n\quad (k\text{: even})
\end{cases},
\end{align}
with
\begin{align}
d_n^{\textrm{odd}}&\sim \text{exp}\Biggl[
3\left(\frac{p^2\zeta(3)}{4k}\right)^{\frac{1}{3}}n^{\frac{2}{3}}
-\frac{(p^2-2)\pi^2}{6(2k^2p^2\zeta(3))^{\frac13}}n^{\frac{1}{3}}
+\frac{p^2}{12k}-\frac{(p^2-2)^2\pi^4}
{432kp^2 \zeta(3)}
\Biggr]
\nonumber\\
&\times 
\frac{
n^{\frac{p^2}{36}\left(k+\frac{4}{k} \right)-1}
}
{
2^{1+\frac{p^2}{36}\left(k+\frac{4}{k}-18 \right)}
3^{\frac12}
\pi^{\frac12(3-p^2)}
A^{\frac{p^2}{k}}
\zeta(3)^{\frac{p^2}{36}\left(k+\frac{4}{k} \right)-\frac12}
k^{\frac{p^2}{36}\left(2k+\frac{11}{k} \right)-\frac12}
p^{\frac{p^2}{18}\left(k+\frac{4}{k} \right)-1}
}
\nonumber\\
&\times 
\prod_{m=1}^{\frac{k-1}{2}}
\Gamma\left(1+\frac{2m-k}{k} \right)^{-\frac{(2m-k)p^2}{k}}
\Gamma\left(1+\frac{2m-k-1}{k}\right)^{-p^2-\frac{(2m-k-1)p^2}{k}}
\label{p1kHodd}
\end{align}
and 
\begin{align}
d_n^{\textrm{even}}&\sim \text{exp}\Biggl[
3\left(\frac{p^2\zeta(3)}{4k}\right)^{\frac{1}{3}}n^{\frac{2}{3}}
-\frac{(p^2-2)\pi^2}{6(2k^2p^2\zeta(3))^{\frac13}}n^{\frac{1}{3}}
+\frac{p^2}{3k}-\frac{(p^2-2)^2\pi^4}
{432kp^2 \zeta(3)}
\Biggr]
\nonumber\\
&\times 
\frac{
n^{\frac{p^2}{36}\left(k+\frac{4}{k} \right)-1}
}
{
2^{\frac{p^2}{36}\left(k-\frac{8}{k}-18 \right)}
3^{\frac12}
\pi^{\frac12(3-p^2)}
A^{\frac{4p^2}{k}}
\zeta(3)^{\frac{p^2}{36}\left(k+\frac{4}{k} \right)-\frac12}
k^{\frac{p^2}{18}\left(k+\frac{10}{k} \right)-\frac12}
p^{\frac{p^2}{18}\left(k+\frac{4}{k} \right)-1}
}
\nonumber\\
&\times 
\prod_{m=1}^{\frac{k}{2}-1}
\Gamma\left(1+\frac{2m-k}{k} \right)^{-p^2-\frac{2(2m-k)}{k}p^2}.
\label{p1kHeven}
\end{align}

We see from (\ref{p1kHodd}) and (\ref{p1kHeven}) that the degeneracy reduces as $k$ grows. 
In the large $k$ limit, we find that
\begin{align}
{\cal I}^{(p,1)_\infty(H)}_{\infty}(\mathfrak{t})=\prod_{n=1}^\infty \frac{1}{(1-\mathfrak{t}^{2n})^{p^2}}.
\end{align}
Applying the Meinardus theorem, we find the asymptotic coefficient
\begin{align}
{\cal I}^{(p,1)_\infty(H)}_{\infty}(\mathfrak{t})=\sum_{n\in 2\mathbb{Z}_{\ge 0}}d_n\mathfrak{t}^n,\quad
d_n\sim 
\Bigl(\frac{p^2}{12}\Bigr)^{\frac{1}{4}+\frac{p^2}{4}}n^{-\frac{3}{4}-\frac{p^2}{4}}\text{exp}\Bigl[\frac{\pi p}{\sqrt{3}}n^{\frac{1}{2}}\Bigr].
\label{p1modellargek}
\end{align}
Here is the comparison between the atcual degeneracy and the analytic result \eqref{p1modellargek} for $p=2$:
\begin{align}
\label{21largek_asy_table}
\begin{array}{c|c|c|c} 
n    &d_n                   &d_n^{\textrm{asym}}   &d_n/d_n^{\textrm{asym}} \\ \hline 
10   &252                   &432.112               &0.583182\\
100  &3.81188\times 10^{11} &4.55050\times 10^{11} &0.837683\\
1000 &8.88977\times 10^{43} &9.40977\times 10^{43} &0.944738\\
10000&8.71694\times 10^{149}&8.87584\times 10^{149}&0.982097\\
\end{array}.
\end{align}

\subsection{Full indices of $U(N)_k\times U(N)_{-k}$ ABJM theory}
\label{sec_ABJMfull}
More generally, the large $N$ limit of the supersymmetric full indices 
(a.k.a. superconformal indices \cite{Bhattacharya:2008zy,Bhattacharya:2008bja,Kim:2009wb,Imamura:2011su,Kapustin:2011jm,Dimofte:2011py})
for the $\mathcal{N}\ge 4$ SCFTs describing $N$ M2-branes 
enumerate the number $d_n$ of the Coulomb branch operators, Higgs branch operators and mixed branch operators with scaling dimension $n/2$.
In this section we study the large $N$ full indices of the $U(N)_{k} \times U(N)_{-k}$ ABJM theory which arises from M2-branes probing $\mathbb{C}^4/\mathbb{Z}_k$. 

The full supersymmetric indices in the large $N$ limit can be obtained from the holographic picture.
The index in the limit $N\to \infty$ can be computed by \cite{Bhattacharya:2008zy, Arai:2020uwd}
\begin{align}\label{ABJM_PE}
I^{\text{ABJM}_k}_\infty(q)=\lim_{N\rightarrow\infty}I^{U(N)_k\times U(N)_{-k}}(q)=\text{PE}[i^{(k)}_{KK}(u_1,u_2,u_3,u_4;q)],
\end{align}
where
\begin{align}
i^{(k)}_{KK}(u_1,u_2,u_3,u_4;q)=\frac{1}{k}\sum_{j=0}^{k-1}i_{KK}\left(\omega_k^ju_1, \omega_k^ju_2, \omega_k^{-j}u_3, \omega_k^{-j}u_4\right),
\label{Zkneutralproj}
\end{align}
with $\omega_k=e^{\frac{2\pi \mathrm{i}}{k}}$ and $i_{KK}(u_1, u_2, u_3, u_4)$ is the single letter index for Kaluza-Klein modes in $AdS_4\times S^7$
\begin{align}
&i_{KK}(u_1, u_2, u_3, u_4)\nonumber \\
&=\frac{\left(1 - q^{\frac{3}{4}}u_1^{-1}\right)\left(1 - q^{\frac{3}{4}}u_2^{-1}\right)\left(1 - q^{\frac{3}{4}}u_3^{-1}\right)\left(1 - q^{\frac{3}{4}}u_4^{-1}\right)}{\left(1 - q^{\frac{1}{4}}u_1\right)\left(1 - q^{\frac{1}{4}}u_2\right)\left(1 - q^{\frac{1}{4}}u_3\right)\left(1 - q^{\frac{1}{4}}u_4\right)(1-q)^2} - \frac{1-q + q^2}{(1-q)^2}. \label{ikk1}
\end{align}
Here $u_1, u_2, u_3, u_4$ are the flavor fugacities
\begin{align}
u_1=t^{-1}z,\quad
u_2=tx,\quad
u_3=t^{-1}z^{-1},\quad
u_4=tx^{-1},
\end{align}
which satisfy $u_1u_2u_3u_4=1$.
The average \eqref{Zkneutralproj} can be calculated by expanding the summand in $q$ and keeping only the terms proportional to $\omega_k^{nj}$ with $n\in k\mathbb{Z}$.
As a result we obtain
\begin{align}
i_{KK}^{(k)}(x,z,t;q)&=
-1+\frac{1}{1-q}
\Bigl[
\sum_{a=1}^2
\Bigl(\prod_{\substack{b=1,2\\ (b\neq a)}}\frac{1-u_au_b^{-1}q}{1-u_a^{-1}u_b}\Bigr)
\Bigl(\prod_{b=3}^4\frac{1-u_a^{-1}u_b^{-1}q^{\frac{1}{2}}}{1-u_au_bq^{\frac{1}{2}}}\Bigr)
\frac{1}{1-u_a^kq^{\frac{k}{4}}}\nonumber \\
&
+\sum_{a=3}^4
\Bigl(\prod_{\substack{b=3,4\\ (b\neq a)}}\frac{1-u_au_b^{-1}q}{1-u_a^{-1}u_b}\Bigr)
\Bigl(\prod_{b=1}^2\frac{1-u_a^{-1}u_b^{-1}q^{\frac{1}{2}}}{1-u_au_bq^{\frac{1}{2}}}\Bigr)
\frac{u_a^kq^{\frac{k}{4}}}{1-u_a^kq^{\frac{k}{4}}}
\Bigr].
\end{align}
In particular, for $x=z=1$ we have
\begin{align}
i_{KK}^{(k)}(t;q)&=\frac{2q^{\frac{1}{2}}}{1-q}
+\frac{2(q^{\frac{k}{2}}+\sum_{n=1}^{k-1}t^{2n-k}q^{\frac{k}{4}})}{(1-t^kq^{\frac{k}{4}})(1-t^{-k}q^{\frac{k}{4}})}
\sum_\pm\Bigl[
\frac{t^{\pm 2}q^{\frac{1}{2}}}{1-t^{\pm 2}q^{\frac{1}{2}}}
+\frac{2t^{\pm k}q^{\frac{1}{2}+\frac{k}{4}}}{(1-q)(1-t^{\pm k}q^{\frac{k}{4}})}\nonumber \\
&
+\frac{2t^{\pm k}q^{\frac{k}{4}}}{(1-t^{\pm 2}q^{\frac{1}{2}})(1-t^{\pm k}q^{\frac{k}{4}})}
\Bigr],
\label{ABJMfullfinitekwitht}
\end{align}
from which it is clear that $i_{KK}^{(k)}(t,q)$ reduces, in the Coulomb limit $q,t\rightarrow 0$ with $\mathfrak{t}=t^{-1}q^{\frac{1}{4}}$ fixed, to the single letter index of the Coulomb index of $U(N)$ ADHM theory with $k$ flavors.

In the following we set $x=z=t=1$ for simplicity.
In this case $i^{(k)}_{KK}(u_1,u_2,u_3,u_4;q)$ simplifies as
\begin{align}
&i_{KK}^{(k)}(q)\nonumber \\
&=
\frac{2q^{\frac{1}{2}}}{1-q^{\frac{1}{2}}}
+\frac{2q^{\frac{1}{2}}}{1-q}
+\frac{2(k-1)q^{\frac{k}{4}}}{1-q^{\frac{k}{4}}}
+\frac{4q^{\frac{k}{4}}}{(1-q^{\frac{1}{2}})(1-q^{\frac{k}{4}})}
+\frac{4q^{\frac{1}{2}+\frac{k}{4}}}{(1-q)(1-q^{\frac{k}{4}})}
+\frac{2kq^{\frac{k}{2}}}{(1-q^{\frac{k}{4}})^2}\nonumber \\
&=\frac{2kq^{\frac{k}{4}}}{(1-q^{\frac{k}{4}})^2} +\frac{\sum_{m=1}^3\left(\left(4\sum_{l=1}^{k-1}q^{l+\frac{mk}{4}}\right) + \left(8\sum_{l=1}^kq^{l+\frac{mk}{4}-\frac{1}{2}}\right)\right)}{(1-q^{k})^2}\nonumber \\
&\hspace{4cm}+\frac{\left(4\sum_{l=1}^{2k}q^{l-\frac{1}{2}}\right)+\left(2\sum_{l=1}^{2k-1}q^{l}\right) + \left(2\sum_{m=1}^7q^{\frac{mk}{4}}\right)}{(1-q^k)^2}.\label{ikk}
\end{align}
Then the large $N$ index \eqref{ABJM_PE} of the ABJM theory becomes 
\begin{align}
&I^{\text{ABJM}_k}_\infty(q)\cr
&=\text{PE}[i^{(k)}_{KK}(q)]\cr
&=\prod_{n=1}^{\infty}\frac{1}{\left(1-q^{\frac{kn}{4}}\right)^{2kn}}\nonumber \\
&\quad \times \prod_{m=1}^3\left\{\left(\prod_{l=1}^{k-1}\frac{1}{\left(1-q^{k(n-1)+l+\frac{mk}{4}}\right)^{4n}}\right)\left(\prod_{l=1}^{k}\frac{1}{\left(1-q^{k(n-1)+l+\frac{mk}{4}-2}\right)^{8n}}\right)\right\}\cr
&\quad\times\left(\prod_{l=1}^{2k}\frac{1}{\left(1-q^{k(n-1)+l-\frac{1}{2}}\right)^{4n}}\right)\left(\prod_{l=1}^{2k-1}\frac{1}{\left(1-q^{k(n-1)+l}\right)^{2n}}\right)\left(\prod_{m=1}^{7}\frac{1}{\left(1-q^{k(n-1)+\frac{mk}{4}}\right)^{2n}}\right),\cr\label{inf_ABJM}
\end{align}
which consists of $10k+4$ infinite products of the form $\prod_{n=1}^\infty\frac{1}{(1-q^{\mathfrak{c}n+\mathfrak{d}})^{\mathfrak{a}n+\mathfrak{b}}}$ \eqref{inf_prod_abcd}.

\subsubsection{$k=1$}
When a stack of $N$ coincident M2-branes probes the flat space $\mathbb{C}^4$, 
the low-energy effective description of the M2-branes is given by the $U(N)_1\times U(N)_{-1}$ ABJM theory 
or equivalently by the $U(N)$ ADHM theory with a fundamental hypermultiplet. 

In the large $N$ limit the full index is given by \eqref{inf_ABJM} with $k=1$. It is possible to rewrite it with fewer factors as
\begin{align}
&I^{\text{ABJM}_1}_\infty(q)=
I^{U(\infty)+\textrm{adj.}-[1]}(q)
\nonumber\\
&=\text{PE}\left[\frac{2q^{\frac{1}{4}}}{\left(1-q^{\frac{1}{4}}\right)^2} +  \frac{2q^{\frac{1}{4}}+6q^{\frac{1}{2}}+10q^{\frac{3}{4}}+12q+10q^{\frac{5}{4}}+6q^{\frac{3}{2}}+2q^{\frac{7}{4}}}{\left(1-q\right)^2} \right]\cr
&=\text{PE}\left[\frac{2q^{\frac{1}{4}}\left(1+q^{\frac{1}{4}}+q^{\frac{1}{2}}+q^{\frac{3}{4}}\right)^2}{\left(1-q\right)^2} +  \frac{2q^{\frac{1}{4}}+6q^{\frac{1}{2}}+10q^{\frac{3}{4}}+12q+10q^{\frac{5}{4}}+6q^{\frac{3}{2}}+2q^{\frac{7}{4}}}{\left(1-q\right)^2}\right]\cr
&=\text{PE}\left[\frac{20q^{\frac{1}{4}}}{\left(1-q\right)^2}+\frac{20q^{\frac{1}{2}}}{\left(1-q\right)^2}+\frac{20q^{\frac{3}{4}}}{\left(1-q\right)^2}+\frac{20q}{\left(1-q\right)^2}-\frac{16q^{\frac{1}{4}}}{\left(1-q\right)}-\frac{10q^{\frac{1}{2}}}{\left(1-q\right)}-\frac{4q^{\frac{3}{4}}}{\left(1-q\right)}\right]\cr
&=\prod_{n=1}^{\infty}\frac{1}{\left(1-q^{n-\frac{3}{4}}\right)^{20n-16}\left(1-q^{n-\frac{1}{2}}\right)^{20n-10}\left(1-q^{n-\frac{1}{4}}\right)^{20n-4}\left(1-q^{n}\right)^{20n}}\label{inf_full_k1}.
\end{align}
We can further put together some factors in the denominator of \eqref{inf_full_k1} and obtain
\begin{align}
\label{inf_fullk1_2}
&
I^{\text{ABJM}_1}_\infty(q)=
I^{U(\infty)+\textrm{adj.}-[1]}(q)\nonumber \\
&=
\prod_{n=1}^{\infty}
\frac{1}{(1-q^{\frac{n}{2}})^{10n} (1-q^{n-\frac34})^{20n-16} (1-q^{n-\frac14})^{20n-4}}
\nonumber\\ 
&=
\prod_{n=1}^{\infty}
\frac{1}{(1-q^{\frac{n}{4}})^{5n-1} }
\prod_{n=1}^{\infty}
\frac{1}{(1-q^{\frac{n}{2}}) (1-q^{n-\frac14})^{2}}. 
\end{align}

Note that the full index should be compared with the Coulomb/Higgs indices with $\mathfrak{t}$ substituted with $q^{\frac{1}{4}}$.
For example, the index counting the degeneracy of mixed operators is given by $\frac{I(q)}{{\cal I}^{(C)}(\mathfrak{t}=q^{\frac{1}{4}}){\cal I}^{(H)}(\mathfrak{t}=q^{\frac{1}{4}})}$.
Applying the Meinardus theorem and the convolutions, we find that 
the coefficients $d_n$ of $q^{\frac{n}{4}}$ of the large $N$ index (\ref{inf_full_k1}) has the asymptotic behavior
\begin{align}
\label{asymp_ABJM1}
d_n&\sim 
\frac{2^{\frac{35}{36}} (5\zeta(3))^{\frac{17}{36}} }{3^{\frac12}\pi^{\frac12} A^5 n^{\frac{35}{36}}}
\frac{\Gamma\left(\frac34 \right)}
{\Gamma\left(\frac14 \right)}
\exp\left[
3\left(
\frac{5\zeta(3)}{4}
\right)^{\frac13}n^{\frac23}
+\frac{5}{12}
\right].
\end{align}
Interestingly, the coefficient of $n^{1/3}$ in the exponential vanishes. 
Actually this is true also for $k\ge 2$, which can be seen from the structure of the infinite product \eqref{inf_ABJM} as follows.
Since the asmyptotic formula for $d_n$ is independent of the order of the convolutions, let us perform the convolution first for the following subfactors in $I_\infty^{\text{ABJM}_k}(q)$ separately:
\begin{align}
&\prod_{n=1}^{\infty}\frac{1}{\left(1-q^{\frac{kn}{4}}\right)^{2kn}}\quad (\text{no convolution}),\label{ABJMfull1stset} \\
&\prod_{m=1}^3
\prod_{l=1}^{k-1}
\prod_{n=1}^{\infty}
\frac{1}{\left(1-q^{k(n-1)+l+\frac{mk}{4}}\right)^{4n}},\label{ABJMfull2ndset}\\
&\prod_{m=1}^3
\prod_{l=1}^{k}
\prod_{n=1}^{\infty}
\frac{1}{\left(1-q^{k(n-1)+l+\frac{mk}{4}-2}\right)^{8n}},\\
&\prod_{l=1}^{2k}
\prod_{n=1}^{\infty}
\frac{1}{\left(1-q^{k(n-1)+l-\frac{1}{2}}\right)^{4n}},\\
&\prod_{l=1}^{2k-1}
\prod_{n=1}^{\infty}
\frac{1}{\left(1-q^{k(n-1)+l}\right)^{2n}},\\
&\prod_{m=1}^{7}
\prod_{n=1}^{\infty}
\frac{1}{\left(1-q^{k(n-1)+\frac{mk}{4}}\right)^{2n}},
\end{align}
and then convoluting the six asymptotic degeneracies.
Now let us calculate $\beta$ for each set of subfactors.
Since $\mathfrak{a}$ and $\mathfrak{c}$ are common and $\mathfrak{b}=0$ within each set of subfactors, the asymptotic formula \eqref{asy_coeff0} and the convolution formula \eqref{Lconv} imply that $\beta$ for each set satisfies
\begin{align}
\beta\propto \sum_{i=1}^L \mathfrak{d}_i.
\end{align}
This turns out to vanish for each set.
For example, for the first set \eqref{ABJMfull1stset} we have a single factor with $\mathfrak{d}=0$.
For the second set \eqref{ABJMfull2ndset} we have ($L=3(k-1)$)
\begin{align}
\{\mathfrak{d}_i\}_{i=1}^L=\Bigl\{-k+l+\frac{mk}{4}\Bigr\}_{1\le m\le 3,1\le l\le k-1},
\end{align}
which satisfies
\begin{align}
\sum_{i=1}^L\mathfrak{d}_i=\sum_{m=1}^3\sum_{l=1}^{k-1}\Bigl(-k+l+\frac{mk}{4}\Bigr)=0,
\end{align}
and so on.
Since we thus find $\beta=0$ for all six sets of subfactors, we find the full index obtained by the convolution of these sets also has $\beta=0$.

The actual coefficients $d_n$ and the analytic values for $k=1$ are given by
\begin{align}
\label{ABJMk1_asy_table}
\begin{array}{c|c|c|c} 
n&d_n&d_n^{\textrm{asym}}&d_n/d^{\textrm{asym}}_n \\ \hline 
10&178578&198322&0.900444 \\
100&3.45762\times 10^{29} &3.53844\times 10^{29}&0.977160\\
1000&4.49045\times 10^{145} &4.51297\times 10^{145}&0.995010\\
10000&1.28563\times 10^{688} &1.28702\times 10^{688}&0.998922 \\
\end{array}.
\end{align}
We see that the asymptotic formula (\ref{asymp_ABJM1}) approaches the actual value as $n$ becomes large.

\subsubsection{$k=2$}
For the $U(N)_2\times U(N)_{-2}$ ABJM theory the large $N$ full index can be obtained from \eqref{inf_ABJM} with $k=2$. The expression can be simplified and it becomes
\begin{align}
\label{inf_fullk2}
I^{\text{ABJM}_2}_\infty(q)
&=\text{PE}\left[\frac{4q^{\frac{1}{4}}}{\left(1-q^{\frac{1}{2}}\right)^2}+\frac{6q^{\frac{1}{2}}}{\left(1-q^{2}\right)^2}+\frac{12q}{\left(1-q^{2}\right)^2}+\frac{18q^{\frac{3}{2}}}{\left(1-q^{2}\right)^2}\right.\cr
&\hspace{4cm}\left.+\frac{24q^2}{\left(1-q^{2}\right)^2}+\frac{18q^{\frac{5}{2}}}{\left(1-q^{2}\right)^2}+\frac{12q^3}{\left(1-q^{2}\right)^2}+\frac{6q^{\frac{7}{2}}}{\left(1-q^{2}\right)^2}\right]\cr
&\text{PE}\left[\frac{4q^{\frac{1}{4}}\left(1+q^{\frac{1}{2}}\right)^2\left(1+q\right)^2}{\left(1-q^2\right)^2}+\frac{6q^{\frac{1}{2}}}{\left(1-q^{2}\right)^2}+\frac{12q}{\left(1-q^{2}\right)^2}+\frac{18q^{\frac{3}{2}}}{\left(1-q^{2}\right)^2}\right.\cr
&\hspace{4cm}\left.+\frac{24q^2}{\left(1-q^{2}\right)^2}+\frac{18q^{\frac{5}{2}}}{\left(1-q^{2}\right)^2}+\frac{12q^3}{\left(1-q^{2}\right)^2}+\frac{6q^{\frac{7}{2}}}{\left(1-q^{2}\right)^2}\right]\cr
&=\text{PE}\left[\frac{10q^{\frac{1}{2}}}{\left(1-q^{\frac{1}{2}}\right)^2}\right]\cr
&=\prod_{n=1}^{\infty}
\frac{1}{(1-q^{\frac{n}{2}})^{10n}}. 
\end{align}
The asymptotic degeneracy can be obtained without convolutions. 
By applying the Meinardus theorem (\ref{Meinardus_thm}), we obtain
\begin{align}
\label{asymp_ABJM2}
I^{\text{ABJM}_2}_\infty(q)=\sum_{n\in 2\mathbb{Z}_{\ge 0}}d_nq^{\frac{n}{4}},\quad
d_n\sim 
\frac{2^{\frac43}(5\zeta(3))^{\frac49}}
{3^{\frac12}\pi^{\frac12}A^{10}n^{\frac{17}{18}}}
\exp\left[
3\left(
\frac{5\zeta(3)}{4\cdot 2}
\right)^{\frac13}n^{\frac23}
+\frac56
\right]. 
\end{align}
We show the actual coefficients $d_n$ and 
the analytic values $d_n^{\text{asym}}$ evaluated from the asymptotic formula (\ref{asymp_ABJM2})
\begin{align}
\label{ABJMk2_asy_table}
\begin{array}{c|c|c|c} 
n&d_n&d_n^{\textrm{asym}}&d_n/d_n^{\text{asym}} \\ \hline 
10&11202&12449.8&0.899771 \\
100&1.4483\times 10^{23} &1.48168\times 10^{23}&0.977471 \\
1000&1.40911\times 10^{115} &1.41605\times 10^{115}&0.995098 \\
10000&3.34768\times 10^{545} &3.35122\times 10^{545}&0.998942\\
\end{array}.
\end{align}
We see that the asymptotic formula (\ref{asymp_ABJM2}) approaches the actual value as $n$ becomes large. 

\subsubsection{$k=3$}
When $k\ge3$ the ABJM theory preserves $\mathcal{N}=6$ supersymmetry. 
The large $N$ full index of the $U(N)_3\times U(N)_{-3}$ ABJM theory is is given by \eqref{inf_ABJM} with $k=3$.
It is possible to rewrite it with fewer factors as
\begin{align}
\label{asymp_ABJM3}
&I^{\text{ABJM}_3}_\infty(q)\cr
&=\text{PE}\left[\frac{6q^{\frac{3}{4}}}{\left(1-q^{\frac{3}{4}}\right)^2}+\frac{1}{\left(1-q^{3}\right)^2}\left(4q^{\frac{1}{2}}+2q^{\frac{3}{4}}+2q+8q^{\frac{5}{4}}+6q^{\frac{3}{2}}\right.\right.\cr
&\hspace{2cm}\left.\left.+4q^{\frac{7}{4}}+10q^2+10q^{\frac{9}{4}}+8q^{\frac{5}{2}}+12q^{\frac{11}{4}}+12q^3+\cdots\left(\text{palindrome}\right)\cdots+4q^{\frac{11}{2}}\right)\right]\cr
&=\text{PE}\left[\frac{6q^{\frac{3}{4}}\left(1+q^{\frac{3}{4}}+q^{\frac{3}{2}}+q^{\frac{9}{4}}\right)^2}{\left(1-q^3\right)^2}+\frac{1}{\left(1-q^{3}\right)^2}\left(4q^{\frac{1}{2}}+2q^{\frac{3}{4}}+2q+8q^{\frac{5}{4}}+6q^{\frac{3}{2}}\right.\right.\cr
&\hspace{2cm}\left.\left.+4q^{\frac{7}{4}}+10q^2+10q^{\frac{9}{4}}+8q^{\frac{5}{2}}+12q^{\frac{11}{4}}+12q^3+\cdots\left(\text{palindrome}\right)\cdots+4q^{\frac{11}{2}}\right)\right]\cr
&=\text{PE}\left[\frac{12\left(q^{\frac{1}{4}}+q^{\frac{1}{2}}+3q^{\frac{3}{4}}+q+q^{\frac{5}{4}}+3q^{\frac{3}{2}}+q^{\frac{7}{4}}+q^2+3q^{\frac{9}{4}}+q^{\frac{5}{2}}+q^{\frac{11}{4}}+3q^3\right)}{\left(1-q^3\right)^2}\right.\cr
&\hspace{3cm}\left.-\frac{12q^{\frac{1}{4}}+8q^{\frac{1}{2}}+28q^{\frac{3}{4}}+10q+4q^{\frac{5}{4}}+18q^{\frac{3}{2}}+8q^{\frac{7}{4}}+2q^2+8q^{\frac{9}{4}}+4q^{\frac{5}{2}}}{\left(1-q^3\right)}\right]\cr
&=\prod_{n=1}^{\infty}\frac{1}{\left(1-q^{3n-\frac{11}{4}}\right)^{12n-12}\left(1-q^{3n-\frac{5}{2}}\right)^{12n-8}\left(1-q^{3n-\frac{9}{4}}\right)^{36n-28}\left(1-q^{3n-2}\right)^{12n-10}}\cr
&\quad\times\frac{1}{\left(1-q^{3n-\frac{7}{4}}\right)^{12n-4}\left(1-q^{3n-\frac{3}{2}}\right)^{36n-18}\left(1-q^{3n-\frac{5}{4}}\right)^{12n-8}\left(1-q^{3n-1}\right)^{12n-2}}\cr
&\quad\times\frac{1}{\left(1-q^{3n-\frac{3}{4}}\right)^{36n-8}\left(1-q^{3n-\frac{1}{2}}\right)^{12n-4}\left(1-q^{3n-\frac{1}{4}}\right)^{12n}\left(1-q^{3n}\right)^{36n}}.
\end{align}
By applying the Meinardus theorem and convolution theorem, 
we find the asymptotic degeneracy 
\begin{align}
\label{asymp_ABJM32}
&I^{\text{ABJM}_3}_\infty(q)=\sum_{n\ge 0}d_nq^{\frac{n}{4}},\nonumber \\
&d_n\sim 
\frac{
(5\zeta(3))^{\frac{5}{12}}
}
{
2^{\frac{1}{12}} 3^{\frac14} \pi^{\frac12} A^7 n^{\frac{11}{12}}
}
\frac{
\Gamma\left(\frac{3}{4} \right)
\Gamma\left(\frac{5}{12} \right)^{2}
}
{
\Gamma\left(\frac{1}{4} \right)
\Gamma\left(\frac{7}{12} \right)^{2}
}
\exp\left[
3 \left(
\frac{5 \zeta(3)}{4\cdot 3}
\right)^{\frac13} n^{\frac23}
+\frac{7}{12}
\right]. 
\end{align}
The actual coefficients and the values evaluated from (\ref{asymp_ABJM32}) are 
\begin{align}
\label{ABJMk3_asy_table}
\begin{array}{c|c|c|c} 
n&d_n&d_n^{\textrm{asym}}&d_n/d^{\textrm{asym}}_n \\ \hline 
10&1128&1351.02&0.834926 \\
100&4.92925\times 10^{19} &5.04784\times 10^{19}&0.976506\\
1000&9.15118 \times 10^{99} &9.19818\times 10^{99}&0.994890 \\
10000&6.88703\times 10^{475} &6.89464\times 10^{475}&0.998897 \\
\end{array}.
\end{align}

\subsubsection{$k=4$}
The large $N$ limit of the $U(N)_{4}\times U(N)_{-4}$ ABJM theory can be obtained from \eqref{inf_ABJM} with $k=4$. In this case the infinite product also admits a simpler expression
\begin{align}
\label{inf_fullk4}
&I^{\text{ABJM}_4}_\infty(q)\cr
&=\text{PE}\left[\frac{8q}{\left(1-q\right)^2}+\frac{1}{\left(1-q^4\right)^2}\left(4q^{\frac{1}{2}}+4q+12q^{\frac{3}{2}}+8q^2+20q^{\frac{5}{2}}\right.\right.\cr
&\hspace{5cm}\left.\left.+12q^3+28q^{\frac{7}{2}}+16q^4+\cdots\left(\text{parlindrome}\right)\cdots+4q^{\frac{15}{2}}\right)\right]\cr
&=\text{PE}\left[\frac{8q\left(1+q\right)^2\left(1+q^2\right)^2}{\left(1-q^4\right)^2}+\frac{1}{\left(1-q^4\right)^2}\left(4q^{\frac{1}{2}}+4q+12q^{\frac{3}{2}}+8q^2+20q^{\frac{5}{2}}\right.\right.\cr
&\hspace{5cm}\left.\left.+12q^3+28q^{\frac{7}{2}}+16q^4+\cdots\left(\text{parlindrome}\right)\cdots+4q^{\frac{15}{2}}\right)\right]\cr
&=\text{PE}\left[\frac{4q^{\frac{1}{2}}}{(1-q)^2} +\frac{12q}{(1-q)^2}+\frac{4q^{\frac{3}{2}}}{\left(1-q\right)^2}\right]\cr
&=\text{PE}\left[\frac{8q^{\frac{1}{2}}}{(1-q)^2} +\frac{12q}{(1-q)^2} -\frac{4q^{\frac{1}{2}}}{1-q}\right]\cr
&=
\prod_{n=1}^{\infty}
\frac{1}{(1-q^{n-\frac12})^{8n-4} (1-q^{n})^{12n}}. 
\end{align}
According to the Meinardus theorem (\ref{Meinardus_thm}) and the convolution theorem (\ref{conv_thm}), 
we obtain the asymptotic behavior 
\begin{align}
\label{asymp_ABJM4}
I^{\text{ABJM}_4}_\infty(q)=\sum_{n\in 2\mathbb{Z}_{\ge 0}}d_nq^{\frac{n}{4}},\quad
d_n\sim 
\frac{
2^{\frac{10}{9}} 
(5\zeta(3))^{\frac{7}{18}}
}
{3^{\frac12}\pi^{\frac12} A^{8} n^{\frac{8}{9}}}
\exp\left[
3\left( \frac{5 \zeta(3)}{4\cdot 4} \right)^{\frac13}n^{\frac23}
+\frac23
\right].
\end{align}
The actual coefficients and the values evaluated from (\ref{asymp_ABJM4}) are 
\begin{align}
\label{ABJMk4_asy_table2}
\begin{array}{c|c|c|c} 
n&d_n&d_n^{\textrm{asym}}&d_n/d^{\textrm{asym}}_n \\ \hline 
10&988&1122.23&0.880392 \\
100&1.09744\times 10^{18} &1.12487\times 10^{18}&0.975608 \\
1000&8.20995 \times 10^{90} &8.25378\times 10^{90}&0.994689 \\
10000&2.31761\times 10^{432} &2.32027\times 10^{432}&0.998853 \\
\end{array}.
\end{align}

We further consider the explicit comparison between the actual coefficients and the asymptotic coefficients for the expansion of the large $N$ full indices of the ABJM theories for $k=5, 6, 7, 8$ in Appendix \ref{app_ABJMfull}.

\subsubsection{General $k$}

Also for general $k$, we can rewrite $I_\infty^{\text{ABJM}_k}(q)$ \eqref{inf_ABJM} for each $k$ mod 4 into the expressions with a fewer set of the infinite products $\prod_{n=1}^\infty\frac{1}{(1-q^{\mathfrak{c}n+\mathfrak{d}})^{\mathfrak{a}n+\mathfrak{b}}}$.
Hence we can calculate the asymtotics of the total degeneracy with fewer convolutions.
For $k=0$ mod 4 we can simplify \eqref{inf_ABJM} as
\begin{align}
\label{inf_fullk4l}
I^{\text{ABJM}_k}_\infty(q)
&=
\prod_{n=1}^{\infty}
\prod_{m=1}^{\frac{k}{2}}
\frac{1}{
(1-q^{\frac{kn}{4}+m-\frac{k}{4}-\frac12})^{4n}
}
\nonumber\\
&\quad \times 
\prod_{n=1}^{\infty}
\prod_{m=1}^{\frac{k}{4}-1}
\frac{1}{
(1-q^{\frac{kn}{4}+m-\frac{k}{4}})^{2n}
(1-q^{\frac{kn}{4}+m})^{2n}
}
\nonumber\\
&\quad \times 
\prod_{n=1}^{\infty}
\frac{1}
{
(1-q^{\frac{kn}{4}})^{(2k+4)n}
},
\end{align}
where the number of infinite products is $k-1$.
For $k=1$ mod 4, we can simplify \eqref{inf_ABJM} as
\begin{align}
&I_\infty^{\text{ABJM}_k}(q)\nonumber \\
&=
\prod_{n=1}^\infty
\frac{1}{
(1-q^{kn-\frac{3k}{4}})^{(8k+12)n-6k-10}
(1-q^{kn-\frac{k}{2}})^{(8k+12)n-4k-6}
(1-q^{kn-\frac{k}{4}})^{(8k+12)n-2k-2}
}\nonumber \\
&\quad \times \frac{1}{
(1-q^{kn})^{(8k+12)n}
}\nonumber \\
&\quad \times \prod_{c=1}^{\frac{k-1}{4}}\frac{1}{
(1-q^{kn-k-\frac{3}{4}+c})^{12n-12}
(1-q^{kn-k-\frac{1}{2}+c})^{12n-8}
(1-q^{kn-k-\frac{1}{4}+c})^{12n-12}
}\nonumber \\
&\quad \times \frac{1}{
(1-q^{kn-k+c})^{12n-10}
(1-q^{kn-\frac{3k}{4}-\frac{3}{4}+c})^{12n-8}
(1-q^{kn-\frac{3k}{4}-\frac{1}{2}+c})^{12n-4}
}\nonumber \\
&\quad \times \frac{1}{
(1-q^{kn-\frac{3k}{4}-\frac{1}{4}+c})^{12n-10}
(1-q^{kn-\frac{3k}{4}+c})^{12n-8}
(1-q^{kn-\frac{k}{2}-\frac{3}{4}+c})^{12n-4}
}\nonumber \\
&\quad \times \frac{1}{
(1-q^{kn-\frac{k}{2}-\frac{1}{2}+c})^{12n-2}
(1-q^{kn-\frac{k}{2}-\frac{1}{4}+c})^{12n-8}
(1-q^{kn-\frac{k}{2}+c})^{12n-4}
}\nonumber \\
&\quad \times \frac{1}{
(1-q^{kn-\frac{k}{4}-\frac{3}{4}+c})^{12n-2}
(1-q^{kn-\frac{k}{4}-\frac{1}{2}+c})^{12n}
(1-q^{kn-\frac{k}{4}-\frac{1}{4}+c})^{12n-4}
(1-q^{kn-\frac{k}{4}+c})^{12n}
},\label{inf_fullk4l+1}
\end{align}
where the number of infinite products is $4k$.
For $k=2$ mod 4 we can simplify \eqref{inf_ABJM} as
\begin{align}
\label{inf_fullk4l-2}
I^{\text{ABJM}_k}_\infty(q)
&=
\prod_{n=1}^{\infty}
\prod_{m=1}^{\frac{k-2}{2}}
\frac{1}{
(1-q^{\frac{kn}{4} +m-\frac{k}{4}})^{2n}
}\nonumber \\
&\quad \times \prod_{n=1}^{\infty}
\prod_{m=1}^{\frac{k-2}{4}}
\frac{1}{
(1-q^{\frac{kn}{4} +m-\frac{k+2}{4}})^{4n}
(1-q^{\frac{kn}{4} +m})^{4n}
}\nonumber \\
&\quad \times \prod_{n=1}^{\infty}
\frac{1}
{
(1-q^{\frac{kn}{4}})^{(2k+6)n}
}, 
\end{align}
where the number of infinite products is $k-1$.
For $k=3$ mod 4 we can simplify \eqref{inf_ABJM} as
\begin{align}
&I_\infty^{\text{ABJM}_k}(q)\nonumber \\
&=\prod_{n=1}^\infty \frac{1}{
(1-q^{kn-\frac{3k}{4}})^{(8k+12)n-6k-10}
(1-q^{kn-\frac{k}{2}})^{(8k+12)n-4k-6}
(1-q^{kn-\frac{k}{4}})^{(8k+12)n-2k-2}
}\nonumber \\
&\quad \times \frac{1}{
(1-q^{kn})^{(8k+12)n}
(1-q^{kn-k+\frac{1}{4}})^{12n-12}
(1-q^{kn-k+\frac{1}{2}})^{12n-8}
(1-q^{kn-\frac{k}{2}-\frac{1}{2}})^{12n-10}
}\nonumber \\
&\quad \times \frac{1}{
(1-q^{kn-\frac{k}{2}-\frac{1}{4}})^{12n-4}
(1-q^{kn-\frac{k}{2}+\frac{1}{4}})^{12n-8}
(1-q^{kn-\frac{k}{2}+\frac{1}{2}})^{12n-2}
(1-q^{kn-\frac{1}{2}})^{12n-4}
}\nonumber \\
&\quad \times \frac{1}{
(1-q^{kn-\frac{1}{4}})^{12n}
}\nonumber \\
&\quad \times \prod_{c=1}^{\frac{k-3}{4}}\frac{1}{
(1-q^{kn-k-\frac{1}{4}+c})^{12n-12}
(1-q^{kn-k+c})^{12n-10}
(1-q^{kn-k+\frac{1}{4}+c})^{12n-12}
}\nonumber \\
&\quad \times \frac{1}{
(1-q^{kn-k+\frac{1}{2}+c})^{12n-8}
(1-q^{kn-\frac{3k}{4}-\frac{3}{4}+c})^{12n-10}
(1-q^{kn-\frac{3k}{4}-\frac{1}{2}+c})^{12n-4}
}\nonumber \\
&\quad \times \frac{1}{
(1-q^{kn-\frac{3k}{4}-\frac{1}{4}+c})^{12n-8}
(1-q^{kn-\frac{3k}{4}+c})^{12n-8}
(1-q^{kn-\frac{k}{2}-\frac{1}{4}+c})^{12n-4}
(1-q^{kn-\frac{k}{2}+c})^{12n-4}
}\nonumber \\
&\quad \times \frac{1}{
(1-q^{kn-\frac{k}{2}+\frac{1}{4}+c})^{12n-8}
(1-q^{kn-\frac{k}{2}+\frac{1}{2}+c})^{12n-2}
(1-q^{kn-\frac{k}{4}-\frac{3}{4}+c})^{12n-4}
(1-q^{kn-\frac{k}{4}-\frac{1}{2}+c})^{12n}
}\nonumber \\
&\quad \times \frac{1}{
(1-q^{kn-\frac{k}{4}-\frac{1}{4}+c})^{12n-2}
(1-q^{kn-\frac{k}{4}+c})^{12n}
},\label{inf_fullk4l-1}
\end{align}
where the number of infinite products is $4k$.
Making use of the Meinardus theorem and the convolutions, we find the asymptotic growth
\begin{align}
\label{asymp_ABJMk}
I^{\text{ABJM}_k}_\infty(q)=\sum_n d_nq^{\frac{n}{4}},\quad
d_n=
n^{-\frac{36-k}{36}}
\exp\left[
3\left( \frac{5 \zeta(3)}{4\cdot k} \right)^{\frac13}n^{\frac23}
+0\cdot n^{\frac{1}{3}}+\cdots \right],
\end{align}
where the coefficient of $n^{1/3}$ in the exponential precisely vanishes, as we have shown below \eqref{asymp_ABJM1}.
 where the subleading coefficient precisely vanishes. 
Here $n$ runs over $\mathbb{Z}_{\ge 0}$ for $k\in 2\mathbb{N}-1$, while $n$ runs over $2\mathbb{Z}_{\ge 0}$ for $k\in 2\mathbb{Z}$.

When we view the $S^7$ of the holographic dual geometry $AdS_4\times S^7$ as an $S^1$ Hopf fibration over $\mathbb{CP}^3$, 
the circle has a constant radius of the order $N/k^5$ in such a way that the M-theory description is valid when $k^5\ll N$. 
When $k$ becomes large, the radius becomes small \cite{Nilsson:1984bj}, 
which will lead to Type IIA string theory on $AdS_4\times \mathbb{CP}^3$ \cite{Aharony:2008ug}. 
From (\ref{asymp_ABJMk}) we see that the asymptotic coefficient decreases as $k$ increases. 
In the large $k$ limit we find
\begin{align}
i_{KK}^{(k=\infty)}(x,z,t;q)=-\frac{2q}{1-q}+\sum_\pm \Bigl[\frac{t^{\pm 2}q^{\frac{1}{2}}}{1-t^{\pm 2}q^{\frac{1}{2}}}+\frac{x^{\pm 1}z^{\mp 1}q^{\frac{1}{2}}}{1-x^{\pm 1}z^{\mp 1}q^{\frac{1}{2}}}\Bigr],
\end{align}
hence the large $N$ index is
\begin{align}
\lim_{k\rightarrow\infty}I^{\text{ABJM}_k}_\infty(x,z,t,q)=\frac{(q;q)_\infty^2}{\prod_\pm (t^{\pm 2}q^{\frac{1}{2}};t^{\pm 2}q^{\frac{1}{2}})_\infty
(x^{\pm 1}z^{\mp 1}q^{\frac{1}{2}};x^{\pm 1}z^{\mp 1}q^{\frac{1}{2}})_\infty
}.
\end{align}
In particular, for $x=z=t=1$, this can be written with the Jacobi theta function $\vartheta_4(z;q)$ with argument $z=0$
\begin{align}
\vartheta_4(0;q)
=\sum_{n\in \mathbb{Z}} (-1)^n q^{\frac{n^2}{2}}
=(q;q)_{\infty} (q^{\frac12};q)_{\infty}^2,
\label{theta4}
\end{align}
as
\begin{align}
&
\lim_{k\rightarrow\infty}I^{\text{ABJM}_k}_\infty(q)
=\vartheta_4(0;q)^{-2}
=1+4q^{\frac{1}{2}}+12q+32q^{\frac{3}{2}}+76q^2+168q^{\frac{5}{2}}
+352q^3\nonumber \\
&\quad\quad\quad\quad\quad\quad\quad\quad\quad\quad\quad\quad\quad\quad\quad\quad +704q^{\frac{7}{2}}+1356q^4+2532q^{\frac{9}{2}}
+4600q^5+\cdots.
\label{inf_fullkinf}
\end{align}

Making use of the Meinardus theorem, we obtain the asymptotic coefficient of the large $N$ and $k$ index (\ref{inf_fullkinf})
\begin{align}
\label{asymp_largek}
\lim_{k\rightarrow\infty}I^{\text{ABJM}_k}_\infty(q)=\sum_{n\in 2\mathbb{Z}_{\ge 0}}d_nq^{\frac{n}{4}},\quad
d_n\sim 
\frac{\exp
\left[
(3 \cdot 2\zeta(2))^{\frac12} n^{\frac12}
\right]
}{2^{\frac52} n^{\frac54}}.
\end{align}
The actual coefficients $d_n$ and the values evaluated from (\ref{asymp_largek}) are 
\begin{align}
\label{kinf_asy_table}
\begin{array}{c|c|c|c} 
n&d_n&d_n^{\textrm{asym}}&d_n/d^{\textrm{asym}}_n \\ \hline 
10&168 &205.098&0.819120\\
100&2.3166\times 10^{10}&2.46144\times 10^{10}&0.941158\\
1000&4.31078\times 10^{38}&4.39333\times 10^{38}&0.981210\\
10000&4.81354\times 10^{130} &4.8424\times 10^{130}&0.994040 \\
\end{array}.
\end{align}

\section{Discussion}
\label{sec_semi1}

\subsection{Semiclassical quantization of membrane}
\label{sec_semi_membrane}
In the preceding sections we have observed that the degeneracies of the indices of M2-brane SCFTs enjoy a universal rule of the large charge asymptotics $\log d_n\sim n^{2/3}$.
We have also observed that this reduces to $\log d_n\sim n^{1/2}$ in the type IIA limit where M2-branes reduce to the fundamental strings. 
Intriguingly, these degeneracies of indices also arise in the semiclassical quantization of an M2-brane (and fundamental string) compactified on torus, as we review below.

In the semiclassical quantization of the M2-brane \cite{Duff:1987cs,Mezincescu:1987kj,Bergshoeff:1987qx}, the fluctuation spectrum of the bosonic sector of the M2-brane compactified on a torus and propagating in $\mathbb{R}^{8}$ is given by the eigenvalue of the total ``number'' operator
\begin{align}
\hat{N}&=
\sum_{j=1}^{8}
\sum_{(n_1,n_2)\in \mathbb{Z}^2\setminus \{(0,0)\}}\omega_{n_1,n_2} {\hat{a}^{\dag j}_{n_1,n_2}} \hat{a}_{n_1,n_2}^{j},
\end{align}
where
\begin{align}
\omega_{n_1,n_2}
&=(n_1^2+n_2^2)^{\frac12}
\end{align}
is the frequency and ${\hat{a}^{\dag j}}_{n_1,n_2}$ and $\hat{a}_{n_1,n_2}^j$ are creation and annihilation operators corresponding to the fluctuation modes along the $j$-th direction of $\mathbb{R}^8$. 
For simplicity we have set the M2-brane tension and the radii of the compactified torus to unity. 
Let us consider a generating function 
\begin{align}
Z^{\textrm{semi}}(\beta)&=
\Tr e^{-\beta \hat{N}}\nonumber\\
&=
\prod_{j=1}^{8}
\prod_{(n_1,n_2)\in \mathbb{Z}^2\setminus\{(0,0)\}}
\sum_{N_{n_1,n_2}^j=0}^\infty
\langle N_{n_1,n_2}^j| e^{-\beta \omega_{n_1,n_2} \hat{N}_{n_1,n_2}^{j} } |N_{n_1,n_2}^j \rangle\nonumber\\
&=
\prod_{(n_1,n_2)\in \mathbb{Z}^2\setminus\{(0,0)\}}
\frac{1}{(1-e^{-\beta \omega_{n_1,n_2}})^{8}},
\label{semi_pfn}
\end{align}
where $\hat{N}_{n_1,n_2}^{j}$ $:=$ ${\hat{a}^{\dag j}}_{n_1,n_2}\hat{a}_{n_1,n_2}^j$ is the number operator. 

Let us consider the degeneracy $d_n$ of the states which is given by the expansion coefficients of the generating function (\ref{semi_pfn})
\begin{align}
Z^{\textrm{semi}}(q)&=\sum_{n=0}^{\infty} d_n q^n,
\end{align}
where $q=e^{-\beta}$.
To find the asymptotics of $d_n$ we can follow Meinardus' approach.
In this case, since the infinite product \eqref{semi_pfn} contains two components $n_1$ and $n_2$, the auxiliary Dirichlet series $D(s)$ \eqref{D_series} is given by the Epstein zeta function \cite{MR1511190}
\begin{align}
D(s)&=\sum_{(n_1,n_2)\in \mathbb{Z}^2 \setminus\{(0,0)\}} 
\frac{8}{(n_1^2+n_2^2)^{\frac{s}{2}}}.\label{DM2_series}
\end{align}
Since this $D(s)$ has a pole at $s=2$ of order one with residue $R_0$ $=$ $8\cdot 2\pi$, we obtain the leading asymptotic degeneracy
\begin{align}
d_n&=
\exp\left[
3 \left(\alpha \frac{\zeta(3)}{4}
\right)^{\frac13} n^{\frac{2}{3}}
+\cdots \right],
\label{M2_asymp}
\end{align}
where 
\begin{align}
\alpha&=16\pi. 
\end{align}
The asymptotic degeneracy for a $p$-brane can be evaluated in the same way as $\log d_n\sim n^{p/(p+1)}$ \cite{Fubini:1972mf, Dethlefsen:1974dr,Strumia:1975rd,Alvarez:1991qs,Harms:1992jt}.
In particular, for the fundamental strings ($p=1$) we have $\log d_n\sim n^{1/2}$ \cite{Hagedorn:1965st,Fubini:1969qb,Huang:1970iq,Frautschi:1971ij,Carlitz:1972uf}.

Note that while (\ref{M2_asymp}) exhibits the same asymptotic form of the degeneracy of M2-brane indices ($\log d_n\sim n^{\frac{2}{3}}$), 
the connection with our observations in the previous sections 
is not entirely clear yet.
In the previous sections we studied the theories of $N$ M2-branes probing various geometries in the large $N$ limit and considered the supersymmetric indices where there are cancellations of the degrees of freedom between bosonic modes and fermionic modes.
However, in this section the analysis does not focus on a stack of $N$ coincident M2-branes, 
but rather on a single M2-brane propagating a flat space.
Also it only takes into account the bosonic modes in the semiclassical approximation.
It would be interesting to find a more appropriate physical interpretation for the asymptotic degeneracy $\log d_n\sim n^{2/3}$ of the M2-branes as well as the asymptotic coefficients. 

\subsection{Thermodynamic functions}
\label{sec_thermo}
The supersymmetric indices as well as 
the function (\ref{semi_pfn}) can be viewed as certain partition functions where $\beta$ is identified with the inverse temperature. 
The partition function leads us to define the ``free energy'' $F$,  the ``entropy'' $S$ and the ``internal energy'' $U$ by
\begin{align}
F&=-\frac{1}{\beta}\log Z^{\textrm{semi}}(\beta),\label{fene} \\
S&=\beta^2\frac{\partial}{\partial \beta} F,\label{entropy} \\
U&=-\frac{\partial}{\partial \beta}\log Z^{\textrm{semi}}(\beta).\label{intene}
\end{align}
The leading coefficient $\alpha$ in $d_n$ is also associated to the ``thermodynamic functions'' in the high-temperature limit. 
As $n\rightarrow \infty$, we are led to the high-temperature limit $\beta\rightarrow 0$ or $T\rightarrow \infty$ with $\beta=1/T$. 
Given the asymptotic form
\begin{align}
d_n&\sim 
\exp\left[
\alpha n^{\frac{\delta}{\delta+1}}+\cdots
\right], 
\end{align}
the free energy (\ref{fene}), the entropy (\ref{entropy}) and the internal energy (\ref{intene}) in the high-temperature limit are given by
\begin{align}
F&\sim -\alpha T^{\delta+1},\label{fene_asym} \\
S&\sim (\delta+1)\alpha T^{\delta},\label{entropy_asym} \\
U&\sim \delta \alpha T^{\delta+1}.\label{U_asym}
\end{align}

It follows from the result (\ref{asymp_ABJMk}) that the ``free energy'' of the large $N$ ABJM theory with level $k$ in the high-temperature limit takes the form 
\begin{align}
\label{abjm_htemp_F}
F&\sim -3 \left( \frac{5\zeta(3)}{4k} \right)T^3. 
\end{align}
This has the same form as that of an ideal gas of massless particles in two dimensions. 
On the other hand, from the asymptotic form (\ref{asymp_largek}), the large $N$ and large $k$ free energy in the high-temperature limits is 
\begin{align}
\label{abjmlargek_htemp_F}
F\sim -\pi T^2, 
\end{align} 
which behaves as that of an ideal gas in one dimension. 
Accordingly, the free energies (\ref{abjm_htemp_F}) and (\ref{abjmlargek_htemp_F}) show that 
the large $N$ ABJM theory contains the characteristic 
$F\propto T^3$ for a membrane-like behavior as well as $F\propto T^2$ for a string-like behavior, as argued in \cite{Klebanov:1996ag}. 

\subsection{Future directions}
\begin{itemize}
\item
In the case of the Coulomb limit, the indices for general $N\ge 2$ and the large $N$ limit are obtained by the symmetric product (Plethystic exponential) of the index at $N=1$.
This fact can be understood as the generating function $\Xi(\mu)$ for the Coulomb index having a simple structure \eqref{XiCgeneralstructure}.
It would interesting to understand the generating function for the Higgs indices or the full indices of the same theories.
Comparing these indices for $N=1$ and those for $N>1$ we find that the grand partition function does not enjoy the simple structure \eqref{XiCgeneralstructure}.
Nevertheless, the fact that the large $N$ indices take simple expressions might suggest that the generating function for these indices also have some simple closed form expressions.
In particular for the $U(N)$ ADHM theory with $l$ flavors, a new general expression in $N$ was obtained which is written in terms of Hall-Littlewood polynomials \cite{Crew:2020psc}.
These expressions for finite $N$ might be useful in guessing the generating function of the Higgs indices.
\item
Although in this paper we have focused on the large $N$ indices with $|\mathfrak{t}|<1$ (or $|q|<1$ in section \ref{sec_ABJMfull}) and their degeneracy in large charge limit, there are various different ways to take the limit of the indices.
For example, the limit $|\mathfrak{t}|\rightarrow 1-0$ (or $|q|\rightarrow 1-0$) does not commute with the large $N$ limit \cite{Choi:2019zpz}.
It would be interesting to study the structure of the indices in such limit and their M-theoretic interpretation. 
\item 
The indices can be decorated by the BPS Wilson/vortex line operators \cite{Dimofte:2011ju,Drukker:2012sr}. 
It would be interesting to study their effect on the asymptotic growth of degeneracies. 
Such indices should be much richer as one can also take the large representation limit.
We hope to report more details in the future work. 
\item 
We are also interested in further examining the asymptotic behaviors of other large $N$ indices of 3d supersymmetric gauge theories, including the $\mathcal{N}=4$ SYM theory with other kinds of rank 2 hypers, the ABJ theories \cite{Aharony:2008gk}, $U(N)_k\times U(N)_0^{\otimes (p-1)}\times U(N)_{-k}\times U(N)_0^{\otimes (q-1)}$ circular quiver Chern-Simons matter theories \cite{Imamura:2008nn} and ${\widehat D},{\widehat E}$-type quiver Chern-Simons matter theories \cite{Gulotta:2011vp} as well as those with rank deformations.
It would also be interesting to study the asymptotic degeneracies in the three-dimensional supersymmetric gauge theories which are not necessarily associated with M2-branes, e.g.~the SQCD and $T[G]$ theory \cite{Gaiotto:2008ak}, the Gaiotto-Tomasiello theory \cite{Gaiotto:2009mv} and the Assel-Tachikawa-Tomasiello theory \cite{Assel:2022row}. 
\end{itemize}

\subsection*{Acknowledgements}
The authors would like to thank Kimyeong Lee, Hai Lin and Minwoo Suh for useful discussions and comments. 
Part of the exact expansion coefficients of the superconformal indices was obtained 
by using the high performance computing facility provided by Yukawa Institute for Theoretical Physics (Sushiki server). The work of H.H. is supported in part by JSPS KAKENHI Grant Number JP18K13543 and JP23K03396.
The work of T.O. is supported by the Startup Funding no.~4007012317 of the Southeast University.
The data used for the current study will be made available on reasonable request.


\appendix

\section{Formulas for $\zeta'(-1,a)$}
\label{app_HZformulas}
The Hurwitz zeta function $\zeta(s,a)$ is defined as
\begin{align}
&\zeta(s,a)=\sum_{n=0}^\infty\frac{1}{(n+a)^s}.
\label{HZdef}
\end{align}
Let us denote $\partial_s\zeta(s,a)$ as $\zeta'(s,a)$, which satisfies the following relations:
\begin{align}
&\zeta'(-1,1)=\zeta'(-1)=\frac{1}{12}-\log A,\label{HZrelation1} \\
&\zeta'(-1,a+1)=\zeta'(-1,a)+a\log a,\label{HZrelation2} \\
&\sum_{j=1}^k\zeta'\Bigl(-1,\frac{j}{k}\Bigr)=\frac{1}{12k}-\frac{\log k}{12k}-\frac{\log A}{k}.\label{HZrelation3}
\end{align}
Here $A$ is Glaisher-Kinkelin constant.
Note that \eqref{HZrelation3} can be obtained by decomposing the summation in \eqref{HZdef} as $\sum_{\ell=0}^\infty\sum_{m=1}^k(n=k\ell+m-1)$ and using \eqref{HZrelation1}.
From these relations we can also show the following formulas
\begin{align}
&\zeta'(-1,n)=\frac{1}{12}-\log A+\log H_{n-1},\quad (n\in\mathbb{N})\label{HZformula1} \\
&\zeta'\Bigl(-1,n+\frac{1}{2}\Bigr)=-\frac{1}{24}+\frac{1}{2}\log A+\Bigl(-n^2-\frac{n}{2}-\frac{1}{24}\Bigr)\log 2+\frac{1}{2}\log\frac{H_{2n}}{(H_{n})^2},\quad (n\in\mathbb{Z}_{\ge 0})\label{HZformula2}
\end{align}
Here $H_n$ is hyperfactorial $H_n=\prod_{k=1}^nk^k$.

For all of the supersymmetric indices considered in section \ref{sec_largeN} and in appendix \ref{app_numerical}, we observe that the sum of the derivatives of Hurwitz zeta function $\zeta'(-1,a)$ in $\gamma$ \eqref{coeff_repeat3} can be arranged, after shifting each $a$ with some integer by using \eqref{HZrelation2}, into a finite number of terms of the form \eqref{HZformula1} or \eqref{HZformula2} and the rest which can be put together into the form \eqref{HZrelation3} with some $k$.
Hence after the simplification, $\gamma$ are written without using $\zeta'(-1,a)$ explicitly.

\section{Numerical results}
\label{app_numerical}
We list more results of the explicit comparison between the actual coefficients and the asymptotic coefficients for the expansion of the large $N$ indices of the M2-brane SCFTs in this appendix. 
\subsection{Coulomb indices}

\subsubsection{$\mathbb{C}^2/\mathbb{Z}_l$}
\label{app:Coulomb_C2Zl}
For $l=5$ we find
\begin{align}
\label{U_C5_asy_table}
\begin{array}{c|c|c|c} 
n&d_n&d_n^{\textrm{asym}}&d_n/d_n^{\textrm{asym}} \\ \hline 
10   &12                &12.0738                &0.993891 \\ 
100  &2.08112\times 10^{9} &2.10340\times 10^{9} &0.989408\\ 
1000 &3.18302\times 10^{49}&3.17211\times 10^{49}&1.00344\\ 
10000&4.15842\times 10^{236}&4.14390\times 10^{236}&1.00350\\ 
\end{array}. 
\end{align}
For $l=6$ we find
\begin{align}
\label{U_C6_asy_table}
\begin{array}{c|c|c|c} 
n&d_n&d_n^{\textrm{asym}}&d_n/d_n^{\textrm{asym}} \\ \hline 
10   &15                &18.1211               &0.827763 \\ 
100  &8.99625\times 10^{8} &9.17867\times 10^{8} &0.980126\\ 
1000 &4.91979\times 10^{46}&4.92619\times 10^{46}&0.998702\\ 
10000&4.30315\times 10^{222}&4.29786\times 10^{222}&1.00123\\ 
\end{array}. 
\end{align}
For $l=7$ we get
\begin{align}
\label{U_C7_asy_table}
\begin{array}{c|c|c|c} 
n&d_n&d_n^{\textrm{asym}}&d_n/d_n^{\textrm{asym}} \\ \hline 
10   &7                &7.37615                &0.949040 \\ 
100  &1.37938\times 10^{8} &1.41822\times 10^{8} &0.972614\\ 
1000 &8.56185\times 10^{43}&8.60800\times 10^{43}&0.994639\\ 
10000&9.09503\times 10^{210}&9.10200\times 10^{210}&0.999234\\ 
\end{array}. 
\end{align}
For $l=8$ we find
\begin{align}
\label{U_C8_asy_table}
\begin{array}{c|c|c|c} 
n&d_n&d_n^{\textrm{asym}}&d_n/d_n^{\textrm{asym}} \\ \hline 
10   &11                &12.6747               &0.867874 \\ 
100  &1.07451\times 10^{8} &1.11192\times 10^{8} &0.966353\\ 
1000 &1.68438\times 10^{42}&1.69954\times 10^{42}&0.991076\\ 
10000&8.00790\times 10^{201}&8.02840\times 10^{201}&0.997447\\
\end{array}. 
\end{align}
For $l=9$ we obtain
\begin{align}
\label{U_C9_asy_table}
\begin{array}{c|c|c|c} 
n&d_n&d_n^{\textrm{asym}}&d_n/d_n^{\textrm{asym}} \\ \hline 
10   &7                &5.65292               &1.23830 \\ 
100  &2.48563\times 10^{7} &2.58611\times 10^{7} &0.961146\\ 
1000 &1.76155\times 10^{40}&1.78313\times 10^{40}&0.987898\\ 
10000&5.20003\times 10^{193}&5.22185\times 10^{193}&0.995820\\ 
\end{array}. 
\end{align}

\subsubsection{$\mathbb{C}^2/\widehat{D}_l$}
\label{app:Coulomb_C2Dl}
For $l=4$ we have
\begin{align}
\label{OSp_C4_asy_table}
\begin{array}{c|c|c|c} 
n&d_n&d_n^{\textrm{asym}}&d_n/d_n^{\textrm{asym}} \\ \hline 
10&1&2.80528&0.356471\\ 
100&501813&515396&0.973645\\ 
1000&3.40006\times 10^{32}&3.42707\times 10^{32}&0.992120\\
10000&6.66171\times 10^{158}&6.67079\times 10^{158}&0.99864\\ 
\end{array}. 
\end{align}
For $l=5$ we have
\begin{align}
\label{OSp_C5_asy_table}
\begin{array}{c|c|c|c} 
n&d_n&d_n^{\textrm{asym}}&d_n/d_n^{\textrm{asym}} \\ \hline 
10   &1                &2.22957                &0.448517 \\ 
100  &150026 &154982 &0.968019\\ 
1000 &1.05908 \times 10^{30}&1.07039\times 10^{30}&0.989434\\ 
10000&1.47757\times 10^{147}&1.48159\times 10^{147}&0.997292\\ 
\end{array}. 
\end{align}
For $l=6$ we have
\begin{align}
\label{OSp_C6_asy_table}
\begin{array}{c|c|c|c} 
n&d_n&d_n^{\textrm{asym}}&d_n/d_n^{\textrm{asym}} \\ \hline 
12   &4                &2.58351                &1.54828 \\ 
100  &64132&64412.0 &0.995653\\ 
1000 &1.3736 \times 10^{28}&1.39125\times 10^{28}&0.987309\\ 
10000&2.04534\times 10^{138}&2.05323\times 10^{138}&0.996158\\ 
\end{array}. 
\end{align}
For $l=7$ we obtain
\begin{align}
\label{OSp_C7_asy_table}
\begin{array}{c|c|c|c} 
n&d_n&d_n^{\textrm{asym}}&d_n/d_n^{\textrm{asym}} \\ \hline 
12   &3                &2.27679               &1.31765\\ 
100  &32564 &32906.7  &0.989585\\ 
1000 &4.46825 \times 10^{26}&4.53362\times 10^{30}&0.985581\\ 
10000&1.81995\times 10^{131}&1.82877\times 10^{131}&0.995175\\ 
\end{array}. 
\end{align}
For $l=8$ we obtain
\begin{align}
\begin{array}{c|c|c|c} 
n&d_n&d_n^{\textrm{asym}}&d_n/d_n^{\textrm{asym}} \\ \hline 
12   &3                &2.0767               &1.44460\\ 
100  &20196 &19359.7  &1.04320\\ 
1000 &2.74419 \times 10^{25}&2.78839\times 10^{25}&0.984148\\ 
10000&2.90303\times 10^{125}&2.91966\times 10^{125}&0.994304\\ 
\end{array}. 
\end{align}
For $l=9$ we obtain
\begin{align}
\begin{array}{c|c|c|c} 
n&d_n&d_n^{\textrm{asym}}&d_n/d_n^{\textrm{asym}} \\ \hline 
12   &3                &1.93986&1.54651\\ 
100  &13043&12606.3&1.03464\\ 
1000 &2.67459\times 10^{24}&2.72101\times 10^{24}&0.982940\\ 
10000&3.79548\times 10^{120}&3.82024\times 10^{120}&0.993519\\
\end{array}. 
\end{align}
\subsection{Higgs indices}

\subsubsection{$U(N)$ ADHM theories}
\label{app:Higgs_C2Zl}
For $l=4$ we have 
\begin{align}
\label{U_H_asy_table4}
\begin{array}{c|c|c|c} 
n&d_n&d_n^{\textrm{asym}}&d_n/d_n^{\textrm{asym}} \\ \hline 
10   &641477                & 131358                & 4.88343\\ 
100  &3.10318\times 10^{36} & 1.38562\times 10^{36} & 2.23957\\ 
1000 &1.53265\times 10^{194}& 1.03606\times 10^{194}& 1.47930\\ 
10000&8.24259\times 10^{961}& 6.84520\times 10^{961}& 1.20414\\ 
20000&1.73806\times 10^{1545}&1.49892\times 10^{1545}& 1.15955\\
\end{array}. 
\end{align}
For $l=5$ we find
\begin{align}
\label{U_H_asy_table5}
\begin{array}{c|c|c|c} 
n&d_n&d_n^{\textrm{asym}}&d_n/d_n^{\textrm{asym}} \\ \hline 
10   &2.61121\times 10^{6}   &103195.                &25.3036\\
100  &8.02835\times 10^{40}  &1.53109\times 10^{40}  &5.24355\\
1000 &5.40240\times 10^{220} &2.40389\times 10^{220} &2.24736\\
10000&6.97847\times 10^{1103}&4.74614\times 10^{1103}&1.47035\\
20000&6.26202\times 10^{1776}&4.60486\times 10^{1776}&1.35987\\
\end{array}.
\end{align}
For $l=6$ we find
\begin{align}
\label{U_H_asy_table6}
\begin{array}{c|c|c|c} 
n&d_n&d_n^{\textrm{asym}}&d_n/d_n^{\textrm{asym}} \\ \hline 
10&8.96894\times 10^6&36001.1&249.129\\
100&7.01107\times 10^{44}&4.03149\times 10^{43}&17.3908\\
1000&7.35453\times 10^{244}&1.80630\times 10^{244}&4.07160\\
10000&3.18822\times 10^{1234}&1.63047\times 10^{1234}&1.95540\\
20000&3.54996\times 10^{1990}&2.07899\times 10^{1900}&1.70755\\
\end{array}.
\end{align}
For $l=7$ we find
\begin{align}
\label{U_H_asy_table7}
\begin{array}{c|c|c|c} 
n&d_n&d_n^{\textrm{asym}}&d_n/d_n^{\textrm{asym}} \\ \hline 
10   &2.69447\times 10^{7}   &5480.10                &4916.82\\
100  &2.59595\times 10^{48}  &3.06942\times 10^{46}  &84.5746\\
1000 &1.22359\times 10^{267} &1.36426\times 10^{266} &8.96888\\
10000&1.76585\times 10^{1356}&6.17343\times 10^{1355}&2.86041\\
20000&1.70517\times 10^{2190}&7.36856\times 10^{2189}&2.31412\\
\end{array}.
\end{align}
For $l=8$ we find
\begin{align}
\label{U_H_asy_table8}
\begin{array}{c|c|c|c} 
n&d_n&d_n^{\textrm{asym}}&d_n/d_n^{\textrm{asym}} \\ \hline 
10   &7.26789\times 10^{7}   &355.696                &204329.\\
100  &4.78465\times 10^{51}  &7.68166\times 10^{48}  &622.866\\
1000 &5.52213\times 10^{287} &2.25715\times 10^{286} &24.4651\\
10000&4.79100\times 10^{1470}&1.03117\times 10^{1470}&4.64620\\
20000&2.41241\times 10^{2378}&7.07277\times 10^{2377}&3.41084\\
\end{array}.
\end{align}

\subsection{ABJM full indices}
\label{app_ABJMfull}
For $k=5$ the large $N$ index is given by \eqref{inf_fullk4l+1} with $k=5$,
\begin{align}
I^{\text{ABJM}_5}_\infty(q)&=\frac{1}{
(1-q^{5n-\frac{19}{4}})^{12n-12}
(1-q^{5n-\frac{9}{2}})^{12n-8}
(1-q^{5n-\frac{17}{4}})^{12n-12}
(1-q^{5n-4})^{12n-10}
}\nonumber \\
&\quad \times \frac{1}{
(1-q^{5n-\frac{15}{4}})^{52n-40}
(1-q^{5n-\frac{7}{2}})^{12n-8}
(1-q^{5n-\frac{13}{4}})^{12n-4}
(1-q^{5n-3})^{12n-10}
}\nonumber \\
&\quad \times \frac{1}{
(1-q^{5n-\frac{11}{4}})^{12n-8}
(1-q^{5n-\frac{5}{2}})^{52n-26}
(1-q^{5n-\frac{9}{4}})^{12n-4}
(1-q^{5n-2})^{12n-2}
}\nonumber \\
&\quad \times \frac{1}{
(1-q^{5n-\frac{7}{4}})^{12n-8}
(1-q^{5n-\frac{3}{2}})^{12n-4}
(1-q^{5n-\frac{5}{4}})^{52n-12}
(1-q^{5n-1})^{12n-2}
}\nonumber \\
&\quad \times \frac{1}{
(1-q^{5n-\frac{3}{4}})^{12n}
(1-q^{5n-\frac{1}{2}})^{12n-4}
(1-q^{5n-\frac{1}{4}})^{12n}
(1-q^{5n})^{52n}
}.\label{inf_fullk5}
\end{align}
The application of the Meinardus theorem gives
\begin{align}
\label{asymp_ABJM5}
&I^{\text{ABJM}_5}_\infty(q)=\sum_{n\ge 0}d_nq^{\frac{n}{4}},\nonumber \\
&d_n\sim 
\frac{
2^{\frac{559}{900}} 
5^{\frac{8}{15}}\zeta(3)^{\frac{13}{36}}
}
{
3^{\frac12} \pi^{\frac12} A^{\frac{53}{5}} n^{\frac{31}{36}}
}
\exp\left[
3\left( \frac{5 \zeta(3)}{4\cdot 5} \right)^{\frac13}n^{\frac23}
+\frac{53}{60}
\right]
\nonumber\\
&\quad\quad \times 
\left(
\frac{\Gamma\left(\frac{3}{20} \right)}{\Gamma\left(\frac{17}{20} \right)}
\right)^{\frac{1}{5}}
\left(
\frac{\Gamma\left(\frac{1}{5} \right)}{\Gamma\left(\frac{4}{5} \right)}
\right)^{\frac{16}{5}}
\left(
\frac{\Gamma\left(\frac{7}{20} \right)}{\Gamma\left(\frac{13}{20} \right)}
\right)^{\frac{9}{5}}
\left(
\frac{\Gamma\left(\frac{11}{20} \right)}{\Gamma\left(\frac{9}{20} \right)}
\right)
\left(
\frac{\Gamma\left(\frac{7}{10} \right)}{\Gamma\left(\frac{3}{10} \right)}
\right)^{2}
\left(
\frac{\Gamma\left(\frac{3}{4} \right)}{\Gamma\left(\frac{1}{4} \right)}
\right)
\left(
\frac{\Gamma\left(\frac{19}{20} \right)}{\Gamma\left(\frac{1}{20} \right)}
\right). 
\end{align}
This agrees well with the actual coefficients of \eqref{inf_fullk5}:
\begin{align}
\label{ABJMk5_asy_table}
\begin{array}{c|c|c|c} 
n&d_n&d_n^{\textrm{asym}}&d_n/d_n^{\textrm{asym}} \\ \hline 
10&268&319.191&0.839624 \\
100&2.41386\times 10^{16} &2.47643\times 10^{16}&0.974731 \\
1000&9.95106 \times 10^{83} &1.00062\times 10^{84}&0.994487 \\
10000&8.64731\times 10^{400}&8.65762\times 10^{400}&0.998809 \\
\end{array}.
\end{align}
For $k=6$ the large $N$ index is given by \eqref{inf_fullk4l-2} with $k=6$,
\begin{align}
I^{\text{ABJM}_6}_\infty(q)
&=
\prod_{n=1}^{\infty}
\frac{1}{(1-q^{\frac32 n-1})^{4n} (1-q^{\frac32 n-\frac12})^{2n} (1-q^{\frac32 n})^{18n} (1-q^{\frac32 n+\frac12})^{2n} (1-q^{\frac32 n+1})^{4n}}.\label{inf_fullk6} 
\end{align}
The application of the Meinardus theorem gives
\begin{align}
\label{asymp_ABJM6}
&I^{\text{ABJM}_6}_\infty(q)=\sum_{n\in 2\mathbb{Z}_{\ge 0}}d_nq^{\frac{n}{4}},\nonumber \\
&d_n\sim 
\frac{(5\zeta(3))^{\frac13} }{2^{\frac13} 3^{\frac12} \pi^{\frac32} A^{14} n^{\frac{5}{6}}}
\Gamma\left(
\frac16
\right)^2
\exp
\left[
3\left(\frac{5\zeta(3)}{4\cdot 6} \right)^{\frac13}
n^{\frac23}+\frac76
\right].
\end{align}
This agrees well with the actual coefficients obtained from \eqref{inf_fullk6}:
\begin{align}
\label{abjmk6_asy_table}
\begin{array}{c|c|c|c} 
n&d_n&d_n^{\textrm{asym}}&d_n/d_n^{\textrm{asym}} \\ \hline 
10   &372   &435.366  &0.54454 \\
100  &4.73244 \times 10^{15}  &4.86001 \times 10^{15}  &0.973751\\
1000 &1.91090\times 10^{79} &1.92194 \times 10^{79} &0.994257\\
10000&3.23192\times 10^{377}&3.23594 \times 10^{377}&0.998759\\
\end{array}.
\end{align}
For $k=7$ the large $N$ index is given by \eqref{inf_fullk4l-1} with $k=7$,
\begin{align}
I^{\text{ABJM}_7}_\infty(q)
&=\frac{1}{
(1-q^{7n-\frac{27}{4}})^{12n-12}
(1-q^{7n-\frac{13}{2}})^{12n-8}
(1-q^{7n-\frac{25}{4}})^{12n-12}
(1-q^{7n-6})^{12n-10}
}\nonumber \\
&\quad \times \frac{1}{
(1-q^{7n-\frac{23}{4}})^{12n-12}
(1-q^{7n-\frac{11}{2}})^{12n-8}
(1-q^{7n-\frac{21}{4}})^{68n-52}
(1-q^{7n-5})^{12n-10}
}\nonumber \\
&\quad \times \frac{1}{
(1-q^{7n-\frac{19}{4}})^{12n-4}
(1-q^{7n-\frac{9}{2}})^{12n-8}
(1-q^{7n-\frac{17}{4}})^{12n-8}
(1-q^{7n-4})^{12n-10}
}\nonumber \\
&\quad \times \frac{1}{
(1-q^{7n-\frac{15}{4}})^{12n-4}
(1-q^{7n-\frac{7}{2}})^{68n-34}
(1-q^{7n-\frac{13}{4}})^{12n-8}
(1-q^{7n-3})^{12n-2}
}\nonumber \\
&\quad \times \frac{1}{
(1-q^{7n-\frac{11}{4}})^{12n-4}
(1-q^{7n-\frac{5}{2}})^{12n-4}
(1-q^{7n-\frac{9}{4}})^{12n-8}
(1-q^{7n-2})^{12n-2}
}\nonumber \\
&\quad \times \frac{1}{
(1-q^{7n-\frac{7}{4}})^{68n-16}
(1-q^{7n-\frac{3}{2}})^{12n-4}
(1-q^{7n-\frac{5}{4}})^{12n}
(1-q^{7n-1})^{12n-2}
}\nonumber \\
&\quad \times \frac{1}{
(1-q^{7n-\frac{3}{4}})^{12n}
(1-q^{7n-\frac{1}{2}})^{12n-4}
(1-q^{7n-\frac{1}{4}})^{12n}
(1-q^{7n})^{68n}
}.\label{inf_fullk7}
\end{align}
The application of the Meinardus theorem gives
\begin{align}
\label{asymp_ABJM7}
&I^{\text{ABJM}_7}_\infty(q)=\sum_{n\ge 0}d_nq^{\frac{n}{4}},\nonumber \\
&d_n\sim 
\frac{
2^{\frac{3293}{1764}} 
(5\zeta(3))^{\frac{11}{36}} 7^{\frac{19}{252}} \pi^{\frac{25}{14}}
}
{
3^{\frac12} A^{\frac{101}{7}} n^{\frac{29}{36}}
}
\exp\left[
3\left( \frac{5 \zeta(3)}{4\cdot 7} \right)^{\frac13}n^{\frac23}
+\frac{101}{84}
\right]
\nonumber\\
&\quad\quad \times 
\Gamma \left(\frac{1}{7}\right)^{\frac{22}{7}} \Gamma \left(\frac{9}{28}\right)^{\frac{29}{7}}
\Gamma \left(\frac{13}{28}\right)^{\frac{17}{7}} \Gamma \left(\frac{17}{28}\right)^{\frac{5}{7}}
\Gamma \left(\frac{3}{4}\right) \Gamma \left(\frac{23}{28}\right)^{15/7} \Gamma 
\left(\frac{25}{28}\right)^{\frac{9}{7}} \Gamma \left(\frac{27}{28}\right)^{\frac{3}{7}}
\nonumber\\
&\quad\quad \times 
\frac{1}{
\Gamma \left(\frac{1}{28}\right)^{\frac{3}{7}}
\Gamma \left(\frac{3}{28}\right)^{\frac{9}{7}}
\Gamma\left(\frac{5}{28}\right)^{\frac{15}{7}}
\Gamma \left(\frac{1}{4}\right)
\Gamma \left(\frac{2}{7}\right)^{\frac{8}{7}}
\Gamma \left(\frac{11}{28}\right)^{\frac{5}{7}}
}
\nonumber\\
&\quad\quad \times 
\frac{1}{
\Gamma \left(\frac{3}{7}\right)^{\frac{12}{7}}
\Gamma \left(\frac{15}{28}\right)^{\frac{17}{7}}
\Gamma \left(\frac{19}{28}\right)^{\frac{29}{7}}
\Gamma \left(\frac{5}{7}\right)^{\frac{2}{7}}
\Gamma \left(\frac{11}{14}\right)^{\frac{10}{7}}
\Gamma \left(\frac{13}{14}\right)^{\frac{22}{7}}
}.
\end{align}
This agrees with the actual coefficients of \eqref{inf_fullk7}:
\begin{align}
\label{ABJMk7_asy_table}
\begin{array}{c|c|c|c} 
n&d_n&d_n^{\textrm{asym}}&d_n/d_n^{\textrm{asym}} \\ \hline 
10&168&166.707&1.00775 \\
100&3.89220\times 10^{14}&4.00193\times 10^{14} &0.972579\\
1000&9.94215\times 10^{74} &1.00024\times 10^{75}&0.993977\\
10000&1.88344\times 10^{358}&1.88590\times 10^{358}&0.998697 \\
\end{array}.
\end{align}
For $k=8$ the large $N$ index is given by \eqref{inf_fullk4l} with $k=8$,
\begin{align}
    I^{\text{ABJM}_8}_\infty(q)
&=
\prod_{n=1}^{\infty}
\frac{1}{(1-q^{2n-\frac32})^{4n} (1-q^{2n-1})^{2n} (1-q^{2n-\frac12})^{4n} }
\nonumber\\
&\quad \times 
\frac{1}{(1-q^{2n})^{20n} (1-q^{2n+\frac12})^{4n} (1-q^{2n+1})^{2n} (1-q^{2n+\frac32})^{4n}}.\label{inf_fullk8}
\end{align}
The application of the Meinardus theorem gives
\begin{align}
\label{asymp_ABJM8}
I^{\text{ABJM}_8}_\infty(q)=\sum_{n\in 2\mathbb{Z}_{\ge 0}}d_nq^{\frac{n}{4}},\quad
d_n\sim 
\frac{
2^{\frac79} (5\zeta(3))^{\frac{5}{18}} }{3^{\frac12} \pi^{\frac12} A^{16} n^{\frac{7}{9}}}
\frac{
\Gamma\left(\frac14 \right)^{2}
}
{
\Gamma\left(\frac34 \right)^{2}
}
\exp
\left[
3\left(\frac{5\zeta(3)}{4\cdot 8} \right)^{\frac13}
n^{\frac23}+\frac43
\right].
\end{align}
This agrees with the actual coefficients of \eqref{inf_fullk8}:
\begin{align}
\label{abjmk8_asy_table}
\begin{array}{c|c|c|c} 
n&d_n&d_n^{\textrm{asym}}&d_n/d_n^{\textrm{asym}} \\ \hline 
10   &248   &276.432  &0.897146\\
100  &1.83189 \times 10^{14}  &1.88644 \times 10^{14}  &0.971085\\
1000 &1.08150\times 10^{72} &1.08843 \times 10^{72} &0.993629\\
10000&9.37769\times 10^{342} &9.39065 \times 10^{342}&0.998621\\
\end{array}.
\end{align}


\bibliographystyle{utphys}
\bibliography{ref}
\end{document}